%% file: MTEmultiple.tex
\documentclass[11pt]{article}
\usepackage{amsfonts,amsmath,amssymb,amsthm}
\allowdisplaybreaks
\usepackage[pdftex]{graphicx}
\usepackage[hmargin=1in,vmargin=1in]{geometry}
\usepackage{natbib}
\usepackage[colorlinks,citecolor=blue]{hyperref}
\usepackage{enumerate}
\usepackage{caption}
\usepackage[usenames,dvipsnames]{xcolor}
\usepackage{footmisc,fancyvrb}
\usepackage{xfrac}
\usepackage{cleveref}
\usepackage{multirow}
\usepackage{threeparttable}
\usepackage{mathtools}
\usepackage{array}
\newcommand{\PreserveBackslash}[1]{\let\temp=\\#1\let\\=\temp}
\newcolumntype{C}[1]{>{\PreserveBackslash\centering}p{#1}}
\newcolumntype{R}[1]{>{\PreserveBackslash\raggedleft}p{#1}}
\newcolumntype{L}[1]{>{\PreserveBackslash\raggedright}p{#1}}
\usepackage{standalone}
\usepackage{booktabs}
\usepackage{soul}
\setuldepth{Berlin}
\usepackage[position=below]{subcaption}
\usepackage{float}
\usepackage{pdflscape}
\usepackage{afterpage}
\usepackage{rotating}
\usepackage{tabularx}

\VerbatimFootnotes
\def\qed{\rule{2mm}{2mm}}

\DeclareMathOperator*{\argmax}{arg\,max}

\newcommand\posscite[1]{\citeauthor{#1}' (\citeyear{#1})}

\usepackage{tikz,graphicx,pgfplots,pgfkeys,subcaption}   

\def\addlegendimage{\csname pgfplots@addlegendimage\endcsname}
\usetikzlibrary{decorations.pathreplacing,angles,quotes}

\usetikzlibrary{external}
\usepackage[pagewise,mathlines]{lineno}
\mathchardef\dash="2D

\theoremstyle{definition}

\newtheorem{proposition}{Proposition}

\newtheorem{assump}{Assumption}
\newenvironment{assumption2}[2][]
{\renewcommand\theassump{#2}\begin{assump}[#1]}
{\end{assump}}
\newtheorem{defi}{Definition}
\newenvironment{definition2}[2][]
{\renewcommand\thedefi{#2}\begin{defi}[#1]}
{\end{defi}}

\usepackage{etoolbox} 
\AtEndEnvironment{remark}{~\qed}%
\AtEndEnvironment{example}{~\qed}%

\usepackage{thmtools}
\newtheorem{example}{Example}
\renewcommand\thmcontinues[1]{Continued}

\begin{document}

\author{
Vishal Kamat \\
School of Economics and Finance\\
Queen Mary University of London\\
\url{v.kamat@qmul.ac.uk}
\and
Samuel Norris\\
Department of Economics\\
University of British Columbia\\
\url{sam.norris@ubc.ca}
\and
Matthew Pecenco\\
Department of Economics\\
Brown University\\
\url{matthew_pecenco@brown.edu} \vspace{0.5cm}
}

\title{\vspace{-1cm}Identification in Multiple Treatment Models under Discrete Variation\thanks{We thank Toru Kitagawa, Soonwoo Kwon, Max Tabord-Meehan, and participants at several seminars and conferences for useful comments. First version: arXiv:2307.06174 dated July 13, 2023.}}

\maketitle

\vspace{-0.5cm}
\begin{abstract}
We develop a marginal treatment effect based method to learn about causal effects in multiple treatment models with discrete instruments. We allow selection into treatment to be governed by a general class of threshold crossing models that permit multidimensional unobserved heterogeneity. An inherent complication is that the primitives characterizing the selection model are not generally point-identified. Allowing these primitives to be point-identified up to a finite-dimensional parameter, we show how a two-step computational program can be used to obtain sharp bounds for a number of treatment effect parameters when the marginal treatment response functions are allowed to satisfy only nonparametric shape restrictions or are additionally parameterized. We demonstrate the benefits of our method by revisiting \posscite{kline/walters:16} empirical analysis of the Head Start program. Our approach relaxes their point-identifying assumptions on the selection model and marginal treatment response functions, allowing us to assess the robustness of their conclusions.
\end{abstract}

\thispagestyle{empty}



\newpage
\setcounter{page}{1}

\section{Introduction}\label{sec:introduction}

In the analysis of treatment effects using instrumental variables, a threshold crossing selection model with a single dimension of unobserved heterogeneity underlies the predominant framework in the literature \citep{imbens/angrist:94, heckman/vytlacil:05, vytlacil:02}. Designed for the canonical binary treatment setup, it presumes that individuals choose between only two treatments. In many scenarios, however, individuals choose between multiple treatments. Such cases naturally give rise to selection models with multiple dimensions of unobserved heterogeneity \citep{heckman/etal:06,heckman/etal:08,lee/salanie:18}.

In this paper, we develop a method to learn about treatment effects in multiple treatment setups under a general class of selection models with multidimensional unobserved heterogeneity. Specifically, we consider the class of multidimensional threshold crossing models from \cite{lee/salanie:18}. As highlighted in Section \ref{sec:selection}, this class nests the standard binary treatment model and also covers many empirically relevant settings with multidimensional unobserved heterogeneity, including multinomial, sequential, multi-stage, and double-hurdle choice models.

In this class of models, our objective is to show how to learn about various treatment effects that can be written as weighted averages of the marginal treatment response functions (MTRs) \citep{heckman/vytlacil:99,heckman/vytlacil:05,mogstad/etal:18}, which capture the mean potential outcomes conditional on the unobserved heterogeneity governing the selection model. As in the standard binary treatment case, MTRs provide a unified framework to express a number of parameters of interest such as the average treatment effect as well as other averages that target policy relevant subgroups.

To analogously learn about such parameters, \cite{lee/salanie:18} assume availability of data with continuous variation in the instrument. They show how a sufficient amount of such variation allows nonparametric identification of the selection model and the MTRs over their entire support, and in turn all the parameters of interest---see also \cite{heckman/etal:06, heckman/etal:08}, \cite{mountjoy:22} and \cite{tsuda:23} for related arguments in specific models. However, in many empirical scenarios, instruments are only discrete-valued. In such cases, the parameters are usually partially identified as there generally exist multiple admissible values of the MTRs and primitives characterizing the selection model that are consistent with the data.

We show how to compute the identified set for our parameters of interest in these cases. In particular, we build on \cite{mogstad/etal:18}, who study this identification problem in the special case of a binary treatment, where a single dimension of unobserved heterogeneity underlies selection. In this case, the primitives characterizing the selection model are point identified and the problem is linear in the remaining primitives, namely the MTRs. Imposing a linear parametrization on the MTRs, \cite{mogstad/etal:18} show how these features allow computing sharp bounds by solving two linear programs. They also show that with carefully chosen basis functions over a unidimensional partition, such parametrizations can accommodate fully nonparametric MTRs satisfying only shape restrictions. However, two inherent challenges arise when multidimensional unobserved heterogeneity underlies selection: the primitives characterizing the selection model are not point-identified, and the MTRs are no longer unidimensional.

We propose a two-step solution that accounts for these complications. In an inner step, for a given selection primitive, we first exploit the fact that the problem remains linear in the MTRs. Similarly linearly parametrizing the MTRs, we show this implies that we can continue to use linear programs to compute sharp bounds given the selection primitives. Moreover, we introduce novel arguments to show that such a result can also continue to accommodate entirely nonparametric MTRs. Specifically, as expanded in Section \ref{sec:identifiedset_NP}, we show how to leverage the rectangular structure underlying our general selection model and construct a new partition that generalizes the unidimensional one from \cite{mogstad/etal:18} to the multidimensional case.

In an outer step, we then introduce alternative arguments to account for the fact that the selection model is not point-identified. Specifically, this step takes the unions of the first step bounds across admissible values of the selection primitives. To do so feasibly, we assume that these primitives can be point identified up to a finite dimensional parameter. We demonstrate in several empirically relevant models how this assumption can be satisfied under familiar parametric assumptions on the distribution of the unobserved heterogeneity. In summary, our method translates to solving two linear programs to compute the identified set at the point identified value of the selection model across various values of the finite dimensional parameter, and then taking the union of these sets. We show how this two-step characterization also translates into tractable two-step strategies to estimate the identified set and construct confidence intervals.

Our method offers several advantages over recent approaches to point identification in settings with multiple treatments and discrete instruments. A common method is to impose monotonicity assumptions that restrict selection patterns such that the researcher can nonparametrically point-identify treatment effects for certain response types \citep[e.g.,][]{angrist/imbens:95,bhuller/sigstad:24,goff:25,heckman/pinto:18,kline/walters:16, kirkeboen/etal:16,lee/salanie:23,navjeevan/etaL:23,pinto:19}. As we highlight in Section \ref{sec:selection}, many such monotonicity assumptions can be captured by selection models in our general class. In such cases, although our method parametrizes the selection model, sufficiently flexible specifications can exactly replicate the nonparametric estimates these approaches produce.\footnote{\cite{kline/walters:19} and \cite{mogstad/etal:18} similarly note in the binary treatment case that even if the selection model and MTRs are parametrized, one can produce the nonparametric point identified local average treatment effect.} But our method also importantly extends beyond these settings: it allows researchers to learn about effects that are only partially identified and to accommodate richer selection patterns that would otherwise preclude point identification.

A smaller group of papers similarly considers selection models to learn about additional treatment effects, but limits attention to specific models and parameterizations that are point identified by the data \citep[e.g.,][]{hull:20, kline/walters:16,pinto:19}. Our method contributes to these papers by nesting such point identification as a special case while allowing more robust parameterizations of the selection model and MTRs that generally imply bounds on the treatment effects of interest. Moreover, our analysis applies to a general class of selection models. Our method therefore provides a blueprint for robustly learning about treatment effects across a wide range of empirical settings.

Our method also has advantages relative to a complementary literature on partial identification in multiple treatment models. Papers in this literature either make no assumptions on selection \citep[e.g.,][]{manski:97,manski/pepper:00,manski/pepper:09} or impose only nonparametric restrictions \citep[e.g.,][]{bai/etal:2025,kamat:21,lee/salanie:23}. However, because these approaches do not model the MTRs, they cannot accommodate parameters such as policy-relevant treatment effects or assumptions such as parametric restrictions and separability that are common in empirical work. Our method allows for such parameters and assumptions as well as flexible variants that enable researchers to assess the robustness of their conclusions.

To concretely demonstrate these various benefits, we use our method to revisit \posscite{kline/walters:16} empirical analysis of the Head Start preschool program using the Head Start Impact Study. Due to oversubscription, Head Start offers were randomized to applicants in the study, which enabled an experimental evaluation of Head Start by comparing outcomes of children with offers to those without. However, what these experimental effects can identify on the effects of Head Start and their policy relevance are clouded by the fact that many children enroll in competing non-Head Start preschools that provide similar services to Head Start and are also publicly subsidized. To account for this complication, \cite{kline/walters:16} formulate a multinomial selection model of preschool choice and slot provision. But, in estimating this model, they impose two simplifying assumptions to ensure that both the selection model and MTRs are point-identified: competing preschools do not ration slots, and the MTRs are linear and separable.

We illustrate how our method can be used to gradually weaken these assumptions in a step-by-step manner and analyze the sensitivity of their main empirical conclusions to them. Our bounds reveal that their first conclusion that the positive experimental effects are driven by a positive effect from families who enroll in no preschools absent a Head Start offer is robust to these assumptions. In particular, this conclusion can hold under nonparametric and nonseparable MTRs, and when the selection model allows rationing of competing preschools. In contrast, for their second conclusion that marginally expanding Head Start has a positive, statistically significant effect, we find their assumptions are generally necessary unless one is willing to take the relation between test score and earnings to be towards the higher end of values reported in the literature.

The remainder of the paper is organized as follows. Section \ref{sec:setup} introduces the class of models and treatment effect parameters we analyze. Section \ref{sec:identification} presents our arguments to compute the identified set. Section \ref{sec:estimation_inference} discusses how these arguments can be used to estimate the identified set and construct confidence intervals. Section \ref{sec:application} presents our empirical application. Section \ref{sec:conclusion} concludes. Proofs of all results are presented in Appendix \ref{Asec:proofs}.

\section{Setup}\label{sec:setup}

\subsection{Observed and Potential Variables}\label{sec:structure}

For each individual, we observe an outcome of interest $Y$, their received treatment $D$, their assigned instrument value $Z$, and baseline covariates $X$. We take the list of possible treatments to be given by a discrete set $\mathcal{D}$, and model the support of the instrument and covariates denoted by $\mathcal{Z}$ and $\mathcal{X}$ to be discrete. We assume the outcome and treatment to be generated by the usual potential outcomes structure. In particular, denoting by $D(z)$ the potential treatment had the individual's assigned instrument value been $z \in \mathcal{Z}$, the observed treatment is given by
\begin{align}\label{eq:D_equation}
D = \sum_{z \in \mathcal{Z}} D(z)1\{Z=z\}~,
\end{align}
and, similarly, denoting by $Y(d)$ the potential outcome had the individual's received treatment been $d \in \mathcal{D}$, the observed outcome is given by
\begin{align}\label{eq:Y_equation}
Y = \sum_{d \in \mathcal{D}} Y(d)1\{D=d\}~.
\end{align}
Moreover, as usual, we assume that the instrument is statistically independent of the potential variables conditional on the covariates as follows.

\begin{assumption2}{E}{(Exogeneity)}\label{ass:E}
$(\{Y(d) : d \in \mathcal{D}\}, \{D(z) : z \in \mathcal{Z}\}) \perp Z | X~.$
\end{assumption2}

\subsection{Selection Model}\label{sec:selection}

We assume the potential treatments to be related across instrument values through a selection model. Let $U \equiv (U_1, \ldots, U_J)$ denote a vector of individual unobservables governing selection. Following  \citet[][Assumption 2.1]{lee/salanie:18}, we take potential treatment to be determined by the region in which $U$ falls, where these regions satisfy certain properties in addition to being disjoint to logically ensure that they cannot imply several different treatments.

\begin{assumption2}{SM}{(Selection Model)}\label{ass:SM}
Conditional on $X = x \in \mathcal{X}$, let $U$ be continuously distributed with support contained in $[0,1]^J$ with uniform marginal distributions, and let
\begin{align}\label{eq:SM}
  D(z) = \sum_{d \in \mathcal{D}} d 1\{U \in \mathcal{U}^{\text{sm}}_{d,z|x} \}
\end{align}
for $z \in \mathcal{Z}$, where $\{\mathcal{U}^{\text{sm}}_{d,z|x} : d \in \mathcal{D}\}$ denotes a collection of disjoint subsets of $[0,1]^J$ that are members of the $\sigma$-field generated by the sets $\{\{u \in [0,1]^J : u_j \leq g_{j,z|x}\} : j =1,\ldots,J\}$ for some unknown threshold values $\{g_{j,z|x} : j = 1, \ldots, J\}$.
\end{assumption2}

\noindent As the generated $\sigma$-field is obtained by taking unions, intersections and complements of the sets in $\{\{u \in [0,1]^J : u_j \leq g_{j,z|x}\} : j =1,\ldots,J\}$, Assumption \ref{ass:SM} requires that each of the regions $\mathcal{U}^{\text{sm}}_{d,z|x}$ determining potential treatment can be written as
\begin{align}\label{eq:rectangle1}
\mathcal{U}^{\text{sm}}_{d,z|x} = \bigcup_{l=1}^{L_{d}} \prod_{j=1}^J[\underline{u}_{j,l,d,z|x},\bar{u}_{j,l,d,z|x}]~,
\end{align}
for some finite $L_{d}$ and $\underline{u}_{j,l,d,z|x},\bar{u}_{j,l,d,z|x} \in \{0,g_{j,z|x},1\}$, i.e. as a finite union of rectangular sets. As usual, the restriction in Assumption \ref{ass:SM} that $U$ has support contained in $[0,1]^J$ with uniform marginal distributions can be viewed as a normalization. To see this, let $\tilde U = (\tilde U_1,\dots,\tilde U_J)$ denote a vector of latent unobservables with continuous marginal distribution functions $\tilde F_{j\mid x}$ given $X=x$. For any thresholds $\underline{\tilde u}_{j,l,d,z\mid x}$ and $\bar{\tilde u}_{j,l,d,z\mid x}$,
\[
1\{\underline{\tilde u}_{j,l,d,z\mid x} \le \tilde U_j \le \bar{\tilde u}_{j,l,d,z\mid x}\}
=
1\{\tilde F_{j\mid x}(\underline{\tilde u}_{j,l,d,z\mid x}) \le
      \tilde F_{j\mid x}(\tilde U_j) \le
      \tilde F_{j\mid x}(\bar{\tilde u}_{j,l,d,z\mid x})\}.
\]
Defining $U_j = \tilde F_{j\mid x}(\tilde U_j)$, $\underline{u}_{j,l,d,z\mid x} = \tilde F_{j\mid x}(\underline{\tilde u}_{j,l,d,z\mid x})$, $
\bar u_{j,l,d,z\mid x} = \tilde F_{j\mid x}(\bar{\tilde u}_{j,l,d,z\mid x})$,
we obtain a normalized vector $U=(U_1,\dots,U_J)$ with support $[0,1]^J$ and uniform marginal distributions conditional on $X=x$, and selection regions of the form in \eqref{eq:rectangle1} with endpoints $\{\underline{u}_{j,l,d,z\mid x},\bar{u}_{j,l,d,z\mid x}\}$. 

As highlighted in \cite{lee/salanie:18}, Assumption \ref{ass:SM} reduces to the standard threshold crossing model \citep{heckman/vytlacil:05, imbens/angrist:94} when we have a binary treatment and a single dimension of unobserved heterogeneity given by
\begin{align}\label{eq:standard}
D(z) = 1\{U_1 \leq g_{1,z|x} \}
\end{align}
for $z \in \mathcal{Z}$ conditional on $X = x \in \mathcal{X}$. But, in cases where $J > 1$, it provides a unified framework to capture a number of additional common, empirically relevant models of selection with multiple as well as binary treatments. Below, we briefly provide examples of several such models. Our first three examples consider multiple treatment models, while our fourth example considers a binary treatment model. For simplicity, when possible, we take the number of treatments in the multiple treatment examples to be solely equal to three.

\begin{example}{(Multinomial Choice)}\label{ex:arum}
Let $\mathcal{D} = \{0,1,2\}$, and, conditional on $X = x \in \mathcal{X}$, let
\begin{align}\label{eq:arum1}
 D(z) = \argmax_{d \in \mathcal{D}} \tilde{g}_{d,z|x} - \tilde{U}_d
\end{align}
for $z \in \mathcal{Z}$, where $(\tilde{g}_{1,z|x},\tilde{g}_{2,z|x})$ is unknown, $(\tilde{U}_1,\tilde{U}_2)$ some continuous unobserved variables, and $\tilde{g}_{0,z|x}$ and $\tilde{U}_0$ are normalized to 0. This is the standard additively separable utility model of multinomial choice as studied in \cite{heckman/etal:06,heckman/etal:08} and is empirically used to model selection such as that of children into schools \citep{abdulkadirouglu/etaL:20, kline/walters:16} or patients into hospitals \citep{hull:20}. Alternatively, several papers consider monotonicity restrictions on potential treatments that can be equivalently captured by this selection model. For instance, taking $\mathcal{D}$ to denote different fields of study and $\mathcal{Z} = \{0,1,2\}$ with $z \in \{1,2\}$ denoting offer to field $d = z$ and $z=0$ denoting no offer, \cite{kirkeboen/etal:16} impose $D(0) \neq D(d) \implies D(d) = d$ for $d \in \{1,2\}$, i.e. if an offer to field $d$ affects decisions then it does so by only encouraging students into that field. As shown in \citet[][Appendix A]{bai/tabord:25}, this corresponds to equivalently assuming \eqref{eq:arum1} with restricting the thresholds to satisfy $\tilde{g}_{1,0|x} = \tilde{g}_{1,2|x}$ and $\tilde{g}_{2,0|x} = \tilde{g}_{2,1|x}$, i.e. the thresholds for the fields are affected only by its offers. In a similar manner, the related monotonicity restriction in \cite{kline/walters:16} can also be equivalently captured by \eqref{eq:arum1}.

To write \eqref{eq:arum1} in terms of Assumption \ref{ass:SM}, following \citet[][Example 5]{lee/salanie:18}, let $U_j = \tilde{F}_{j|x}(\tilde{U}_j)$ and $g_{j,z|x} = \tilde{F}_{j|x}(\tilde{g}_{j,z|x})$ for $j \in \{1,2\}$, where $\tilde{F}_{j|x}$ denotes the conditional on $X=x$ distribution function of $\tilde{U}_j$, and let $U_3 = \tilde{F}_{12|x}(\tilde{U}_1 - \tilde{U}_2)$ and $g_{3,z|x} = \tilde{F}_{12|x}(\tilde{g}_{1,z|x} - \tilde{g}_{2,z|x})$, where  $\tilde{F}_{12|x}$ is the conditional on $X=x$ distribution function of $\tilde{U}_1 - \tilde{U}_2$. Observe that it then follows that \eqref{eq:arum1} can be equivalently written as
\begin{align}\label{eq:arum2}
 D(z) =
 \begin{cases}
   0 &\text{ if } U_1 > g_{1,z|x},~U_2 > g_{2,z|x}~, \\
   1 &\text{ if } U_1 \leq g_{1,z|x},~U_3 \leq g_{3,z|x}~,\\
   2 &\text{ if } U_2 \leq g_{2,z|x},~U_3 > g_{3,z|x}~,
 \end{cases}
\end{align}
\sloppy which can be straightforwardly re-written in terms of Assumption \ref{ass:SM} by taking $\mathcal{U}^{\text{sm}}_{0,z|x} = \{(u_1,u_2,u_3) \in [0,1]^3 : u_1 > g_{1,z|x},~u_2 > g_{2,z|x} \}$, $\mathcal{U}^{\text{sm}}_{1,z|x} = \{(u_1,u_2,u_3) \in [0,1]^3 : u_1 \leq g_{1,z|x},~u_3 \leq g_{3,z|x} \}$, and $\mathcal{U}^{\text{sm}}_{2,z|x} = \{(u_1,u_2,u_3) \in [0,1]^3 : u_2 \leq g_{2,z|x},~u_3 > g_{3,z|x} \}$.
\end{example}

\begin{example}{(Sequential Choice)}\label{ex:sequential}
Let $\mathcal{D} = \{0,1,2\}$, and, conditional on $X = x \in \mathcal{X}$, let
\begin{align}\label{eq:sequential}
D(z) =
\begin{cases}
  0 &\text{ if } U_1 > g_{1,z|x}~, \\
  1 &\text{ if } U_1 \leq g_{1,z|x}~,~U_2 > g_{2,z|x}~,\\
  2 &\text{ if } U_1 \leq g_{1,z|x}~,~U_2 \leq g_{2,z|x}~,
\end{cases}
\end{align}
for $z \in \mathcal{Z}$, where $(g_{1,z|x},g_{2,z|x})$ is unknown. This model imposes a sequential nature to the decisions where individuals first select into treatment 0 or not, and if not, then select into treatments 1 or 2. For instance, such a model has been used empirically to model judicial decisions for defendants, where judges first decide to convict or not and, if convicted, to additionally incarcerate or not \citep{arteaga:21,kamat/etal:22}. Additionally, a special case of this model when $U_1 = U_2$ corresponds to that studied in \cite{heckman/etal:06}, and is also used empirically to model various decisions such as years in preschool or sentence length in prison  \citep{cornelissen/etal:18, rose/shem:21}. Moreover, as shown in \citet[][Appendix A4]{kamat/etal:22}, this special case also corresponds to limiting versions of the monotonicity restrictions in \cite{angrist/imbens:95} and \cite{heckman/pinto:18}. Observe that \eqref{eq:sequential} can be written in terms of Assumption \ref{ass:SM} by taking $\mathcal{U}^{\text{sm}}_{0,z|x} = \{(u_1,u_2) \in [0,1]^2 : u_1 > g_{1,z|x} \}$, $\mathcal{U}^{\text{sm}}_{1,z|x} = \{(u_1,u_2) \in [0,1]^2 : u_1 \leq g_{1,z|x},~u_2 > g_{2,z|x} \}$, and $\mathcal{U}^{\text{sm}}_{2,z|x} = \{(u_1,u_2) \in [0,1]^2 : u_1 \leq g_{1,z|x},~u_2 \leq g_{2,z|x} \}$.
\end{example}

\begin{example}{(Multi-stage or Dynamic Choice)}\label{ex:dynamic}
Assumption \ref{ass:SM} also permits more general versions of Examples \ref{ex:arum}-\ref{ex:sequential} such as multi-stage or dynamic selection models, which can be represented as decision trees and where a sequence of threshold-crossing rules determine the final treatment. For instance, in the reanalysis of the Moving to Opportunity experiment to evaluate the effects of moving to low-poverty neighborhoods and mental health, \cite{navjeevan/etaL:23} take $\mathcal{Z} = \{0,1\}$, where $z$ denotes an indicator for a MTO voucher, and $\mathcal{D} = \{(d_0,d_1) : d_0,d_1 \in \{0,1\} \}$, where $d_0$ denotes an indicator for whether households moved to low-poverty neighborhoods and  $d_1$ denotes an indicator for whether the head of the households reported having positive mental health. Following \cite{vytlacil:02}, the monotonicity restrictions \cite{navjeevan/etaL:23} impose can be stated as a selection model given by
\begin{align}\label{eq:dynamic}
D(z) =
\begin{cases}
  (0,0) &\text{ if } U_1 > g_{1,z|x}~,~U_2 > g_{2,z|x}~, \\
  (0,1) &\text{ if } U_1 > g_{1,z|x}~,~U_2 \leq g_{2,z|x}~,\\
  (1,0) &\text{ if } U_1 \leq g_{1,z|x}~,~U_3 > g_{3,z|x}~, \\
  (1,1) &\text{ if } U_1 \leq g_{1,z|x}~,~U_3 \leq g_{3,z|x}~,
\end{cases}
\end{align}
for $z \in \{0,1\}$, where $(g_{1,z|x},g_{2,z|x},g_{3,z|x})$ is unknown, i.e. a threshold-crossing rule determines whether households relocate in the first stage and another threshold crossing rule then determines their mental health in the second stage, along with the additional restriction that $U_2 = U_3$, i.e. the same index underlies the second stage regardless of the relocation decision, and $g_{2,0|x} = g_{2,1|x}$ and $g_{3,0|x} = g_{3,1|x}$, i.e. the voucher affects only the decision to relocate and not mental health. Observe that \eqref{eq:dynamic} can be written in terms of Assumption \ref{ass:SM} by taking $\mathcal{U}^{\text{sm}}_{(0,0),z|x} = \{(u_1,u_2,u_3) \in [0,1]^3 : u_1 > g_{1,z|x},~u_2 > g_{2,z|x}\}$,~$\mathcal{U}^{\text{sm}}_{(0,1),z|x} = \{(u_1,u_2,u_3) \in [0,1]^3 : u_1 > g_{1,z|x},~u_2 \leq g_{2,z|x}\}$, $\mathcal{U}^{\text{sm}}_{(1,0),z|x} = \{(u_1,u_2,u_3) \in [0,1]^3 : u_1 \leq g_{1,z|x},~u_3 > g_{3,z|x}\}$,~$\mathcal{U}^{\text{sm}}_{(1,1),z|x} = \{(u_1,u_2,u_3) \in [0,1]^3 : u_1 \leq g_{1,z|x},~u_3 \leq g_{3,z|x}\}$. In a similar manner, we note that versions of the dynamic model studied in \cite{heckman/etal:16} can also be captured by Assumption \ref{ass:SM}.
\end{example}

\begin{example}{(Double Hurdle)}\label{ex:double}
Let $\mathcal{D} = \{0,1\}$, and, for each $z \in \mathcal{Z}$ conditional on $X = x \in \mathcal{X}$, let
\begin{align}\label{eq:double}
 D(z) = 1\{U_1 \leq g_{1,z|x},~ U_2 \leq g_{2,z|x}\}~,
\end{align}
where $(g_{1,z|x},g_{2,z|x})$ is unknown, i.e. both unobserved variables have to be below their thresholds for the individual to receive treatment. Such a model can arise in settings where an individual has to make multiple decisions to receive a treatment \citep{poirier:80} or participate in a survey \citep{dutz/etal:21}, and is also related to the selection model that arises under the partial monotonicity assumption of \cite{mogstad/etal:21, mogstad/etal:24} under multiple instruments. Observe that \eqref{eq:double} can be written in terms of Assumption \ref{ass:SM} by taking $\mathcal{U}^{\text{sm}}_{0,z|x} = \{(u_1,u_2) \in [0,1]^2 : u_1 > g_{1,z|x} \} \cup \{(u_1,u_2) \in [0,1]^2 : u_2 > g_{2,z|x} \}$, and $\mathcal{U}^{\text{sm}}_{1,z|x} = \{(u_1,u_2) \in [0,1]^2 : u_1 \leq g_{1,z|x},~u_2 \leq g_{2,z|x} \}$.
\end{example}

\subsection{Parameters of Interest}\label{sec:parameters}

In our model, we take the primitives to consist of three objects. First, the marginal treatment response (MTR) functions \citep{heckman/vytlacil:99,heckman/vytlacil:05,mogstad/etal:18}, defined by $m_{d|x}(u) = E[Y(d)\mid U {=} u, X {=} x]$, which capture the mean treatment response for $d \in \mathcal{D}$ conditional on the unobserved heterogeneity $U {=} u$ and covariates $X {=} x$.  Second, the distribution of these unobservables conditional on $X {=} x$, denoted by $F_x$.  Third, all threshold values determining selection in Assumption~\ref{ass:SM}, denoted by $g = \{g_{j,z|x} : j = 1,\ldots,J,~z \in \mathcal{Z},~x \in \mathcal{X}\}$. Let $\beta \equiv (m, h)$ summarize these primitives, where $m$ collects the MTRs and $h = (F, g)$ comprises the
selection primitives.

Our parameters of interest are functionals of these primitives. In particular, our analysis applies to any parameter that can be written as a weighted sum of integrals of the MTRs over specific regions of the unobserved heterogeneity. Formally, we consider parameters of the form
\begin{align}\label{eq:theta}
\theta(\beta)
= \sum_{x \in \mathcal{X}} \sum_{l \in \mathcal{L}} \sum_{d \in \mathcal{D}}
  w_{d,l|x}(h)
  \int_{\mathcal{U}^{\theta}_{d,l|x}(g)} m_{d|x}(u)\, dF_x~,
\end{align}
where $\mathcal{L}$ is a known index set that we sum across in addition to the covariate value $\mathcal{X}$ and treatment values $\mathcal{D}$, $w_{d,l|x}(h)$ determines the weight each integral receives, and $\mathcal{U}^{\theta}_{d,l|x}(g)$ is the integration region specifying which individuals are included in each weighted average. We assume that the weights $w_{d,l|x}(h)$ are known or identified functions of the selection primitives $h$, and the integration regions  $\mathcal{U}^{\theta}_{d,l|x}(g)$ can be expressed as a finite union of sets of the form
\begin{align}\label{eq:rectangle2}
\prod_{j=1}^J[\underline{u}_{j,d,l|x}(g),\bar{u}_{j,d,l|x}(g)]~,
\end{align}
where the endpoints $\underline{u}_{j,d,l|x}(g)$ and $\bar{u}_{j,d,l|x}(g)$ are known or identified functions of the thresholds $g$. This rectangular structure mirrors that imposed on the selection regions in \eqref{eq:rectangle1}, which plays a key role in our nonparametric analysis in Section \ref{sec:identifiedset_NP}.

As illustrated in Table \ref{tab:theta_exp}, a number of commonly studied treatment effect parameters can be written in terms of \eqref{eq:theta}. These include the average treatment effect (ATE) between two treatments $d',d'' \in \mathcal{D}$ as well as their analogs that condition on $D=d$ to give the average treatment effect on the treated (ATT) or on a response type $D(z) = d_z \in \mathcal{D}$ for $z \in \mathcal{Z}' \subseteq \mathcal{Z}$ to give local average treatment effects (LATEs). Moreover, they also include the class of policy relevant treatment effects \citep{heckman/vytlacil:99,heckman/vytlacil:05} that evaluate the effects of altering the thresholds determining selection. To formally define these latter effects, note that the sets determining treatment in \eqref{eq:SM} can be more explicitly stated as $\mathcal{U}^{\text{sm}}_{d,z|x} \equiv \mathcal{U}^{\text{sm}}_{d,z|x}(g)$, i.e. as functions of the unknown thresholds. A policy $\delta$ can then be equivalently captured by the thresholds $g + \delta \equiv \{g_{j,z|x} + \delta_{j,z|x} : j = 1, \ldots, J,~z \in \mathcal{Z},~x \in \mathcal{X}\}$, i.e. changing the original thresholds by some known values $\delta \equiv \{\delta_{j,z|x} : j = 1, \ldots, J,~z \in \mathcal{Z},~x \in \mathcal{X}\}$. Let $D^{\delta} = \sum_{d \in \mathcal{D}}d 1\{ U \in \mathcal{U}^{\text{sm}}_{d,Z|X}(g+\delta)\}$ and $Y^{\delta} = Y(D^{\delta})$ denote the individual's treatment and outcome under this policy, respectively. The policy relevant treatment effect (PRTE) between two policies $\delta'$ and $\delta''$ can then be defined by $\text{PRTE}^{\delta',\delta''} = E[Y^{\delta'} - Y^{\delta''}|D^{\delta'} \neq D^{\delta''}]$, i.e. the effect of the counterfactual change in outcomes between the two altered values of thresholds for those who are affected by this change.

\subsection{Identified Set}\label{sec:identifiedset}

The objective of our analysis is to learn about a pre-specified parameter of interest $\theta(\beta)$ given by \eqref{eq:theta}. As the function $\theta$ is known, what we can learn about the parameter depends on what we know about the primitives $\beta$. To this end, denoting by $\mathbf{M}^{\dagger}$ and $\mathbf{H}^{\dagger}$ the space of all possible $m$ and $h$, let $\mathbf{B} \subseteq \mathbf{M}^{\dagger} \times \mathbf{H}^{\dagger}$ be the admissible space that $\beta$ is restricted to lie in, which is determined by the various assumptions we may impose on it. The data also provide information on $\beta$. The observed decision and outcome through \eqref{eq:D_equation} and \eqref{eq:Y_equation} along with the selection model in \eqref{eq:SM} and Assumption \ref{ass:E} provide the following moments
\begin{align}
 P[D=d|Z=z,X=x] \equiv P_{d|z,x} &= \int\limits_{\mathcal{U}_{d,z|x}^{\text{sm}}(g)} dF_{x}~, \label{eq:D_moments} \\
 E[Y1\{D=d\}|Z=z,X=x] \equiv E_{d|z,x} &= \int\limits_{\mathcal{U}^{\text{sm}}_{d,z|x}(g)} m_{d|x}(u) dF_{x}~, \label{eq:Y_moments}
\end{align}
for $d \in \mathcal{D}$, $z \in \mathcal{Z}$ and $x \in \mathcal{X}$. Denoting by $\mathbf{B}^* = \{\beta \in \mathbf{B} : \beta \text{ satisfies } \eqref{eq:D_moments}-\eqref{eq:Y_moments}\}$ the admissible set of primitives satisfying the assumptions and the moments, what we can learn about the parameter of interest can then be formally captured by the identified set defined by
\begin{align}\label{eq:identifiedset}
\Theta \equiv \theta(\mathbf{B}^*) = \{\theta_0 \in \mathbf{R} : \theta(\beta) = \theta_0 \text{ for some } \beta \in \mathbf{B}^*\}~,
\end{align}
i.e. the image of the admissible set $\mathbf{B}^*$ under the function $\theta$.

\section{Identification}\label{sec:identification}

We start by showing how to compute the identified set when the moments of the data in the left hand sides of \eqref{eq:D_moments} and \eqref{eq:Y_moments} are assumed to be known without uncertainty. The subsequent section shows how this analysis translates into a strategy to estimate the identified set and construct confidence intervals.

\subsection{Two-Step Approach}\label{sec:twostep}

From the definition in \eqref{eq:identifiedset}, observe that computing the identified set requires searching through the space $\mathbf{B}^*$ and taking the image under the function $\theta$. The challenge is how to perform this search in a tractable manner.

To do so, we build on the approach of \cite{mogstad/etal:18}, who study this identification problem in the special case of a binary treatment when selection is given by \eqref{eq:standard}. In this case, because a single dimension of unobserved heterogeneity governs selection and is normalized to be uniformly distributed on $[0,1]$, the distribution $F$ is known. Given $F$ is known, $h$ is then point-identified as the threshold function $g$ is directly point-identified by the treatment shares via \eqref{eq:D_moments}. This allows simplifying the search problem as it remains to search only across the admissible values of the MTRs. Specifically, for a given $h \in \mathbf{H}^\dagger$, let $\mathbf{M}(h) \equiv \{m \in \mathbf{M}^\dagger : (m,h) \in \mathbf{B}\}$ denote the set of admissible MTRs and let  $\mathbf{M}^*(h) = \{m \in \mathbf{M}(h) : (m,h) \text{ satisfies } \eqref{eq:Y_moments} \}$ denote those additionally consistent with the outcome moments. The identified set then simplifies to $\theta(\mathbf{M}^*(h^*),h^*) \equiv \{\theta(m,h^*) : m \in \mathbf{M}^*(h^*) \}$, where $h^*$ denotes the point identified value of $h$. Assuming $\mathbf{M}(h)$ takes $m$ to be linearly parameterized, \cite{mogstad/etal:18} exploit the linearity of $\theta$ and $\mathbf{M}^*(h)$ in terms of this parametrization to show that $\theta(\mathbf{M}^*(h^*),h^*) = [\theta_L(h^*),\theta_U(h^*)]$, where
\begin{align}\label{eq:LP_binary}
\theta_L(h^*) = \inf_{m \in \mathbf{M}^*(h^*)} \theta(m,h^*)~\text{ and }~\theta_U(h^*) = \sup_{m \in \mathbf{M}^*(h^*)} \theta(m,h^*)~,
\end{align}
and where these optimization problems correspond to linear programs.

However, in the case where multidimensional unobserved heterogeneity governs selection, an inherent challenge is the fact that $h$ is not generally point identified. This is because, while the marginals continue to be normalized to uniform distributions, the dependence between them is not necessarily known, implying that both $F$ and $g$ may not be known or point identified by the data. We propose a two-step solution to account for this complication. Specifically, let $\mathbf{H}$ denote the set of admissible values of $h$ and $\mathbf{H}^* = \{h \in \mathbf{H} : h \text{ satisfies } \eqref{eq:D_moments}\}$ denote those additionally consistent with the selection moments. The identified set in \eqref{eq:identifiedset} can then be stated as
\begin{align}\label{eq:Theta_union}
\Theta = \bigcup_{h \in \mathbf{H}^*} \theta(\mathbf{M}^*(h),h) ~,
\end{align}
i.e. a union of the identified sets given $h$ across the admissible values of $h$ consistent with the selection moments. This suggests a two-step strategy, where in an inner step, as in the unidimensional case, we can continue to exploit the linearity to compute $\theta(\mathbf{M}^*(h),h)$ given $h \in \mathbf{H}^*$, and then in an outer step, we can aggregate them across $\mathbf{H}^*$ to obtain the final identified set.

In what follows, we show how to operationalize this strategy. In Sections \ref{sec:identifiedset_P} and \ref{sec:identifiedset_NP}, we first show how to generalize the arguments from the unidimensional case to that with multidimensional MTRs to tractably compute $\theta(\mathbf{M}^*(h),h)$ as in \eqref{eq:LP_binary} for each $h \in \mathbf{H}$ when $\mathbf{M}(h)$ takes $m$ to be parameterized or remain entirely nonparametric. In Section \ref{sec:selection_parameterization}, we then show how to parametrize $\mathbf{H}^*$ by a finite-dimensional parameter so that we can feasibly take the union across $\mathbf{H}^*$.

\subsection{Identified Set under Parametric MTRs}\label{sec:identifiedset_P}

As in \cite{mogstad/etal:18}, we assume $\mathbf{M}(h)$ to linearly parametrize $m$ along with a parameter space determined by a system of linear inequalities as follows.

\begin{assumption2}{PM}{(Parametrizing MTRs)}\label{ass:PM}
\sloppy $\mathbf{M}(h) = \bigl\{ m \in \mathbf{M}^{\dagger} : m_{d|x}(u) = \sum_{k=0}^{K} \gamma_{k,d|x} b_{k}(h)(u) \text{ for } d \in \mathcal{D},~x \in \mathcal{X}, \text{ and } \gamma \equiv (\gamma_{k,d|x} : 0 \leq k \leq K,~d \in \mathcal{D},~x \in \mathcal{X}) \in \mathbf{\Gamma}(h) \bigr\}$, where $\{b_{k}(h) : 0 \leq k \leq K \}$ are known functions and $\mathbf{\Gamma}(h) \subseteq \mathbf{R}^{\text{dim}(\gamma)}$ is characterized by a system of linear inequalities.
\end{assumption2}

To see how this allows computing $\theta(\mathbf{M}^*(h),h)$ using linear programs as in \eqref{eq:LP_binary} for each $h \in \mathbf{H}$, it is useful to rewrite it in terms of the parametrizing variable $\gamma$. As $\theta$ is linear in $m$ and $m$ is linear in $\gamma$, we can substitute in the relation between $m$ and $\gamma$ from Assumption \ref{ass:PM} into $\theta$ to obtain a function $\theta^{\Gamma}$ linear in $\gamma$ such that $\theta(m,h) = \theta^{\Gamma}(\gamma,h)$. Similarly, rewriting the outcome moments in \eqref{eq:Y_moments} in terms of $\gamma$
\begin{align}\label{eq:Y_moments_alpha}
 \sum_{k=0}^{K} \gamma_{k,d|x} \int\limits_{\mathcal{U}_{d,z|x}^{\text{sm}}(g)} b_{k}(h)(u) dF_x(u) = E_{d|z,x}
\end{align}
for $d \in \mathcal{D}$, $z \in \mathcal{Z}$, $x \in \mathcal{X}$, we can write $\mathbf{M}^*(h)$ in terms of $\gamma$ by $\mathbf{\Gamma}^*(h) = \left\{ \gamma \in \mathbf{\Gamma}(h) :  \gamma \text{ satisfies } \eqref{eq:Y_moments_alpha} \right\}$. The identified set in terms of $\gamma$ then corresponds to $\theta^{\Gamma}(\mathbf{\Gamma}^*(h),h) \equiv \{\theta^{\Gamma}(\gamma,h) : \gamma \in \mathbf{\Gamma}^*(h)\}$. In the following proposition, we show that the linearity implies that $\theta^{\Gamma}(\mathbf{\Gamma}^*(h),h)$ for $h \in \mathbf{H}$ equals an interval with endpoints given by two optimization problems, which correspond to linear programs as $\theta^{\Gamma}$ is a linear function in $\gamma$ and $\mathbf{\Gamma}^*(h)$ is given by a system of linear inequalities.

\begin{proposition}\label{prop:linearprog}
For $h \in \mathbf{H}$, let $\mathbf{M}(h)$ satisfy Assumption \ref{ass:PM}. If $\mathbf{M}^*(h)$ is empty then so is $\theta(\mathbf{M}^*(h),h)$, and if it is not empty, then $\text{closure}(\theta(\mathbf{M}^*(h),h)) = [\theta_L(h),\theta_U(h)]$, where
\begin{align}\label{eq:optimization}
\theta_L(h) = \inf_{\gamma \in \mathbf{\Gamma}^*(h)} \theta^{\Gamma}(\gamma,h)~ \text{ and }~ \theta_U(h) = \sup_{\gamma \in \mathbf{\Gamma}^*(h)} \theta^{\Gamma}(\gamma,h)~.
\end{align}
\end{proposition}

Assumption \ref{ass:PM} requires the restrictions on $\gamma$ to correspond to a system of linear inequalities. As highlighted in Table \ref{tab:alpha_shape}, a number of common shape restrictions from the literature correspond to linear restrictions on $m$ and in turn to a system of linear inequalities on $\gamma$ when considered over a finite grid of points of $u$ in $[0,1]^J$. Specifically, it allows assumptions such as boundedness typically required to ensure the bounds on mean effects are finite \citep{manski:08}, monotonicity in outcomes between two treatments as in the monotone treatment response assumption from \cite{manski:97}, or the difference in mean outcomes between two treatments to be bounded as in the bounded variation assumption from \cite{manski/pepper:18}. It also allows imposing monotonicity across different values of the unobserved heterogeneity in a given dimension to capture how the outcome may vary for those more or less likely to select into treatment as in the monotone treatment selection assumption from \cite{manski/pepper:00}, or separability between the unobserved heterogeneity and the covariates as in \cite{brinch/etal:17}.

\subsection{Identified Set under Nonparametric MTRs}\label{sec:identifiedset_NP}

Although Proposition \ref{prop:linearprog} parameterizes the MTRs via  Assumption \ref{ass:PM}, it allows the researcher to compute the identified set under fully nonparametric MTRs in special cases.  In particular, let $\mathbf{M}_{\text{np}}(h)$ denote a set of nonparametric MTRs. For certain such sets formalized below, we can show $\theta(\mathbf{M}^*_{\text{np}}(h),h) = \theta(\mathbf{M}^*(h),h)$, where $\mathbf{M}(h)$ denotes the subset of $\mathbf{M}_{\text{np}}(h)$ additionally satisfying Assumption \ref{ass:PM} with $\{b_{k}(h) : 0 \leq k \leq K\}$ given by
\begin{align}\label{eq:b_constant}
b_{k}(h)(u) = 1\{u \in \mathcal{U}_k(h)\}
\end{align}
for a specific choice of partition $\mathbb{U}(h) \equiv \{\mathcal{U}_{0}(h), \mathcal{U}_{1}(h), \ldots, \mathcal{U}_{K}(h)\}$ of $[0,1]^J$, and with
\begin{align}\label{eq:A_constant}
\mathbf{\Gamma}(h) = \left\{\gamma \in \mathbf{R}^{\text{dim}(\gamma)} : \left( \sum\limits_{k=0}^{K} \gamma_{k,d|x} b_{k}(h) : d \in \mathcal{D},~x \in \mathcal{X} \right) \in \mathbf{M}_{\text{np}}(h) \right\}~,
\end{align}
which can be equivalently written as a system of linear inequalities. This mirrors \citet[][Proposition 4]{mogstad/etal:18}, which shows such a result in the case with a single dimension of unobserved heterogeneity. In this case, the rectangular sets underlying the selection model and parameters in \eqref{eq:rectangle1} and \eqref{eq:rectangle2} simplify to just intervals. Exploiting this interval structure, they show that a partition consisting of intervals constructed using the end points of those in \eqref{eq:rectangle1} and \eqref{eq:rectangle2} can generate the nonparametric bounds.

We construct a new partition that shows how to exploit the general rectangular structure of \eqref{eq:rectangle1} and \eqref{eq:rectangle2}, and generalize the unidimensional partition to the case with multidimensional unobserved heterogeneity. Let $\mathbb{U}^{*}(h) = \{\mathcal{U}^{\text{sm}}_{d,z|x}(g) : d \in \mathcal{D},~ z \in \mathcal{Z},~x \in \mathcal{X} \} \cup \{\mathcal{U}^{\theta}_{d,l|x}(g) : d \in \mathcal{D},~ l \in \mathcal{L},~x \in \mathcal{X}\}$ denote the collection of sets underlying the data moments in \eqref{eq:Y_moments} and the parameter of interest in \eqref{eq:theta}. We require that the partition is rich enough to generate each set in this collection so that it can capture the information provided by the data and on the parameter of interest and hence produce the nonparametric bounds. We formalize this property in the following definition.\footnote{We note that our definition shares conceptual similarities to multidimensional partitions considered in discrete choice analysis to obtain nonparametric bounds in alternative identification problems, where one similarly requires the partition to be sufficiently rich so that each element of the partition predicts the same choice behavior \citep[e.g.,][]{chesher/etal:13, gu/etal:22, kamat/norris:22, tebaldi/etal:23}.}

\begin{definition2}{P}{(Partition)}\label{def:P}
Let $\mathbb{U}(h)$ be a finite partition of $[0,1]^J$ such that for each $\mathcal{U}^* \in \mathbb{U}^{*}(h)$, we have that $\bigcup_{\mathcal{U} \in \mathbb{U}} \mathcal{U} = \mathcal{U}^*$ for some $\mathbb{U} \subseteq \mathbb{U}(h)$.
\end{definition2}

\begin{figure}[!t]
\centering
\caption{Partition satisfying Definition \ref{def:P} in Example \ref{ex:double} with $\mathcal{Z} = \{z,z'\}$ and $\mathcal{X} = \{x\}$, and where the parameter of interest is the $\text{ATE}$ between two treatments. In Panel (a), note that $\mathcal{U}^{\text{sm}}_{0,z|x}(g) = \mathcal{U}'_3 \cup \mathcal{U}'_4$, $\mathcal{U}^{\text{sm}}_{1,z|x}(g) = \mathcal{U}'_1 \cup \mathcal{U}'_2$, $\mathcal{U}^{\text{sm}}_{0,z'|x}(g) = \mathcal{U}'_2 \cup \mathcal{U}'_4$, and $\mathcal{U}^{\text{sm}}_{1,z'|x}(g) = \mathcal{U}'_1 \cup \mathcal{U}'_3$.}
\makebox[\textwidth]{ \begin{tabular}{c c c}
\subcaptionbox{Sets underlying $\mathbb{U}^{*}(h)$}{ \includegraphics[width=2.75in]{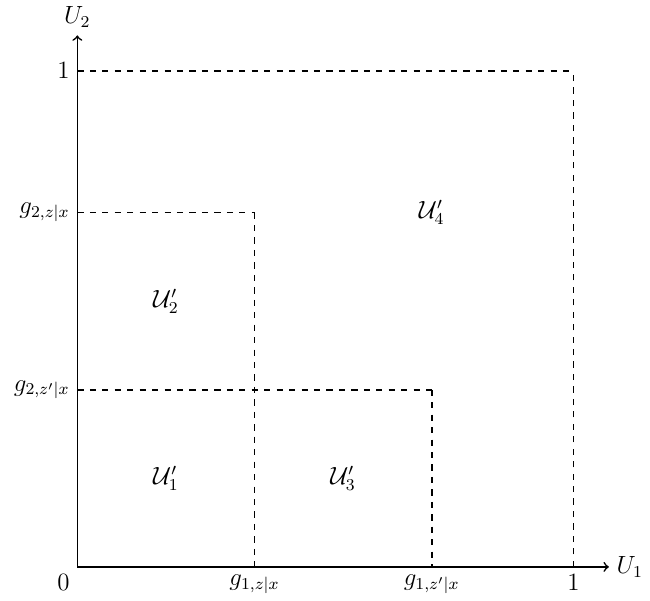}} & ~~~~~~~~~ &
\subcaptionbox{Partition}{\includegraphics[width=2.75in]{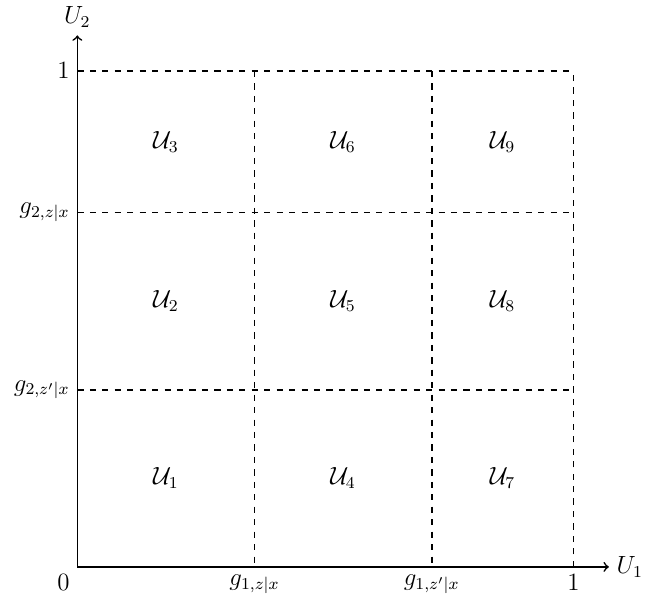}}
\end{tabular}}
\label{fig:sets_example}
\end{figure}

Figure \ref{fig:sets_example} graphically illustrates how we can exploit the rectangular structure underlying the selection model and parameters in \eqref{eq:rectangle1} and \eqref{eq:rectangle2} to construct a partition satisfying Definition \ref{def:P} in the context of a simple version of Example \ref{ex:double}. Figure \ref{fig:sets_example}(a) first presents the sets in $\mathbb{U}^{*}(h)$. Figure \ref{fig:sets_example}(b) then shows how to exploit the rectangular structure of these sets to construct a partition satisfying Definition \ref{def:P}. It takes the endpoints of the various rectangles to form disjoint rectangles that partition the entire space.

This idea can also be applied to construct a partition more generally. Specifically, for each $j=1,\ldots,J$, let $u_{(1),j} \leq \ldots \leq u_{(M_j),j}$ denote the ordered values of all the end points in the $j$th dimension of the underlying rectangular sets in \eqref{eq:rectangle1} and \eqref{eq:rectangle2} that form the sets in $\mathbb{U}^{*}(h)$ together with 0 and 1. Let $\mathbb{U}(h)$ then denote the partition obtained by
\begin{align}\label{eq:parition_rectangular}
\mathbb{U}(h) = \prod_{j=1}^J \left\{ [u_{(1),j},u_{(2),j}) , \ldots, [u_{(M_j - 1),j},u_{(M_j),j}] \right\}~,
\end{align}
i.e. a collection of hyperrectangles constructed by taking the Cartesian product of the set of intervals based on adjacent endpoints in each dimension.

By construction, due to Definition \ref{def:P}, the partition implies that for each $m \in \mathbf{M}^\dagger$ satisfying \eqref{eq:Y_moments}, the parametrized version $m' = \sum_{k=0}^{K} \gamma_{k,d|x} b_{k}(h)$, where $b_k(h)$ satisfies \eqref{eq:b_constant} and $\gamma_{k,d|x} = E[m_{d|x}(U) | U \in \mathcal{U}_k(h)]$, also satisfies \eqref{eq:Y_moments} and also generates the same parameter of interest as $m$ in the sense that $\theta(m,h) = \theta(m',h)$. In turn, provided that
\begin{align}\label{eq:NP_condition}
 (E[m_{d|x}(U) | U \in \mathcal{U}_k(h)] : 0 \leq k \leq K,~d \in \mathcal{D},~x \in \mathcal{X} ) \in \mathbf{\Gamma}(h)
\end{align}
for $\mathbf{\Gamma}(h)$ in \eqref{eq:A_constant} so that the parametric $m'$ also satisfies the assumptions imposed on $m$, it will generate the same identified set for $\theta$ as $m$. We show this in the following proposition.

\begin{proposition}\label{prop:equivalence}
For $h \in \mathbf{H}$ and $\mathbf{M}_{\text{np}}(h)$, let $\mathbf{M}(h) =  \{m \in \mathbf{M}_{\text{np}}(h) : m  = \sum_{k=0}^{K} \gamma_{k,d|x} b_{k}(h),~\gamma \equiv (\gamma_{k,d|x} : 0 \leq k \leq K,~d \in \mathcal{D},~x \in \mathcal{X}) \in \mathbf{\Gamma}(h)\}$ with $\{b_{k}(h) : 0 \leq k \leq K\}$ given by \eqref{eq:b_constant}, where $\mathbb{U}(h)$ satisfies Definition \ref{def:P}, and with $\mathbf{\Gamma}(h)$ given by \eqref{eq:A_constant}. Suppose $\mathbf{\Gamma}(h)$ satisfies \eqref{eq:NP_condition} for every $m \in \mathbf{M}_{\text{np}}(h)$. It then follows that $\theta(\mathbf{M}_{\text{np}}^*(h),h) = \theta(\mathbf{M}^*(h),h)$, where $\theta(\mathbf{M}^*(h),h)$ is characterized in Proposition \ref{prop:linearprog}.
\end{proposition}

Requiring $\mathbf{\Gamma}(h)$ to satisfy \eqref{eq:NP_condition} restricts the sets of nonparametric MTRs to which Proposition \ref{prop:equivalence} applies. However, amongst the linear restrictions in Table \ref{tab:alpha_shape}, the condition in \eqref{eq:NP_condition} fails only when Assumption $\text{M}_j$ is considered. For the remaining restrictions, namely Assumptions B, $\text{M}_{d',d''}$, $\text{BV}_{d',d''}$, and $S$, we can straightforwardly show that for every $m$ satisfying each of these assumptions, $\gamma$ constructed using $m$, i.e. where $\gamma_{k,d|x} = E[m_{d|x}(U) | U \in \mathcal{U}_k(h)]$, satisfies the linear restrictions stated in terms of $\gamma$ given which $\mathbf{\Gamma}(h)$ satisfies \eqref{eq:NP_condition}---see Table \ref{tab:alpha_shape} for the derivations of these linear restrictions. For such restrictions imposed on the MTRs, nonparametric bounds on the parameter of interest can continue to be computed using Proposition \ref{prop:linearprog}.

\subsection{Parametrization of Selection Model Primitives}\label{sec:selection_parameterization}

Having computed $\theta(\mathbf{M}^*(h))$ given $h \in \mathbf{H}$, we next show how to ensure that we can feasibly take their unions over $h \in \mathbf{H}^*$ and compute the identified set in \eqref{eq:Theta_union}. To do so, we make the following assumption on $\mathbf{H}$.

\begin{assumption2}{PS}{(Parametrizing Selection Primitives)}\label{ass:PS}
$\mathbf{H} = \bigcup_{\lambda \in \mathbf{L}} \mathbf{H}(\lambda)$, where $\mathbf{L} \subseteq \mathbf{R}^{d_{\lambda}}$ is known, and $\mathbf{H}^*(\lambda) = \{h \in \mathbf{H}(\lambda) : h \text{ satisfies \eqref{eq:D_moments}}\}$ is a singleton or an empty set for each $\lambda \in \mathbf{L}$.
\end{assumption2}

Recall in the unidimensional case $h$ is point-identified as $F$ is known and $g$ can be identified by the moments in \eqref{eq:D_moments}. Assumption \ref{ass:PS} aims to extend this idea to the general case with multidimensional unobserved heterogeneity by assuming $h$ can be parameterized by $\lambda \in \mathbf{L}$ such that an analogous argument applies for each $\lambda \in \mathbf{L}$. In particular, for each $\lambda \in \mathbf{L}$, it requires $\mathbf{H}(\lambda)$ and the data variation in \eqref{eq:D_moments} to be such that there exists a unique value of $h$ consistent with the data so that it is point identified or that no such $h$ exists and the model is misspecified.

For the purposes of computation, Assumption \ref{ass:PS} along with Proposition \ref{prop:linearprog} allows us to compute the identified set using a two-step strategy performed across $\lambda \in \mathbf{L}$, which we can implement in practice using a grid of values in $\mathbf{L}$. To see this, observe that \eqref{eq:Theta_union} can be rewritten as
\begin{align}\label{eq:Theta_union_alpha}
 \Theta = \bigcup_{\lambda \in \mathbf{L}^*} [\theta_L(h^*(\lambda)),\theta_U(h^*(\lambda))]
\end{align}
where $h^*(\lambda)$ is the point-identified value of $h$ when $\mathbf{H}^*(\lambda)$ is non-empty, $\theta_L$ and $\theta_U$ are defined as in \eqref{eq:optimization}, and $\mathbf{L}^* = \{\lambda \in \mathbf{L} : \mathbf{H}^*(\lambda) \neq \emptyset,~\mathbf{\Gamma}^{*}(h^*(\lambda)) \neq \emptyset \}$ is the set of $\lambda$ for which the model is not misspecified. In turn, in an inner step for each $\lambda \in \mathbf{L}$, we can compute $\theta^{\Gamma}(\mathbf{\Gamma}^*(h^*(\lambda)),h^*(\lambda))$ by solving the linear programs in \eqref{eq:optimization}, where, as elaborated in Section \ref{sec:estimation_inference}, the point-identified value $h^*(\lambda)$ can be computed by solving a minimization problem based on the distance of the parameters to the data in the moments in \eqref{eq:D_moments}. In an outer step, we can then take the union of these sets when non-empty across $\lambda \in \mathbf{L}$ to obtain the identified set $\Theta$.

We conclude this section by revisiting the examples in Section \ref{sec:selection} and highlighting how Assumption \ref{ass:PS} can be naturally satisfied in them under common parametric assumptions on $F$.

\begin{example}[continues=ex:arum]\label{ex:arum2}
In this model, it is more convenient to work with the representation in \eqref{eq:arum1}. We first show that $\tilde{g} \equiv (\tilde{g}_{d,z|x} : d \in \{1,2\},~z \in \mathcal{Z},~x \in \mathcal{X})$ can be point identified given a known $\tilde{F} \equiv (\tilde{F}_{x} : x \in \mathcal{X})$, where $\tilde{F}_x$ denotes the conditional on $X=x \in \mathcal{X}$ distribution of $(\tilde{U}_1,\tilde{U}_2)$. Observe that the non-redundant moments in \eqref{eq:D_moments}, where the moments for one treatment are dropped given that they sum to one, in terms of $\tilde{g}$ and $\tilde{F}$ can be written as
\begin{align}
P(D=1|Z=z,X=x) &= \int 1\{\tilde{u}_1 < \tilde{g}_{1,z|x}, \tilde{u}_2 - \tilde{u}_1 > \tilde{g}_{2,z|x} - \tilde{g}_{1,z|x}\}d\tilde{F}_x(\tilde{u}_1,\tilde{u}_2)~, \label{eq:D_moments_arum1} \\
P(D=2|Z=z,X=x) &= \int 1\{\tilde{u}_2 < \tilde{g}_{2,z|x}, \tilde{u}_1 - \tilde{u}_2 > \tilde{g}_{1,z|x} - \tilde{g}_{2,z|x}\}d\tilde{F}_x(\tilde{u}_1,\tilde{u}_2)~, \label{eq:D_moments_arum2}
\end{align}
for $z \in \mathcal{Z},~x \in \mathcal{X}$. The following proposition shows that if $\tilde{F}_x$ has a strictly positive density for $x \in \mathcal{X}$, then there exists at most a value of $\tilde{g}$ consistent with these moments.

\begin{proposition}\label{prop:arum}
If $\tilde{F}_x$ is known, and has a strictly positive density on $\mathbf{R}^2$ for each $x \in \mathcal{X}$, then there exists at most a single value of $\tilde{g}$ that satisfies \eqref{eq:D_moments_arum1} and \eqref{eq:D_moments_arum2}.
\end{proposition}

Let $\tilde{\mathbf{H}}(\lambda)$ denote the set of all values of $(\tilde{g},\tilde{F})$, where $\tilde{F}$ equals a known parametric distribution $\bar{F}(\lambda)$ with a strictly positive density and parametrized by $\lambda \in \mathbf{L}$, and let $\phi$ denote the relation between $(g,F)$ and $(\tilde{g},\tilde{F})$ implied by $U_1 = \tilde{F}_{1|x}(\tilde{U}_1)$, $U_2 = \tilde{F}_{2|x}(\tilde{U}_2)$ and $U_3 = \tilde{F}_{12|x}(\tilde{U}_1-\tilde{U}_2)$ such that $(g,F) = \phi((\tilde{g},\tilde{F}))$. For example, we can take $\bar{F}(\lambda)$ with $\lambda = (\lambda_x : x \in \mathcal{X}) \in \mathbf{L} = [-1,1]^{|\mathcal{X}|}$ such that
\begin{align}\label{eq:probit}
\bar{F}_x(\lambda)(\tilde{u}_1,\tilde{u}_2) = \Phi_{\lambda_x}(\tilde{u}_1,\tilde{u}_2)~,
\end{align}
where $\Phi_{\lambda_x}$ denotes the distribution function of a bivariate standard normal distribution with correlation coefficient $\lambda_x \in (-1,1)$ as in a probit model---see \citet[][Chapter 2.4]{train:09} for a discussion of various other common distributional parameterizations. To ensure Assumption \ref{ass:PS}, we can then take $\mathbf{H}(\lambda) = \phi(\tilde{\mathbf{H}}(\lambda))$ for $\lambda \in \mathbf{L}$. It follows from Proposition \ref{prop:arum} that $\mathbf{H}^*(\lambda)$ is a singleton for each $\lambda \in \mathbf{L}$ as $\tilde{F} = \bar{F}(\lambda)$ is known and $\tilde{g}$ is point identified by \eqref{eq:D_moments_arum1}-\eqref{eq:D_moments_arum2}, given which both $F$ and $g$ are known and point identified by $(g,F) = \phi((\tilde{g},\tilde{F}))$. We note that in cases where there exist restrictions on $\tilde{g}$, such as those discussed in Section \ref{sec:selection}, we can impose them through $\tilde{\mathbf{H}}(\lambda)$, and in which case we can potentially also identify some parameters of $\tilde{F}$ and reduce the dimension of $\lambda$---see Section \ref{sec:preschoolmodel} for an example in the context of our empirical application.
\end{example}

\begin{example}[continues=ex:sequential]\label{ex:sequential2}
Observe that the non-redundant moments in \eqref{eq:D_moments} in this model can be written as
\begin{align}
P(D=0|Z=z,X=x) &= 1 - g_{1,z|x}~, \label{eq:D_moments_sequential1}\\
P(D=1|Z=z,X=x) &= g_{1,z|x} - F_x(g_{1,z|x},g_{2,z|x})~, \label{eq:D_moments_sequential2}
\end{align}
for $z \in \mathcal{Z}$ and $x \in \mathcal{X}$. The following proposition shows that if $F_x$ is strictly increasing in one of the dimensions for $x \in \mathcal{X}$, then there exists a unique value of $g$ consistent with these moments.

\begin{proposition}\label{prop:sequential}
If $F_x(u_1,u_2)$ is known and strictly increasing in $u_2$ for $u_1 > 0$ for each $x \in \mathcal{X}$, and $g_{1,z|x} > 0$ for each $z \in \mathcal{Z}$ and $x \in \mathcal{X}$, then there exists at most a single value of $g$ that satisfies \eqref{eq:D_moments_sequential1} and \eqref{eq:D_moments_sequential2}.
\end{proposition}

As $U$ has uniformly distributed marginals, let $\bar{F}(\lambda)$ denote a known parametric distribution for $F$, where $\bar{F}_x(\lambda)$ is a continuous copula that is strictly increasing in the second dimension and parametrized by $\lambda \in \mathbf{L}$. For example, we can take $\bar{F}(\lambda)$ with $\lambda = (\lambda_x : x \in \mathcal{X}) \in \mathbf{L} = [-1,1]^{|\mathcal{X}|}$ such that
\begin{align}\label{eq:normal_copula}
\bar{F}_x(\lambda)(u_1,u_2) = \Phi_{\lambda_x}(\Phi^{-1}(u_1),\Phi^{-1}(u_2))~,
\end{align}
where $\Phi$ additionally denotes the distribution function of a standard normal distribution, i.e. a bivariate normal copula with correlation parameter $\lambda_x \in [-1,1]$---see \cite{nelsen:07} for examples of various other families of copulas. To ensure Assumption \ref{ass:PS}, we can then take $\mathbf{H}(\lambda)$ to equal the set of all values of $(g,F)$, where $F = \bar{F}(\lambda)$ and $g_{1,z|x} > 0$ for $z \in \mathcal{Z}$ and $x \in \mathcal{X}$. It directly follows from Proposition \ref{prop:sequential} that $\mathbf{H}^*(\lambda)$ will be a singleton or empty for each $\lambda \in \mathbf{L}$. We note that a restriction such as $U_1 = U_2$, an empirically relevant case as highlighted in Section \ref{sec:selection}, can be accommodated by further restricting $\bar{F}(\lambda)$ to satisfy this requirement. For example, in the context of \eqref{eq:normal_copula}, this amounts to simply restricting $\lambda_x  = 1$ for $x \in \mathcal{X}$.
\end{example}

\begin{example}[continues=ex:dynamic]\label{ex:dynamic2}
Observe that the non-redundant moments in \eqref{eq:D_moments} in this model can be written as
\begin{align}
 P(D = (0,1) | Z=z,X=x) &= g_{2,z|x} - F_x(g_{1,z|x},g_{2,z|x},1) \label{eq:D_moments_dynamic1} \\
 P(D = (1,0) | Z=z,X=x) &= g_{1,z|x} - F_x(g_{1,z|x},1,g_{3,z|x}) \label{eq:D_moments_dynamic2} \\
 P(D = (1,1) | Z=z,X=x) &= F_x(g_{1,z|x},1,g_{3,z|x}) \label{eq:D_moments_dynamic3}
\end{align}
for $z \in \mathcal{Z}$ and $x \in \mathcal{X}$. The following proposition shows that if $F_x$ is strictly increasing in its third dimension and its derivative with respect to the second dimension is less than one, and $g_{1,z|x} \in (0,1)$ so that both nodes are possible in the second stage, then there exists a unique value of $g$ consistent with these moments.

\begin{proposition}\label{prop:dynamic}
\sloppy If $F_x(u_1,1,u_3)$ is known and strictly increasing in $u_3$ for $u_1 > 0$ and $\partial F_x(u_1,u_2,1) / \partial u_2 < 1$ for $u_1 < 1$ for each $x \in \mathcal{X}$, and $g_{1,z|x} \in (0,1)$ for each $z \in \mathcal{Z}$ and $x \in \mathcal{X}$, then there exists at most a single value of $g$ that satisfies \eqref{eq:D_moments_dynamic1}-\eqref{eq:D_moments_dynamic3}.
\end{proposition}

As $U$ has uniformly distributed marginals, let $\bar{F}(\lambda)$ denote a known parametric distribution for $F$, where $\bar{F}_x(\lambda)$ is a continuous copula that is strictly increasing in the third dimension for $u_1 > 0$ and with the property that $\partial F_x(u_1,u_2,1) / \partial u_2 < 1$ for $u_1 < 1$, and parametrized by $\lambda \in \mathbf{L}$. For example, similar to \eqref{eq:probit} and \eqref{eq:normal_copula}, we can take $\bar{F}(\lambda)$ with $\lambda = ((\lambda_{12|x},\lambda_{13|x},\lambda_{23|x}) : x \in \mathcal{X}) \in \mathbf{L} = [-1,1]^{3|\mathcal{X}|}$ such that
\begin{align}\label{eq:normal_copula_multi}
\bar{F}_x(\lambda)(u_1,u_2,u_3) = \Phi_{(\lambda_{12|x},\lambda_{13|x},\lambda_{23|x})}(\Phi^{-1}(u_1),\Phi^{-1}(u_2),\Phi^{-1}(u_3))~,
\end{align}
where $\Phi_{(\lambda_{12|x},\lambda_{13|x},\lambda_{23|x})}$ denotes the distribution function of a trivariate standard normal distribution with correlation between $u_k$ and $u_l$ given by $\lambda_{kl} \in (-1,1)$, i.e. a multivariate normal copula. Indeed, it is increasing in the third dimension, and moreover, as
\begin{align*}
 \frac{\partial}{\partial u_2} \Phi_{(\lambda_{12|x},\lambda_{13|x},\lambda_{23|x})}(\Phi^{-1}(u_1),\Phi^{-1}(u_2),\Phi^{-1}(1)) = \Phi\left( \frac{ \Phi^{-1}(u_1) - \lambda_{12|x} \Phi^{-1}(u_2) }{\sqrt{1 - \lambda^2_{12|x}}} \right)~,
\end{align*}
we also have that it satisfies the property that $\partial F_x(u_1,u_2,1) / \partial u_2 < 1$ for $u_1 < 1$. To ensure Assumption \ref{ass:PS}, we can then take $\mathbf{H}(\lambda)$ to equal the set of all values of $(g,F)$, where $F = \bar{F}(\lambda)$ and $g_{1,z|x} \in (0,1)$ for $z \in \mathcal{Z}$ and $x \in \mathcal{X}$. It directly follows from Proposition \ref{prop:dynamic} that $\mathbf{H}^*(\lambda)$ will be a singleton or empty for each $\lambda \in \mathbf{L}$. As in the above examples, we note that additional restrictions on $g$ and $U$ such as those discussed in Section \ref{sec:selection} can be accommodated through $\mathbf{H}(\lambda)$.
\end{example}

\begin{example}[continues=ex:double]\label{ex:double2}
Observe that the non-redundant moments in \eqref{eq:D_moments} in this model can be written as
\begin{align}\label{eq:D_moments_double}
P(D=1|Z=z,X=x) &= F_x(g_{1,z|x},g_{2,z|x})~,
\end{align}
for $z \in \mathcal{Z}$ and $x \in \mathcal{X}$. Here, unlike Examples \ref{ex:arum2} and \ref{ex:sequential2} above, there are two unknown thresholds, $g_{1,z|x}$ and $g_{2,z|x}$, but only a single moment for each $x \in \mathcal{X}$. In turn, even when $F_x$ is known, it is generally difficult to identify the remaining two unknowns using a single moment. As in \citet[][Section 4.2]{lee/salanie:18}, it is therefore useful to suppose that $\mathcal{Z} = \mathcal{Z}_1 \times \mathcal{Z}_2$ and $g$ to be restricted such that $g_{1,z|x} \equiv g_{1,z_1|x}$ and $g_{2,z|x} \equiv g_{2,z_2|x}$ for all $z = (z_1,z_2) \in \mathcal{Z}$ and $x \in \mathcal{X}$, i.e. there exist two instruments and an exclusion restriction imposing that each instrument affects only one of the thresholds, and that $g_{1,z_1'|x}$ is known for some $z_1' \in \mathcal{Z}_1$. The following proposition shows that if $F_x$ is assumed to be strictly increasing in both its dimensions then at most a single value of $g$ can be consistent with the moments.

\begin{proposition}\label{prop:double}
For each $x \in \mathcal{X}$, let $\mathcal{Z} = \mathcal{Z}_1 \times \mathcal{Z}_2$ be such that $g_{1,z|x} \equiv g_{1,z_1|x} \in (0,1)$ and $g_{2,z|x} \equiv g_{2,z_2|x} \in (0,1)$ for all $z = (z_1,z_2) \in \mathcal{Z}$, and $g_{1,z'_1|x}$ be known for some $z'_{1} \in \mathcal{Z}_1$. If $F_x$ is strictly increasing in both its dimensions when $u_1,u_2 > 0$ and for each $x \in \mathcal{X}$, then there exists at most a single value of $g$ that satisfies \eqref{eq:D_moments_double}.
\end{proposition}

Unlike Propositions \ref{prop:arum} and \ref{prop:sequential} above, Proposition \ref{prop:double} requires $g_{1,z'_1|x}$ to be known. This captures the fact that $g_{1,z|x}$ and $g_{2,z|x}$ can generally be point identified only up to a constant---see also \citet[][Section 4.2]{lee/salanie:18} that show this holds even in the presence of continuous variation in the thresholds. For the purposes of Assumption \ref{ass:PS}, for $\lambda_{1} = (\lambda_{1|x} : x \in \mathcal{X}) \in \mathbf{L}_1 = (0,1)^{|\mathcal{X}|}$, let $\mathbf{G}(\lambda_1)$ equal the set of all $g$ such that, for each $x \in \mathcal{X}$, $g_{1,z|x} \equiv g_{1,z_1|x} \in (0,1)$ and $g_{2,z|x} \equiv g_{2,z_2|x} \in (0,1)$ for $z = (z_1,z_2) \in \mathcal{Z}_1 \times \mathcal{Z}_2 \equiv \mathcal{Z}$, and $g_{1,z'_1|x} = \lambda_{1|x}$ for a given $z_1' \in \mathcal{Z}_1$, i.e. thresholds that takes each instrument to affect only one of the thresholds and where $g_{1,z_1'|x}$ is known and equal to $\lambda_{1|x}$ for a given $z_1' \in \mathcal{Z}_1$. Moreover, as $U$ has uniformly distributed marginals, let $\bar{F}(\lambda_2)$ denote a known parametric distribution for $F$, where $\bar{F}_x(\lambda_2)$ is a continuous copula that is strictly increasing in both dimensions and parametrized by $\lambda_2 \in \mathbf{L}_2$. To ensure Assumption \ref{ass:PS}, we can then take $\mathbf{H}(\lambda) = \{(g,F) : g \in \mathbf{G}(\lambda_1),~F = \bar{F}(\lambda_2)\}$ for $\lambda = (\lambda_1,\lambda_2) \in \mathbf{L}_1 \times \mathbf{L}_2$. It directly follows from Proposition \ref{prop:double} that $\mathbf{H}^*(\lambda)$ will be a singleton or empty for each $\lambda \in \mathbf{L}$. In practice, one could take $\bar{F}(\lambda_2)$ and $\mathbf{L}_2$ similar to that in \eqref{eq:normal_copula}.
\end{example}

\section{Estimation and Statistical Inference}\label{sec:estimation_inference}

We next show how the two-step characterization of the identified set in \eqref{eq:Theta_union_alpha} lends itself to a tractable strategy to estimate the identified set and construct confidence intervals when the moments of the data in \eqref{eq:D_moments} and \eqref{eq:Y_moments_alpha} can only be estimated using sample data.

\subsection{Estimation}\label{sec:estimation}

To estimate the identified set, a natural starting point is the plug-in estimator that replaces $P_{d|z,x}$ and $E_{d|z,x}$ in \eqref{eq:D_moments} and \eqref{eq:Y_moments_alpha} with their sample analogs. However, this approach can deliver an empty estimated set in practice, because the moments may not be exactly satisfied even when the true identified set is non-empty. Following \citet{mogstad/etal:18}, we instead base our estimator on a criterion that measures how well different values of the underlying primitives match the observed moments and retains those that best match the moments.

To describe the estimation procedure, it is useful to first restate the identified set in terms of such criteria. For the purposes of the point-identified value $h^*(\lambda)$, let
\begin{align}\label{eq:Q_sel}
 Q_{\text{sel}}(h) = \min_{\pi} (P - \pi)'\Omega_{S} (P - \pi)
\end{align}
subject to $\int_{\mathcal{U}_{d,z|x}^{\text{sm}}(g)} dF_{x} = \pi_{d|z,x}$ for $d \in \mathcal{D}$, $z \in \mathcal{Z}$, $x \in \mathcal{X}$, where $P$ and $\pi$ denote $(P_{d|z,x} : d \in \mathcal{D},~z \in \mathcal{Z},~x \in \mathcal{X})$ and $(\pi_{d|z,x} : d \in \mathcal{D},~z \in \mathcal{Z},~x \in \mathcal{X})$ in vector notation, and $\Omega_{S}$ is a positive definite weight matrix. This criterion measures the squared distance of the model implied moments to the data selection moments in \eqref{eq:D_moments}. As $Q_{\text{sel}}(h) = 0$ whenever $h$ matches these moments, $h^*(\lambda)$ for each $\lambda \in \mathbf{L}$ if it exists can then be defined by
\begin{align}\label{eq:h_pointidentifed}
 h^*(\lambda) = \text{arg}\min_{h \in \mathbf{H}(\lambda)} Q_{\text{sel}}(h)~.
\end{align}
For the criterion that captures the condition that $\mathbf{H}^*(\lambda)$ and $\mathbf{\Gamma}^*(h^*(\lambda))$ are non-empty, we first note that this can be equivalently stated as there existing $\gamma \in \mathbf{R}^{\text{dim}(\gamma)}$ such that
\begin{align}\label{eq:matrix_est}
 C_1(\lambda) \mu(\lambda) + (C_1(\lambda) C_2(\lambda) + C_3(\lambda)) \gamma \leq c(\lambda)~,
\end{align}
where $\mu(\lambda)$ and $C_2(\lambda)$ are estimable quantities that depend on the right hand sides of the moments in \eqref{eq:D_moments} and \eqref{eq:Y_moments_alpha} evaluated at $h$ equal to $h^*(\lambda)$, and $C_1(\lambda)$, $C_3(\lambda)$ and $c(\lambda)$ are known quantities underlying the linear restrictions $\mathbf{\Gamma}^*(h^*(\lambda))$ and the condition that $h^*(\lambda) \in \mathbf{H}^*(\lambda)$. For brevity, we present the exact forms of these vectors and matrices in Appendix \ref{Asec:Bformulation}. As in \eqref{eq:Q_sel}, let
\begin{align}\label{eq:Q_main}
 Q(\gamma,\lambda) = \min_{\bar{\mu}}(\mu(\lambda) - \bar{\mu})' \Omega_M(\lambda) (\mu(\lambda) - \bar{\mu})
\end{align}
subject to $C_1(\lambda) \bar{\mu} + (C_1(\lambda) C_2(\lambda) + C_3(\lambda)) \gamma \leq c(\lambda)$, where $\Omega_M(\lambda)$ is a positive definite weight matrix, i.e. the minimum possible squared distance of the model implied moments to the data moments in \eqref{eq:D_moments} and \eqref{eq:Y_moments_alpha} for a given $(\gamma,\lambda)$. Again, as $Q(\gamma,\lambda) = 0$ whenever $\gamma \in \mathbf{\Gamma}^*(h^*(\lambda))$ and $h^*(\lambda) \in \mathbf{H}^*(\lambda)$, $\mathbf{L}^*$ and $\mathbf{\Gamma}^*(h^*(\lambda))$ in \eqref{eq:Theta_union_alpha} can then be defined by
\begin{align}
 \mathbf{L}^* &= \{\lambda \in \mathbf{L} : \min_{\gamma \in \mathbf{\Gamma}(h^*(\lambda))} Q(\gamma,\lambda) = 0\}~, \label{eq:L_criterion}\\
 \mathbf{\Gamma}^*(h^*(\lambda)) &= \{\gamma \in \mathbf{\Gamma}(h^*(\lambda)) : Q(\gamma,\lambda) = 0\}~. \label{eq:A_criterion}
\end{align}

The estimated identified set $\widehat{\Theta}$ is obtained by substituting in the estimated versions of \eqref{eq:h_pointidentifed}, \eqref{eq:L_criterion} and \eqref{eq:A_criterion}, where the criterions in their definitions are replaced by their sample analogs. In particular, the estimated version of $h^*(\lambda)$ in \eqref{eq:h_pointidentifed} is given by
\begin{align}\label{eq:h_pointidentified_est}
 \widehat{h}^*(\lambda) = \text{arg}\min_{h \in \mathbf{H}(\lambda), \pi} (\widehat{P} - \pi)'\widehat{\Omega}_{S} (\widehat{P} - \pi)
\end{align}
subject to $\int_{\mathcal{U}_{d,z|x}^{\text{sm}}(g)} dF_{x} = \pi_{d|z,x}$ for $d \in \mathcal{D}$, $z \in \mathcal{Z}$, $x \in \mathcal{X}$, where $\widehat{P}$ is the sample analog estimator of $P$ and $\widehat{\Omega}_{S}$ is a consistent estimator of $\Omega_{S}$. Similarly, the estimated value of $Q(\gamma,\lambda)$ used to obtain estimates of \eqref{eq:L_criterion} and \eqref{eq:A_criterion} is given by
\begin{align}\label{eq:Q_main_est}
 \widehat{Q}(\gamma,\lambda) = \min_{\bar{\mu}}(\widehat{\mu}(\lambda) - \bar{\mu})' \widehat{\Omega}_M(\lambda) (\widehat{\mu}(\lambda) - \bar{\mu})
\end{align}
subject to $C_1(\lambda) \bar{\mu} + (C_1(\lambda) \widehat{C}_2(\lambda) + C_3(\lambda)) \gamma \leq c(\lambda)$, where $\widehat{C}_2(\lambda)$ and $\widehat{\mu}(\lambda)$ are estimates of $C_2(\lambda)$ and $\mu(\lambda)$ obtained by replacing $P_{d|z,x}$, $E_{d|z,x}$ and $h^*(\lambda)$ in the expression provided in Appendix \ref{Asec:Bformulation} by their estimated analogs $\widehat{P}_{d|z,x}$, $\widehat{E}_{d|z,x}$ and $\widehat{h}^*(\lambda)$, and $\widehat{\Omega}_M(\lambda)$ is a consistent estimator of $\Omega_M(\lambda)$. However, simply using $\widehat{Q}$ to obtain the estimated versions of \eqref{eq:L_criterion} and \eqref{eq:A_criterion}, and hence the identified set, can result in an empty set as $\min_{\gamma \in \mathbf{\Gamma}(\widehat{h}^*(\lambda))}\widehat{Q}(\gamma,\lambda)$ may not equal 0 for any $\lambda \in \mathbf{L}$. Following \cite{mogstad/etal:18}, we therefore instead take the estimated versions of \eqref{eq:L_criterion} and \eqref{eq:A_criterion} to be
\begin{align}
 \widehat{\mathbf{L}}^* &= \{\lambda \in \mathbf{L} : \min_{\gamma \in \mathbf{\Gamma}(\widehat{h}^*(\lambda))}\widehat{Q}(\gamma,\lambda) \leq Q^* + \kappa\}~, \label{eq:L_hat}\\
 \widehat{\mathbf{\Gamma}}^*(\widehat{h}^*(\lambda)) &= \{\gamma \in \mathbf{\Gamma}(\widehat{h}^*(\lambda)) : \widehat{Q}(\gamma,\lambda) \leq Q^* + \kappa\}~, \label{eq:A_hat}
\end{align}
where $Q^* = \min_{\lambda \in \mathbf{L}} \min_{\gamma \in \mathbf{\Gamma}(\widehat{h}^*(\lambda))}\widehat{Q}(\gamma,\lambda)$, i.e. the subsets of $\mathbf{L}$ and $\mathbf{\Gamma}(\widehat{h}^*(\lambda))$ that come closest to satisfying the conditions on the estimated criterions up to a tolerance $\kappa$. By construction, these sets, and hence the estimated identified set, are never empty. Here $\kappa$ is a tuning parameter that goes to 0 with sample size at a specific rate and is necessary to ensure that the estimated set is consistent. In cases where $\mathbf{L}$ is a singleton and there exists a unique value of $\gamma$ that ensures $Q(\gamma,h^*(\lambda))$ = 0 so that the model is point identified, we can take $\kappa$ to equal 0.

The above procedure can be summarized by the following algorithm, which can in practice be implemented using a grid of values in $\mathbf{L}$:

\begin{tabularx}{\linewidth}{lX}
Step 1: & For each $\lambda \in \mathbf{L}$, compute $\widehat{h}^*(\lambda)$ by solving \eqref{eq:h_pointidentified_est}.  \\
Step 2: & For each $\lambda \in \mathbf{L}$, compute $\min\limits_{\gamma \in \mathbf{\Gamma}(\widehat{h}^*(\lambda))}\widehat{Q}(\gamma,\lambda)$, where $\widehat{Q}$ is defined in \eqref{eq:Q_main_est}. \\
Step 3: & Compute $\widehat{\mathbf{L}}^*$ as in \eqref{eq:L_hat}. \\
Step 4: & For each $\lambda \in \widehat{\mathbf{L}}^*$, compute $\theta_L(\widehat{h}^*(\lambda))$ and $\theta_U(\widehat{h}^*(\lambda))$ in \eqref{eq:optimization} using $\widehat{\mathbf{\Gamma}}^*(\widehat{h}^*(\lambda))$ in \eqref{eq:A_hat}. \\
 Step 5: & Take $\widehat{\Theta} = \bigcup_{\lambda \in \widehat{\mathbf{L}}^* } [ \widehat{\theta}_L(\widehat{h}^*(\lambda)), \widehat{\theta}_U(\widehat{h}^*(\lambda))]$.
\end{tabularx}

In terms of computation, Step 1 is potentially the most expensive step as \eqref{eq:h_pointidentified_est} can be a nonlinear optimization problem depending on how $h$ enters the problem. For the parametric distributions considered for the examples in Section \ref{sec:selection_parameterization}, these nonlinear problems are similar to those encountered in the estimation of parametric discrete choice problems \citep[e.g.,][Chapter 8]{train:09}. In contrast, as in the computations in \eqref{eq:optimization}, the linearity of $\mathbf{\Gamma}$ and $\theta^{\Gamma}$ and the quadraticity of the sample analog $\widehat{Q}$ in \eqref{eq:Q_main} in terms of $\gamma$ given $\lambda \in \mathbf{L}$ implies that Steps 2 and 4 are more structured and correspond to convex quadratic and quadratically-constrained quadratic programs in $(\bar{\mu},\gamma)$, respectively. In cases where we know the model is point identified and $\kappa$ is taken to be 0, Step 4 can be further simplified. In particular, let $\gamma^* = \text{arg}\min_{\gamma \in \mathbf{\Gamma}(\widehat{h}^*(\lambda))}\widehat{Q}(\gamma,\lambda)$ be the unique solution in Step 2 and the unique value in $\widehat{\mathbf{\Gamma}}^*(\widehat{h}^*(\lambda))$. In Step 4, we then simply have $\widehat{\theta}_L(\widehat{h}^*(\lambda)) = \widehat{\theta}_U(\widehat{h}^*(\lambda)) = \theta^{\Gamma}(\gamma^*,\widehat{h}^*(\lambda))$.

\subsection{Statistical Inference}\label{sec:inference}

Confidence intervals can be constructed by test inversion, i.e. test at level $\alpha \in (0,1)$ the null hypothesis $H_0 : \theta_0 \in \Theta$ and then construct $100(1-\alpha)\%$-confidence intervals by collecting the values of $\theta_0$ that are not rejected. To do so in a computationally tractable manner, we again exploit, as in our estimation procedure, the fact the $h$ is point identified by $h^*(\lambda)$ in \eqref{eq:h_pointidentifed} and the linearity of $\theta^{\Gamma}$ and $\mathbf{\Gamma}^*(h^*(\lambda))$ in $\gamma$ to test $H_{0}(\lambda): \theta_0 \in \theta^{\Gamma}(\mathbf{\Gamma}^*(h^*(\lambda)),h^*(\lambda)),~ h^*(\lambda) \in \mathbf{H}^*(\lambda)$. We then show how to aggregate these tests across $\lambda \in \mathbf{L}$.

For each $\lambda \in \mathbf{L}$, we exploit the observation that, similar to \eqref{eq:matrix_est}, the null hypothesis can be reformulated as
\begin{align}\label{eq:nulllinear}
 H_0(\lambda) : \gamma \in \left\{ a \in \mathbf{R}^{d_{\gamma}} : C_1(\lambda) \mu(\lambda) + (C_1(\lambda) C_2(\lambda) + C_3(\lambda)) a \leq c(\lambda) \right\}
\end{align}
i.e. testing whether $\gamma$ satisfies a linear system of inequalities, where $\mu(\lambda)$, $C_1(\lambda)$, $C_2(\lambda)$, $C_3(\lambda)$ and $c(\lambda)$ are specified as in \eqref{eq:matrix_est}, but augmented to include the additional condition that $\theta^{\Gamma}(\gamma,h^*(\lambda)) = \theta_0$---see Appendix \ref{Asec:Bformulation} for their exact forms. This formulation allows us to use the procedure from \cite{cox/etal:24}, who show how to test null hypotheses that can be stated as \eqref{eq:nulllinear}. Similar to the criterion in \eqref{eq:Q_main_est}, it is based on a Quasi-Likelihood Ratio test statistic given by
\begin{align}\label{eq:TS}
TS(\lambda) = \min_{\bar{\mu}, \gamma}(\widehat{\mu}(\lambda) - \bar{\mu})'\widehat{\Sigma}^{-1}(\lambda)(\widehat{\mu}(\lambda) - \bar{\mu})
\end{align}
subject to $C_1(\lambda) \bar{\mu} + (C_1(\lambda) \widehat{C}_2(\lambda) + C_3(\lambda)) \gamma \leq c(\lambda)$, where $\widehat{\mu}(\lambda)$ and $\widehat{C}_2(\lambda)$ denote estimates of $\mu(\lambda)$ and $C_2(\lambda)$, and $\widehat{\Sigma}(\lambda)$ denotes an estimate of the asymptotic variance of $(\widehat{\mu}(\lambda) - \mu(\lambda) + (\widehat{C}_2(\lambda) - C_2(\lambda))\gamma^*)$ with
\begin{align}\label{eq:TS_V_min}
 \gamma^* = \text{arg}\min_{\gamma} \gamma'\gamma
\end{align}
subject to  $C_1(\lambda) \mu(\lambda) + (C_1(\lambda) C_2(\lambda) + C_3(\lambda)) \gamma \leq c(\lambda)$. To compute the critical value, let $\widehat{\bar{\mu}}$ and $\widehat{\gamma}$ denote the minimizers of the problem in \eqref{eq:TS}, and let $\widehat{K}$ denote the set of row indices corresponding to the binding inequalities in $C_1(\lambda) \widehat{\bar{\mu}} + (C_1(\lambda) \widehat{C}_2(\lambda) + C_3(\lambda)) \widehat{\gamma} \leq c(\lambda)$. Moreover, let
\begin{align}\label{eq:r_hat}
 \widehat{r} = \text{rank}([C_1(\lambda),C_3(\lambda)]_{\widehat{K}}) - \text{rank}((C_1(\lambda) \widehat{C}_2(\lambda) + C_3(\lambda))_{\widehat{K}})~,
\end{align}
where $[C_1(\lambda),C_3(\lambda)]_{\widehat{K}}$ and $(C_1(\lambda) \widehat{C}_2(\lambda) + C_3(\lambda))_{\widehat{K}}$ denote the submatrices of $[C_1(\lambda),C_3(\lambda)]$ and $(C_1(\lambda) \widehat{C}_2(\lambda) + C_3(\lambda))$ containing only the rows in $\widehat{K}$. The critical value $cv(\lambda,1-\alpha)$ is given by the $(1-\alpha)$-quantile of a chi-squared distribution with $\widehat{r}$ degrees of freedom. The test then equals $\phi(\lambda) = 1\{TS(\lambda) > cv(\lambda,1-\alpha)\}$, i.e. we reject if the test statistic is above the critical value and do not reject otherwise. Provided that $\widehat{\mu}(\lambda)$ is asymptotically normal and under weak assumptions on the consistency of $\widehat{C}_2(\lambda)$, \cite{cox/etal:24} show that this test controls size.

To aggregate these tests across $\lambda \in \mathbf{L}$ and test $H_0$, we exploit the principle underlying the so-called recycling test from \citet[][Section 5]{bugni/etal:15}. Let $\mathbf{L}_0 = \{\lambda \in \mathbf{L} : H_0(\lambda) \text{ is true} \}$ denote the set of $\lambda$ where $H_0(\lambda)$ is satisfied under $H_0$ and, similar to \eqref{eq:L_hat}-\eqref{eq:A_hat}, let $\widehat{\mathbf{L}}_0 = \{\lambda \in \mathbf{L} : TS(\lambda) \leq \min_{\lambda \in \mathbf{L}}TS(\lambda) + \kappa \}$ denote a consistent estimate of this set, where $\kappa$ is again a tuning parameter necessary to ensure that the estimated set is consistent. The aggregated test then equals
\begin{align}\label{eq:test}
 \phi = 1\{\min_{\lambda \in \mathbf{L}}TS(\lambda) > \min_{\lambda \in \widehat{\mathbf{L}}_0}cv(\lambda,1-\alpha)\}~,
\end{align}
i.e. we reject if the minimum test statistic across $\mathbf{L}$ is greater than the minimum critical value across $\widehat{\mathbf{L}}_0$. In Appendix \ref{Asec:prop_recycle}, we show that if $\phi(\lambda)$ controls asymptotic size for each $\lambda \in \mathbf{L}_0$ and $\widehat{\mathbf{L}}_0$ is a consistent estimate for $\mathbf{L}_0$, then the test in \eqref{eq:test} also controls asymptotic size.

As in our estimation procedure, we can summarize the test for a pre-specified choice of $\alpha \in (0,1)$ by the following algorithm:

\begin{tabularx}{\linewidth}{lX}
Step 1: & For each $\lambda \in \mathbf{L}$, compute $\widehat{\mu}(\lambda)$ and $\widehat{C}_2(\lambda)$.  \\
Step 2: & For each $\lambda \in \mathbf{L}$, compute $TS(\lambda)$ in \eqref{eq:TS}.  \\
Step 3: & For each $\lambda \in \mathbf{L}$, compute $cv(\lambda,1-\alpha)$ that equals the ($1-\alpha$)-quantile of the chi-squared distribution with degrees of freedom $\widehat{r}(\lambda)$ given by \eqref{eq:r_hat}.  \\
Step 4: & Compute the test in \eqref{eq:test} using $TS(\lambda)$ and $cv(\lambda,1-\alpha)$ computed in Steps 2 and 3. 
\end{tabularx}

For Step 1, we require, as in \eqref{eq:Q_main_est}, replacing $P_{d|z,x}$, $E_{d|z,x}$ and $h^*(\lambda)$ in the expression of $\mu(\lambda)$ and $C_1(\lambda)$ presented in Appendix \ref{Asec:Bformulation} by their estimated analogs $\widehat{P}_{d|z,x}$, $\widehat{E}_{d|z,x}$ and $\widehat{h}^*(\lambda)$. As in Step 1 of the estimation algorithm, this step can be computationally intensive as computing $\widehat{h}^*(\lambda)$ in \eqref{eq:h_pointidentified_est} can be a nonlinear problem. Step 2 is again more structured as the optimization problems in \eqref{eq:TS} and \eqref{eq:TS_V_min} are both convex quadratic problems. Given this algorithm, confidence intervals immediately follow by applying it over a fine grid of values of $\theta_0$ and then collecting the values where we do not reject. In practice, to speed up computation, this part can be easily parallelized across the different values of $\theta_0$.

\section{Estimating the Returns to Head Start}\label{sec:application}

In this section, we use our method to revisit the analysis of \citet[][KW hereafter]{kline/walters:16}, who evaluate Head Start (HS) using data from the Head Start Impact Study (HSIS).

\subsection{Background}\label{sec:background}

HS is a federally funded program in the United States that provides free preschool to three- and four-year-old children from disadvantaged families. The HSIS was a randomized experiment conducted in Fall 2002 to evaluate the program. Due to oversubscription, HS offers were randomized among applicants, enabling experimental evaluation by comparing outcomes of children who received an offer with those who did not \citep{puma_etal:10}.

However, HS is not the only available preschool option; many alternatives provide similar services and also receive public subsidies. A non-trivial proportion of families in the experiment enrolled in such preschools. As highlighted in KW, this feature clouds the interpretation of the experimental effects as they estimate a combination of HS effects relative to enrolling in such preschools and no preschools. Moreover, it complicates a cost-benefit analysis of HS using these effects as enrollment out of these preschools may introduce fiscal savings and as these preschools may adjust the number of slots offered in response to that of HS.

Motivated by this feature, KW formulate a selection model that governs choice into the different preschool options and how preschools allocate offers. But, in estimating this model using the experimental data, they impose various assumptions such that the model is point-identified. In what follows, we illustrate how our method can be used to analyze the robustness of their conclusions on the returns to HS to these assumptions.

\subsection{Setting and Data}\label{sec:setting_data}

We follow KW in mapping the experiment into the treatment effect setup in Section \ref{sec:structure}. Let $Z \in \{0,1\}$ denote an indicator for whether the child was assigned to the treatment group and offered HS, $D \in \mathcal{D}$ and $Y$ denote their preschool choice and outcome of interest in their first year in the experiment, and $X$ denote covariates. As in KW, we take $\mathcal{D} = \{p,c,n\}$, where $p$ denotes HS program preschools, $c$ denotes competing non-HS preschools, and $n$ denotes no preschool; and the outcome of interest to be the average of the Woodcock Johnson III and Peabody Picture and Vocabulary Test scores, with each test score normed to have mean 0 and variance 1 in the control group by cohort. While KW consider a vector of covariates, we take $X$ to be a single binary covariate for simplicity as it is sufficient to present the main ideas. In particular, we take $X \in \{0,1\}$ to be an indicator for whether the child was in the three- or four-year-old cohort. Our analysis sample includes 3,627 children, all three- and four-year-olds in the sample with complete data for all variables mentioned above.

\begin{table}[!t]
\begin{center}
\caption{Descriptive Statistics}
\label{tab:descriptive}
\scalebox{0.8}{
\def\arraystretch{1}
\makebox[\textwidth]{\begin{threeparttable}
\input{table_descriptive.tex}
\begin{tablenotes}[flushleft]
\setlength\labelsep{0pt}
\item \footnotesize Standard errors reported in brackets constructed using the nonparametric bootstrap that draws at the level of HS centers. As in KW, data moments are estimated with observations that are weighted by the inverse probability of the child's experimental assignment, calculated at the level of HS centers.
\end{tablenotes}
\end{threeparttable}}}
\end{center}
\end{table}

Table \ref{tab:descriptive} reports several motivating statistics. Panel (a) presents intention-to-treat (ITT) effects on enrollment into HS and outcomes defined by
\begin{align*}
 \text{ITT}^D &= \sum_{x \in \{0,1\}}P(X=x) \cdot \left(E[1\{D=p\}|Z=1,X=x] - E[1\{D=p\}|Z=0,X=x] \right)~, \\
 \text{ITT}^Y &= \sum_{x \in \{0,1\}}P(X=x) \cdot \left(E[Y|Z=1,X=x] - E[Y|Z=0,X=x]\right)~,
\end{align*}
and the instrumental variable (IV) estimand, defined as the average ratio of the two ITTs by cohort,
\begin{align}\label{eq:IV}
 \text{IV} = \sum_{x \in \{0,1\}}P(X=x) \cdot \frac{E[Y|Z=1,X=x] - E[Y|Z=0,X=x]}{E[1\{D=p\}|Z=1,X=x] - E[1\{D=p\}|Z=0,X=x]}~.
\end{align}
ITT$^Y$ reveals that the offer increased test scores, while the ITT$^D$ reveals noncompliance---not all children with an offer enrolled in HS and some without one did. The IV estimand accounts for this by estimating the effect of the offer for those who comply. We calculate a statistically significant effect of 0.230 standard deviations.\footnote{Our IV estimate is similar to the 0.247 standard deviation effect reported in KW. Minor differences reflect our simplified covariate specification and sample restrictions.}

As noted, evaluating the returns to HS using these experimental effects is complicated by the fact that many children without a HS offer enrolled in competing preschools. Panel (b) presents the proportion of children enrolled in the various care settings by offer status. Close to a third of the children without an offer enrolled in competing preschools, and the offer reduced this share. As the estimates below more precisely capture, this indicates that a non-trivial proportion of children induced into HS by the offer came from competing preschools.

\subsection{Model of Preschool Choice and Provision}\label{sec:preschoolmodel}

We consider the general multinomial selection model outlined in \citet[][Appendix C.2]{kline/walters:16} that allows for the possibility that competing preschools were also oversubscribed and randomized offers similar to HS. Recall that $Z_p \equiv Z \in \{0,1\}$ denotes the observed indicator for whether an offer to HS was randomly assigned. Moreover, as in KW, we introduce $Z_c$ to denote an unobserved indicator for whether an offer to a competing preschool was randomly assigned. Let $\delta_{p|x} \equiv E[Z_p | X{=}x]$ and $\delta_{c|x} \equiv E[Z_c | X{=}x]$ denote the offer probabilities with which HS and competing preschools are rationed, respectively, conditional on the cohort $x \in \{0,1\}$.

Following offers that are distributed by lottery in a first stage, families make choices in a second stage, where those who do not receive an offer from a preschool can enroll there by exerting additional effort. Let $V_d(z_p,z_c)$ define the family's utility from alternative $d \in \mathcal{D}$ had their offer statuses from HS and competing preschools been set to $(z_p, z_c) \in \{0,1\}^2$. The potential preschool choice given offer statuses equals the utility-maximizing choice given by
\begin{align}
 D(z_p,z_c) = \argmax_{d \in \mathcal{D}} V_d(z_p,z_c)~,
\end{align}
and the observed choice is simply the utility-maximizing one given their realized offers, i.e. $D = D(Z_p,Z_c)$. As in KW, we take $V_n(z_p,z_c) = 0$, i.e. normalize the utility of no preschools to 0, and impose the restriction that
\begin{align}
 V_d(z_p,z_c) \equiv V_d(z_d) = \zeta_{d|x} + \zeta_x z_d - \tilde{U}_d~ \text{ for }~d \in \{p,c\}~, \label{eq:U_HS_1}
\end{align}
where $\tilde{U}_d$ is the alternative-specific unobserved heterogeneity and $\zeta_x > 0$, i.e. an offer to a preschool only increases the utility for that preschool and the effect of an offer on utilities is the same for both types of preschool.\footnote{We find that the estimated $\zeta_x$ is greater than 0 and hence that the assumption $\zeta_x > 0$ is satisfied by the model. However, we note that we explicitly make this relevance-type assumption as we require $\zeta_x \neq 0$ for the selection model to be identified—see Proposition \ref{prop:firststage_HS} below.} Moreover, following KW, we assume the distribution on the unobserved heterogeneity conditional on the cohort $x \in \{0,1\}$ to be $\Phi_{\rho_x}$, i.e. a standard bivariate normal distribution with a correlation parameter $\rho_x$ that varies by cohort.\footnote{The restriction on $V_p(z_p,z_c)$ and the distributional assumption on $(\tilde{U}_p,\tilde{U}_c)$ are stated in \citet[][Section VII.B]{kline/walters:16}. The restriction on $V_c(z_p,z_c)$ follows from the discussion in \citet[][Section IX.A]{kline/walters:16}, where they note when using their selection model to compute the proportion induced into competing preschools from an offer they assume the utility value from an offer is comparable to that of a HS offer. }

For the purposes of our method, recall that the selection model is required to satisfy Assumption \ref{ass:PS}, i.e. it needs to be point-identified up to a finite-dimensional parameter. To this end, the above model corresponds to that in Example \ref{ex:arum} for $D(z_p,z_c)$, where the thresholds $\tilde{g} = (\tilde{g}_{d,z_p,z_c|x} : d \in \{p,c\},~z_p,z_c, x \in \{0,1\})$ satisfy the restrictions in \eqref{eq:U_HS_1} and the distribution of the unobservables $(\tilde{U}_p,\tilde{U}_c)$ corresponds to $\Phi_{\rho_x}$, which can be captured by taking $\tilde{g}$ and the distribution of unobservables $\tilde{F} = (\tilde{F}_x : x \in \mathcal{X})$ to satisfy
\begin{align}
 \tilde{g}_{p,z_p,0|x} = \tilde{g}_{p,z_p,1|x} &= \zeta_{p|x} + \zeta_x z_p \text{ for } z_p \in \{0,1\}~, \label{eq:HS_h_rest_1} \\
 \tilde{g}_{c,1,z_c|x} = \tilde{g}_{c,0,z_c|x} &= \zeta_{c|x} + \zeta_x z_c \text{ for } z_c \in \{0,1\}~,\label{eq:HS_h_rest_2} \\
 \tilde{F}_{x}(\tilde{u}_p,\tilde{u}_c) &= \Phi_{\rho_x}(\tilde{u}_p,\tilde{u}_c) \label{eq:HS_h_rest_3}
\end{align}
for some values $(\zeta_{p|x},\zeta_{c|x},\zeta_{x},\rho_x) \in \mathbf{R}^2 \times \mathbf{R}_{++} \times (-1,1)$ for each $x \in \mathcal{X}$. However, unlike showing how Assumption \ref{ass:PS} can be satisfied in Example \ref{ex:arum2} in Section \ref{sec:selection_parameterization}, a complication here is that one of the instruments, namely $Z_c$, is not observed. Nonetheless, given the additional restrictions that reduce the number of unknowns characterizing the thresholds, we can identify $(\zeta_{p|x},\zeta_{c|x},\zeta_{x},\rho_x)$ for a fixed value of the competing preschool offer probability $\delta_{c|x}$. In particular, instead of \eqref{eq:D_moments_arum1}-\eqref{eq:D_moments_arum2}, note that the selection moments under the restrictions in \eqref{eq:HS_h_rest_1}-\eqref{eq:HS_h_rest_3} and accounting for the fact that $Z_c$ is not observed correspond to
\begin{align}
P_{c|z_p,x} &= \sum_{z_c \in \{0,1\}} P_{z_c,c|x} \int_{\tilde{u}_c < \zeta_{c|x} + z_c \zeta_x, \tilde{u}_c - \tilde{u}_p < \zeta_{c|x} + z_c \zeta_x - \zeta_{p|x} - z_p \zeta_x}d\Phi_{\rho_x}(\tilde{u}_p,\tilde{u}_c) \equiv H_{c,z_p}(\sigma_x,\delta_{c|x})~,  \label{eq:D_moments_arum1_hs} \\
P_{p|z_p,x} &= \sum_{z_c \in \{0,1\}} P_{z_c,c|x} \int_{\tilde{u}_p < \zeta_{p|x} + z_p \zeta_x, \tilde{u}_p - \tilde{u}_c < \zeta_{p|x} + z_p \zeta_x - \zeta_{c|x} - z_c \zeta_x}d\Phi_{\rho_x}(\tilde{u}_p,\tilde{u}_c) \equiv H_{p,z_p}(\sigma_x,\delta_{c|x})~,  \label{eq:D_moments_arum2_hs}
\end{align}
\sloppy for $z_p,x \in \{0,1\}$, where $P_{1,c|x} \equiv \delta_{c|x}$ and $P_{0,c|x} \equiv 1 - \delta_{c|x}$, $\sigma_x = (\zeta_{p|x},\zeta_{c|x},\zeta_x,\rho_{x})$,  and $H_{c,z_p}$ and $H_{p,z_p}$ capture the right hand side of these moments in terms of the parameters underlying the restrictions in \eqref{eq:HS_h_rest_1}-\eqref{eq:HS_h_rest_3}. The following proposition states the identification result given a rank condition on $H = (H_{c,0},H_{c,1},H_{p,0},H_{p,1})$.

\begin{proposition}\label{prop:firststage_HS}
Suppose $J(\sigma,\delta_{c}) = \partial H(\sigma,\delta_{c}) / \partial \sigma$ has full rank for all $(\sigma,\delta_c) \in \mathbf{R}^2 \times \mathbf{R}_{++} \times (-1,1) \times [0,1]$. Then, for each $\delta_{c|x} \in [0,1]$, there exists at most a single value of $\sigma_x$ that satisfies \eqref{eq:D_moments_arum1_hs} and \eqref{eq:D_moments_arum2_hs}.
\end{proposition}

Given the rank condition, we can ensure Assumption \ref{ass:PS} by taking $\mathbf{H}(\lambda) = \phi(\tilde{\mathbf{H}})$ for $\lambda = (\delta_{c|0},\delta_{c|1}) \in [0,1]^2 = \mathbf{L}$, where $\tilde{\mathbf{H}}$ is the set of all values of $(\tilde{g},\tilde{F})$ that satisfy \eqref{eq:HS_h_rest_1}-\eqref{eq:HS_h_rest_3} and $\phi$ denotes the relation as defined in Example \ref{ex:arum2} in Section \ref{sec:selection_parameterization} such that $(g,F) = \phi((\tilde{g},\tilde{F}))$. It follows by  Proposition \ref{prop:firststage_HS} that $\mathbf{H}^*(\lambda)$ is a singleton for each $\lambda \in \mathbf{L}$ as $(\tilde{g},\tilde{F})$ can be point identified by \eqref{eq:D_moments_arum1_hs}-\eqref{eq:D_moments_arum2_hs} taking $(\delta_{c|0},\delta_{c|1}) = \lambda$ given which $(g,F)$ can be identified by $\phi(\tilde{g}),\tilde{F}))$. While it is difficult to analytically show that the rank condition holds given the nonlinear structure of the Jacobian, we can verify it numerically in practice using a grid for the values of $(\sigma,\delta_c)$---see Appendix \ref{Asec:rank_jacobian} for details.\footnote{Existing results on analytically showing full rank of the Jacobian in cases with normally distributed errors are present only for the binary choice case, where one exploits related properties to that of the inverse Mills ratio \citep[e.g.,][]{chern/etal:23,chern/etal:24,han/vytlacil:17}. Such results, however, do not straightforwardly extend to the multinomial case, and we hence leave it as an extension to future work.} Doing so we find that the Jacobian does indeed satisfy this condition.

In estimating the above model, KW additionally assume $\delta_{c|x} = 0$ for $x \in \mathcal{X}$.\footnote{While they do not explicitly make this assumption, we note that the special case of the model that they estimate in \citet[][Section VII.B]{kline/walters:16} that does not introduce $Z_c$ is only logically consistent with the discussion of their general model when taking $\delta_{c|x} = 0$.} As formalized by the above proposition, this ensures that the selection model is point-identified. In the context of the above model, this restriction can be interpreted as having no lottery for competing preschools and hence eliminating the possibility that their slots are rationed in a manner similar to HS. However, as highlighted in \citet[][Section IX.A]{kline/walters:16}, it is reasonable to believe that such preschools do indeed ration slots as the majority of states do not have universal preschool mandates and preschools face relatively fixed budgets. As our method does not require such an assumption, we use it below to analyze the robustness of their conclusions to it by allowing $\delta_{c|x}$ to vary in $[0,1]$, in which case the selection model is not necessarily point-identified.

\subsection{Decomposing IV}\label{sec:IVdecomposition}

We focus on two sets of parameters that underlie the main conclusions in KW. The first set of parameters decompose the returns to a HS offer evaluated by the IV estimand into those arising from children drawn into HS from competing preschools versus no preschool. Under our selection model, as $\zeta_x > 0$ for $x \in \mathcal{X}$, we can decompose the IV estimand in \eqref{eq:IV} as follows
\begin{align}
 IV &= \sum_{x \in \{0,1\}}P(X=x) \cdot \left( S_{np}(x) LATE_{np}(x) + (1 - S_{np}(x))LATE_{cp}(x) \right) \nonumber \\
 &= S_{np}LATE_{np} + (1 - S_{np})LATE_{cp}~, \label{eq:IVdecomposition}
\end{align}
where
\begin{align}
 S_{dp}(x) &= \frac{P(D(1,Z_c) = p, D(0,Z_c) = d | X=x)}{P(D(1,Z_c) = p, D(0,Z_c) \neq p | X=x)}~, \nonumber  \\
 LATE_{dp}(x) &= E[Y(p) - Y(d)|D(1,Z_c) = p, D(0,Z_c) = d, X=x]~\nonumber
\end{align}
capture the proportion induced by a HS offer into HS from alternative $d \in \{n,c\}$ and the (local) average effect for this subgroup of children for cohort $x \in \{0,1\}$, respectively, and
\begin{align}
 S_{dp} &= \sum_{x \in \{0,1\}}P(X=x) S_{dp}(x)~,\nonumber \\
 LATE_{dp} &= \frac{\sum\limits_{x \in \{0,1\}}P(X=x) \cdot S_{dp}(x) LATE_{dp}(x) }{\sum\limits_{x \in \{0,1\}}P(X=x) S_{dp}(x) }~,\nonumber
\end{align}
capture this proportion averaged across the two cohorts and the effect weighted by the proportion of the subgroup across the two cohorts, respectively. Here the first line in the decomposition, as outlined in \citet[][Section IV]{kline/walters:16}, follows from the arguments of \cite{imbens/angrist:94} as $\zeta_x \geq 0$ implies the monotonicity condition that if $D(0,Z_c) \neq D(1,Z_c)$ then $D(1,Z_c) = p$, while the second line averages the expressions across the two cohorts.

A challenge is that while $S_{np}$ in the decomposition is identified by the data through
\begin{align}\label{eq:Snh}
 S_{np} = \sum_{x \in \{0,1\}}P(X=x)  \frac{P(D=n|Z_p = 0,X=x) - P(D=n|Z_p = 1,X=x)}{P(D=p|Z_p = 1,X=x) - P(D=p|Z_p = 0,X=x)}
\end{align}
due to the monotonicity condition that also implies \eqref{eq:IVdecomposition}, $LATE_{np}$ and $LATE_{cp}$ are not. KW exploit the point-identified version of the selection model discussed above along with additional assumptions on the Marginal Treatment Response (MTR) functions under this model to identify these effects. We now examine how their conclusions hold under weaker alternatives.

As in Section \ref{sec:parameters}, we define the MTRs here by $m_{d|x}(u_p,u_c) = E[Y(d)|U_p = u_p, U_c = u_c, X=x]$ for $d \in \mathcal{D}$, $x \in \mathcal{X}$, where $U_p = \tilde{F}_{\tilde{U}_p|X}(\tilde{U}_p)$ and $U_c = \tilde{F}_{\tilde{U}_c|X}(\tilde{U}_c)$ correspond to transformed versions of the unobserved heterogeneity as discussed in Example \ref{ex:arum}. Note that we do not condition on $\tilde{F}_{\tilde{U}_p - \tilde{U}_c}(\tilde{U}_p - \tilde{U}_c)$ here as it is deterministically implied by $U_p$ and $U_c$. To ensure point identification, KW assume the MTRs to be linear in $u_p$ and $u_c$ and separable in $x$, i.e.
\begin{align}\label{eq:specKW}
 m_{d|x}(u_p,u_c) = \gamma_{d,1} u_p + \gamma_{d,2} u_c + \mu_{d|x}~.
\end{align}
To see why this implies point identification, observe the outcome moments in \eqref{eq:Y_moments} are given by
\begin{align}\label{eq:Y_moments_HS}
 E[Y 1\{D=d\} | Z_p=z_p, X=x] &= \sum_{z_c \in \{0,1\}} P_{z_c,c|x} \int_{\mathcal{U}^{\text{sm}}_{d,z_p,z_c}(g)} m_{d|x}(u_p,u_c) dF_x
\end{align}
for $d \in \mathcal{D}$, $z_p,x \in \{0,1\}$, where $\mathcal{U}^{\text{sm}}_{d,z_p,z_c}(g)$ is defined as in Example \ref{ex:arum}, and $(g,F)$ are implied by the restrictions on $(\tilde{g},\tilde{F})$ in \eqref{eq:HS_h_rest_1}-\eqref{eq:HS_h_rest_3} and the relation between $(g,F)$ and $(\tilde{g},\tilde{F})$ noted in Section \ref{sec:preschoolmodel}. For each $d$, we have four moments in \eqref{eq:Y_moments_HS} as $z_p$ and $x$ take two values. Moreover, given that KW assume $\delta_{c|x}$ for $x \in \mathcal{X}$ to be known, Proposition \ref{prop:firststage_HS} implies that there exist only four unknowns underlying these moments, namely $(\gamma_{d,1},\gamma_{d,2},\mu_{d|0},\mu_{d|1})$. As these moments are linear in the MTRs and the MTRs are linear in the unknowns, we can invert this linear system to identify them---and thus the LATEs.

To illustrate how our method can assess robustness, Table \ref{tab:IVdecomposition} reports estimates and 95\% confidence intervals for $LATE_{np}$ and $LATE_{cp}$ under KW's specification and a range of alternatives.\footnote{The estimates and confidence intervals are obtained using the algorithms described in Section \ref{sec:estimation_inference}. In our implementation, we take a grid for $\mathbf{L}$ given by $\{0,0.05,\ldots,0.95,1\}^2$, and $\mathcal{U}_{\text{grid}} = \{0,0.25,0.5,0.75,1\}^2$ when imposing the shape restrictions in the parametric case in Table \ref{tab:alpha_shape}. For the estimation algorithm, we take $\widehat{\Omega}_{S}$ to equal the identity matrix, and $\widehat{\Omega}_{M}$ to be computed as in $\widehat{\Sigma}^{-1}(\lambda)$ in \eqref{eq:TS} using $\mu(\lambda)$, $C_1(\lambda)$, $C_2(\lambda)$, $C_3(\lambda)$ and $c(\lambda)$ from \eqref{eq:matrix_est}. We take $\kappa = 0.1$ in both algorithms. To compute $\widehat{\Sigma}(\lambda)$, we use the bootstrap, i.e. we take $\widehat{\Sigma}(\lambda) = 1/B\sum_{b=1}^B (\widehat{\mu}_b(\lambda) - \widehat{\mu}(\lambda) + (\widehat{C}_{2,b}(\lambda) - \widehat{C}_{2}(\lambda))\widehat{\gamma}^*)(\widehat{\mu}_b(\lambda) - \widehat{\mu}(\lambda) + (\widehat{C}_{2,b}(\lambda) - \widehat{C}_{2}(\lambda))\widehat{\gamma}^*)'$, where $\widehat{\gamma}^*$ corresponds to the estimated version of \eqref{eq:TS_V_min} and $\widehat{\mu}_b(\lambda)$ and $\widehat{C}_{2,b}(\lambda)$ are the bootstrap analogs of $\widehat{\mu}(\lambda)$ and $\widehat{C}_{2}(\lambda)$ estimated using the $b$th bootstrap drawn with replacement at the level of the HS centers and where we take $B=200$.} We do so by allowing flexibility across two separate dimensions: the selection model and the MTRs. For each parameter, the upper left estimate reflects our analog to the baseline model in KW, i.e.  the specification in \eqref{eq:specKW} under the assumption that $\delta_{c|x} = 0$.  Moving across the columns highlights flexibility in the specification of the MTRs by relaxing the linearity and separability assumptions. Moving down a row relaxes the selection model assumption that there is no rationing for competing preschools, and instead allows them to be rationed by an unknown $\delta_{c|x} \in [0,1]$. 

\begin{table}[!t]
\begin{center}
\caption{Decomposition of IV estimand}
\label{tab:IVdecomposition}
\scalebox{0.8}{
\def\arraystretch{0.85}
\makebox[\textwidth]{\begin{threeparttable}
\input{table_IVdecomposition.tex}
\begin{tablenotes}[flushleft]
\setlength\labelsep{0pt}
\item \footnotesize For each parameter and specification combination, Est. LB (UB) denote estimated lower (upper) bounds, and Lower (Upper) CI denote lower (upper) values of 95$\%$ confidence intervals. A single point estimate reported if the parameter is point-identified. For the parametric specifications, MTRs are specified as $ m_{d|x}(u_p,u_c) = \sum_{\substack{k,l = 0 \\ k + l \leq K}}^K \gamma_{d,k,l|x} u_p^k u_c^l$. NP denotes the nonparametric specification. For both specifications, assumptions are imposed on $\mathbf{\Gamma}(h)$ as in Table A2. Across all specifications, MTRs are also taken to be bounded as in Assumption B in Table \ref{tab:alpha_shape}, where $L$ and $U$ equal the 0.1 and 0.9 quantiles of the observed outcomes, respectively. As in KW, data moments are estimated with observations that are weighted by the inverse probability of the child's experimental assignment, calculated at the level of HS centers. 
\end{tablenotes}
\end{threeparttable}}}
\end{center}
\end{table}

Under our analog of KW’s baseline specification, the estimates for $LATE_{np}$ and $LATE_{cp}$ reveal that HS increases test scores for those induced from no preschool and competing preschools, respectively, by 0.27 and 0.14. Consistent with KW, however, both of these effects are statistically imprecise. KW considers a number of additional covariates and fixed effects to improve precision. We instead consider shape restrictions to improve precision \citep{chetverikov/etal:18}. To this end, Column (2) imposes a bounded variation assumption as in Table \ref{tab:alpha_shape} given by $|m_{p|x}(u_p,u_c) - m_{n|x}(u_p,u_c)| \leq 0.8$, $|m_{c|x}(u_p,u_c) - m_{n|x}(u_p,u_c)| \leq 0.8$, and  $|m_{p|x}(u_p,u_c) - m_{c|x}(u_p,u_c)| \leq 0.2 $. We choose the 0.8 value because this is close to the largest effect for a preschool program noted in the literature \citep[][Table A.IV]{kline/walters:16}. Similarly, we bound the difference in effects between HS and competing preschools to be at most 0.2, since the two options provide similar services and hence are likely to have more similar effects.

Focusing on $LATE_{np}$ first, we find that its estimate under this specification remains substantively unchanged, capturing that the estimated MTRs satisfy the additional restrictions.  But, consistent with KW’s results under their more precise specifications, its confidence interval shrinks, resulting in a statistically significant effect. In Columns (3)-(7), we analyze the robustness of this positive effect to relaxing the linearity and separability assumptions in \eqref{eq:specKW}. Column (3) allows for a more flexible polynomial specification by adding interactions of  $u_p$ and $u_c$ and their squares to \eqref{eq:specKW}, and Column (4) allows the MTRs to be entirely nonparametric. Columns (5)-(7) further relax these specifications by removing separability. The parameter in these cases is partially identified, but our estimated bounds as well as confidence intervals reveal that neither linearity nor separability are required to conclude that $LATE_{np}$ is positive.

The second row, which allows for rationing of competing preschools ($\delta_{c|x} \in [0,1]$), reveals similar robustness. In this case, the confidence intervals for $LATE_{np}$ in Columns (2)-(7) lie entirely above zero (with lower bounds around 0.04-0.11). Hence the positive effect when competing preschools may be rationed also remains statistically significant even under nonparametric and nonseparable MTRs.

Turning to $LATE_{cp}$, again consistent with KW, the confidence intervals across these specifications cannot rule out a null effect. To better understand this, Table \ref{tab:IVdecomposition} also presents results for $S_{np}$. Here, as it is directly identified by the data moments in \eqref{eq:Snh}, we obtain estimates by simply taking sample analogs and confidence intervals using a standard $t$-test. The estimate suggests that the null finding arises because the data may simply be worse-powered for this effect; roughly only 30\% of those induced into HS come from competing preschools.

To analyze whether stronger restrictions can allow us to reach a more informative conclusion, we consider a final specification in Column (8) that imposes on Column (2) two additional assumptions: (i) a positive selection assumption, similar to the Assumption $M_j$ in Table \ref{tab:alpha_shape}, that takes $|m_{d|x}(u_p,u_c) - m_{n|x}(u_p,u_c)|$ to be decreasing in $u_p$ and $u_c$ for $d \in \{c,n\}$, i.e. those more likely to enroll in a preschool have a higher effect; and (ii) a positive effect assumption as in $M_{d,n}$ in Table \ref{tab:alpha_shape} for $d \in \{p,c\}$, i.e. children have higher test scores under preschool relative to no preschool. Even under these restrictions, the confidence intervals for $LATE_{cp}$ remain unaffected and continue to cover the entire logical region from -0.2 to 0.2 permitted under the bounded variation assumption. This strengthens our takeaway that the data do not rule out the possibility of a null effect.

\subsection{Policy Effects of Expanding Head Start}\label{sec:policy}

The second set of parameters we analyze evaluate the policy returns to marginal expansions of HS by marginally changing the probability of a HS offer, $\delta_{p|x}$. Following KW, we first define the benefits and costs of a HS offer. Let $B$ denote the total after-tax lifetime income for all the children, which relates to test scores by
\begin{align}\label{eq:Benefit}
 B = (1 - \tau) \phi_{\text{ben}} E[Y]~,
\end{align}
where $\phi_{\text{ben}}$ denotes the pre-specified relation between test scores and future income and $\tau$ the pre-specified tax rate faced by the children of eligible households. The net costs to the government of financing preschool are given by
\begin{align}\label{eq:Cost}
 C = \phi_p P(D = p) + \phi_c P(D=c) - \tau \phi_{\text{ben}} E[Y]
\end{align}
where $\phi_p$ and $\phi_c$ denote the pre-specified costs of providing HS and competing preschool services, respectively, and $\tau \phi_{\text{ben}}  E[Y]$ captures the revenue generated by taxes on the adult earnings of HS-eligible children. Exploiting the fact that $E[Y]$ and $P(D=d)$ can be written in terms of $E[Y(D(z_p,z_c))|X=x]$, $P[D(z_p,z_c)=d|X=x]$ for $d\in\{p,c\}$ and $(\delta_{p|x},\delta_{c|x})$, we can differentiate these expressions with respect to $\delta_{p|x}$ for $x \in \{0,1\}$ to obtain the marginal benefits and costs of expanding access to HS---see Appendix \ref{Asec:Mderivation} for details. In doing so, following KW, we either assume that $\delta_{c|x}$ does not adjust to changes in $\delta_{p|x}$, which we refer to as the non-adjusting case \citep[][Section V.C]{kline/walters:16}, or that $\delta_{c|x}$ adjusts such that enrollment in competing preschools stays constant, i.e. $d P(D = c|X=x) / d \delta_{p|x} = 0$, which we refer to as the adjusting case \citep[][Section V.D]{kline/walters:16}. Performing this exercise, we can obtain expressions for the marginal benefits (MB) and costs (MC) under the non-adjusting case given by
\begin{align}
 MB_{\text{non-adj}} &= \sum_{x \in \{0,1\}} P(X=x) [(1-\tau) \phi_{\text{ben}} \Delta_{Y,p|x}] \label{eq:MB_nonrat} \\
 MC_{\text{non-adj}} &= \sum_{x \in \{0,1\}} P(X=x) [\phi_p \Delta_{1\{D=p\},p|x} + \phi_c  \Delta_{1\{D=c\},p|x} - \tau \phi_{\text{ben}} \Delta_{Y,p|x}] \label{eq:MC_nonrat}
\end{align}
and under the adjusting case given by
\begin{align}
 MB_{\text{adj}} &= \sum_{x \in \{0,1\}} P(X=x) [(1-\tau) \phi_{\text{ben}} [\Delta_{Y,p|x} + (d\delta_{c|x}/d\delta_{p|x}) \Delta_{Y,c|x}] ] \label{eq:MB_rat} \\
 MC_{\text{adj}} &= \sum_{x \in \{0,1\}} P(X=x)[ \phi_p [\Delta_{1\{D=p\},p|x} + (d\delta_{c|x}/d\delta_{p|x}) \Delta_{1\{D=p\},c|x}]  \nonumber \\
 &\phantom{= \sum_{x \in \{0,1\}} P(X=x)[} ~~- \tau \phi_{\text{ben}} [\Delta_{Y,p|x} + (d\delta_{c|x}/d\delta_{p|x})  \Delta_{Y,c|x} ] ] \label{eq:MC_rat}
\end{align}
where $\Delta_{R,p|x} = E[R(1,Z_c) - R(0,Z_c)|X=x]$ and $\Delta_{R,c|x} = E[R(Z_p,1) - R(Z_p,0)|X=x]$ for random variable $R$, and $d \delta_{c|x} / d \delta_{p|x} = -\Delta_{1\{D=c\},p|x} / \Delta_{1\{D=c\},c|x}$. The marginal benefits and costs are then compared using the marginal value of public funds (MVPF) \citep{hendren:16} defined by
\begin{align}\label{eq:MVPF}
 MVPF =
 \begin{cases}
   + \infty &\text{ if } MB \geq 0,~MC \leq 0~, \\
   - \infty &\text{ if } MB \leq 0,~MC \geq 0~, \\
   MB / MC &\text{ otherwise }~.
 \end{cases}
\end{align}

Table \ref{tab:policy} reports estimates and 95$\%$ confidence intervals for the MVPF parameter under the non-adjusting case ($\text{MVPF}_{\text{non-adj}}$) and the adjusting case ($\text{MVPF}_{\text{adj}}$), for the specifications considered in Table \ref{tab:IVdecomposition}.\footnote{In unreported results, we also consider an additional MVPF parameter for the adjusting case derived under the assumption that competing preschool offers only draw children from no preschool alternatives and that the effect for these children equals $LATE_{np}$. While this assumption is not logically consistent with the selection model in Section \ref{sec:preschoolmodel}, we consider it for completeness as it is presented in KW. As it produces qualitatively similar to $\text{MVPF}_{\text{adj}}$, we focus on the latter, more logically consistent parameter for brevity.} For the pre-specified values in the expression, following KW, we take $\phi_{\text{ben}} = \kappa_{\text{ben}} E$, where $E$ corresponds to the average present discounted value of lifetime earnings for HS applicants and $\kappa_{\text{ben}}$ is the percentage change in earnings induced by a one standard deviation increase in test scores. We take $\kappa_{\text{ben}} = 0.13$, which corresponds to the estimate found in \cite{chetty/etal:11}---see \citet[][Table A.IV]{kline/walters:16} for an overview of such estimates in the literature.  Moreover, following KW, we take $\tau = 0.35$, $E=\$343,392$, $\phi_p = \$8,000$ and $\phi_c = 0.75 \cdot \phi_p = \$6,000$.

\begin{table}[!t]
\begin{center}
\caption{MVPF of expanding Head Start}
\label{tab:policy}
\scalebox{0.8}{
\def\arraystretch{0.85}
\makebox[\textwidth]{\begin{threeparttable}
\input{table_policy.tex}
\begin{tablenotes}[flushleft]
\setlength\labelsep{0pt}
\item \footnotesize For each parameter and specification combination, Est. LB (UB) denote estimated lower (upper) bounds, and Lower (Upper) CI denote lower (upper) values of 95$\%$ confidence intervals. A single point estimate reported if the parameter is point-identified. For the parametric specifications, MTRs are specified as $ m_{d|x}(u_p,u_c) = \sum_{\substack{k,l = 0 \\ k + l \leq K}}^K \gamma_{d,k,l|x} u_p^k u_c^l$. NP denotes the nonparametric specification. For both specifications, assumptions are imposed on $\mathbf{\Gamma}(h)$ as in Table A2. Across all specifications, MTRs are also taken to be bounded as in Assumption B in Table \ref{tab:alpha_shape}, where $L$ and $U$ equal the 0.1 and 0.9 quantiles of the observed outcomes, respectively. As in KW, data moments are estimated with observations that are weighted by the inverse probability of the child's experimental assignment, calculated at the level of HS centers. 
\end{tablenotes}
\end{threeparttable}}}
\end{center}
\end{table}

Observe that $\text{MVPF}_{\text{non-adj}}$ is directly point identified by moments of the data. This is because $\Delta_{R,p} = E[R|Z_p=1] - E[R|Z_p=0]$, which simply corresponds to the ITT effects on the variable $R$. Similar to the parameter $S_{np}$ in Table \ref{tab:IVdecomposition}, we estimate it by taking sample analogs of the various ITTs, with confidence intervals derived by inverting a standard $t$-test. Consistent with KW, the estimate of $\text{MVPF}_{\text{non-adj}}$ is greater than one and the lower bound of the confidence interval exceeds one, suggesting that, absent adjustment of competing preschools to HS expansion, marginally expanding HS access yields benefits that outweigh the costs.

For $\text{MVPF}_{\text{adj}}$, as in Table \ref{tab:IVdecomposition}, we obtain estimates and confidence intervals following the algorithms in Section \ref{sec:estimation_inference}, but slightly adapted to account for the piecewise definition of the MVPF parameter in \eqref{eq:MVPF}---see Appendix \ref{Asec:comp_prog} for details. Consistent again with KW, we find that under the linear and separable specification on the MTRs in Columns (1)-(2) and when $\delta_{c|x} = 0$, the estimates highlight that $\text{MVPF}_{\text{adj}}$ is also larger than one. Keeping $\delta_{c|x} = 0$, Columns (3), (5), (6), and (8) reveal this finding is robust to alternative parametric specifications, but Columns (4) and (7) show that the lower bound of the identified set falls below one when the MTRs are nonparametric.

When allowing for the possibility that competing preschools are rationed as in HS by allowing $\delta_{c|x} \in [0,1]$, Columns (1)-(7) show that the lower bound on $\text{MVPF}_{\text{adj}}$ falls below one across all specifications. Only under the additional restrictions of positive selection and positive effects in Column (8) does the lower bound of the identified set remain above one. Furthermore, across all specifications, the confidence intervals do not rule out the possibility that $\text{MVPF}_{\text{adj}}$ is smaller than one.

\begin{figure}[!t]
\centering
\caption{$p$-values for null hypothesis that $\text{MVPF}_{\text{adj}}$ is at most 1 for different values of $\kappa_{\text{ben}}$}
\makebox[\textwidth]{\includegraphics[width=2.75in]{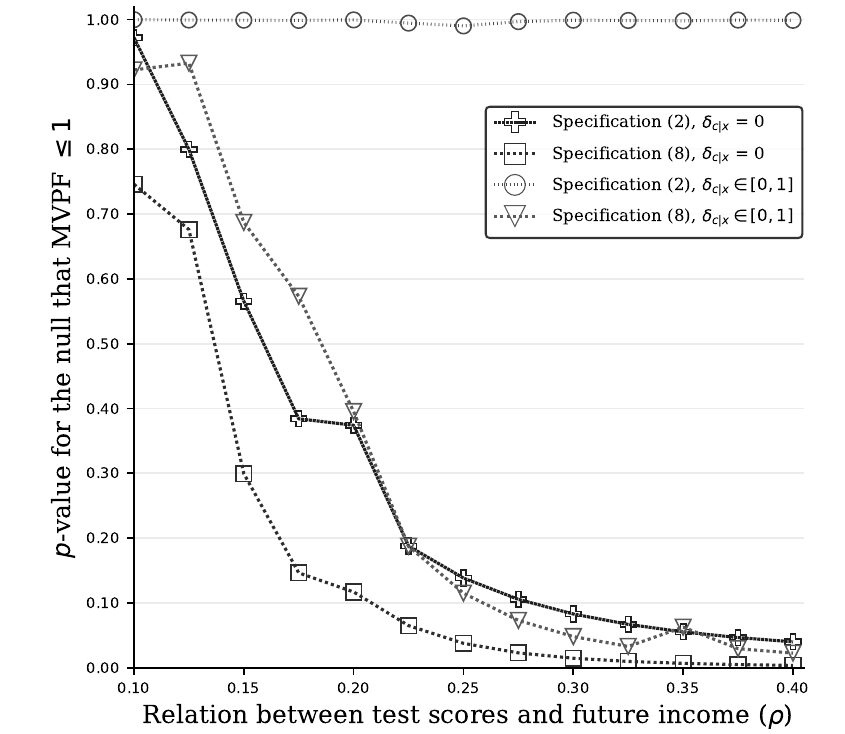}}
\label{fig:policy}
\end{figure}

We conclude our analysis by analyzing whether alternative values for $\kappa_{\text{ben}}$, i.e. the relation between test scores and earnings, can potentially result in $\text{MVPF}_{\text{adj}}$ being statistically greater than one. In Figure \ref{fig:policy}, we report $p$-values for the null hypothesis that $\text{MVPF}_{\text{adj}} \leq 1$ for alternative values of $\kappa_{\text{ben}}$ under Specifications (2) and (8) from Table \ref{tab:policy}. Our results reveal that for $\delta_{c|x} \in [0,1]$ and under additional sufficiently strong restrictions on the MTRs, namely specification (8), we can statistically reject that $\text{MVPF}_{\text{adj}}$ is at most one at a 5$\%$ significance level if we are willing to assume that the percentage increase in earnings induced by a one standard deviation increase in test scores is close to 30$\%$. This value is near the upper bound of estimates found in the literature, comparable to the Perry Preschool Program \citep{heckman/etal:10}.

\section{Conclusion}\label{sec:conclusion}

In cases with multiple treatments, treatment selection models generally exhibit multidimensional unobserved heterogeneity. In this paper, we develop a marginal treatment effect based method to learn about treatment effects in a general class of such models with discrete-valued instruments. Allowing the selection model to be identified up to a finite-dimensional parameter, we show how a two-step computational program can be used to compute the identified set for the treatment effect parameters when the marginal treatment response functions underlying them remain nonparametric or are additionally parameterized. We demonstrate the benefits of our method by revisiting the empirical analysis of the Head Start program by \cite{kline/walters:16}.

\newpage
\appendix
\setcounter{equation}{0}
\renewcommand{\theequation}{\Alph{section}-\arabic{equation}}
\setcounter{table}{0}
\renewcommand{\thetable}{A\arabic{table}}

\section{Restatement in Matrix Notation}\label{Asec:Bformulation}

In this section, we show how we can arrive at the equivalent formulations in \eqref{eq:matrix_est} and \eqref{eq:nulllinear}, and present the  exact forms of the matrices in these formulations. To do so, it is first useful to write the restrictions in \eqref{eq:D_moments} in vector notation by
\begin{align}\label{eq:restriction_sm_mat}
 A_1(h) = P ~,
\end{align}
where $P = (P_{d,z|x} : d \in \mathcal{D} \setminus \{d_0\},~z \in \mathcal{Z},~x \in \mathcal{X})$. Here the moments for some $d_0 \in \mathcal{D}$ are dropped to avoid perfect multicollinearity as they are redundant given that the choice probabilities sum to one. Similarly, write the linear restrictions in terms of $\gamma$ in \eqref{eq:Y_moments_alpha} as
\begin{align}
 A_2(h) \gamma = E~, \label{eq:restriction_data}
\end{align}
where $E = (E_{d,z|x} : d \in \mathcal{D},~z \in \mathcal{Z},~x \in \mathcal{X})$, and finally the linear system of inequality restrictions characterized by $\mathbf{\Gamma}(h)$ by
\begin{align}
 A_3 \gamma \leq a_3~. \label{eq:restriction_shape}
\end{align}
For the restrictions characterizing $\mathbf{\Gamma}(h)$, observe that we have supposed that their restriction matrix does not depend on $h$, which is satisfied by all the restrictions considered in Table \ref{tab:alpha_shape}. In the case that some of these restrictions depend on $h$ and hence correspond to having an estimated restriction matrix in the implementation, they can be accommodated in a similar manner to having an estimated parameter vector as described below.

To show how $\mathbf{\Gamma}^*(h^*(\lambda))$ is non-empty and $h^*(\lambda) \in \mathbf{H}^*(\lambda)$ can be written as \eqref{eq:matrix_est}, note that $h^*(\lambda) \in \mathbf{H}^*(\lambda)$ is equivalent to having that $h^*(\lambda)$ satisfies
\begin{align}\label{eq:restriction_sm_h}
 A_1(h^*(\lambda)) &= P ~,
\end{align}
i.e. $h^*(\lambda)$ satisfies the restrictions in \eqref{eq:D_moments}, and that $\mathbf{\Gamma}^*(h^*(\lambda))$ is non-empty is equivalent to having that there exists $\gamma$ such that
\begin{align}
 A_2(h^*(\lambda)) \gamma &= E~, \label{eq:restriction_data_h}
\end{align}
i.e. $\gamma$ satisfies the moments in \eqref{eq:Y_moments_alpha} when evaluated at $h = h^*(\lambda)$, and the shape restrictions in \eqref{eq:restriction_shape}. To ensure that the estimate of $\mu$ in both the formulations in \eqref{eq:matrix_est} and \eqref{eq:nulllinear} is asymptotically normal, which is particularly important for the validity of procedure in \cite{cox/etal:24}, it is useful to first rewrite the restriction in \eqref{eq:restriction_sm_h}. In particular, note that \eqref{eq:restriction_sm_h} can be equivalently stated as
\begin{align}
 M (A_1(h^*(\lambda)) - P) = 0
\end{align}
where $M = 1 - B(B'B)^{-1}B'$ with $B = \partial A_1(h) / \partial h |_{h = h^*(\lambda)}$, by assuming that an interior solution exists and noting that the unique solution to \eqref{eq:h_pointidentifed} satisfies $B'(A_1(h^*(\lambda)) - P) = 0$. Given that $M$ is an idempotent matrix, its spectral decomposition gives $M = H'D H$, where $D$ is a diagonal matrix with entries that are either 0 or 1 and $HH' = I$. Let $[H(A_1(h^*(\lambda)) - P)]'_{D=1}$ denote the submatrix of $H(A_1(h^*(\lambda)) - P)$ with rows that correspond to where the diagonal entries of $D$ equal to 1. Then, as $(A_1(h^*(\lambda)) - P)'(A_1(h^*(\lambda)) - P) = [H(A_1(h^*(\lambda)) - P)]'_{D=1} [H(A_1(h^*(\lambda)) - P)]_{D=1}$, observe that \eqref{eq:restriction_sm_h} can be equivalently rewritten as
\begin{align}
 [H(A_1(h^*(\lambda)) - P)]_{D=1} = 0~.
\end{align}
Here observe that while $A_1(\widehat{h}^*(\lambda)) - \widehat{P}$ is not generally asymptotically normal, we have that
\begin{align*}
 [H(A_1(\widehat{h}^*(\lambda)) - \widehat{P})]_{D=1} = [H(A_1(h^*(\lambda)) - \widehat{P})]_{D=1}
\end{align*}
with probability approaching 1, and hence which is asymptotically normal provided $P$ is.

Given the above, we can now show how \eqref{eq:restriction_shape}, \eqref{eq:restriction_sm_h} and \eqref{eq:restriction_data_h} can be stated as \eqref{eq:matrix_est}. In particular, this can be done by taking
\begin{align*}
 \mu(\lambda) = \begin{pmatrix} [H(A_1(h^*(\lambda)) - P)]_{D=1} \\ E \end{pmatrix} ~&,~c(\lambda) = \begin{pmatrix} 0_{2(d_1 + d_2) \times 1} \\ a_3  \end{pmatrix}~, \\
 C_1(\lambda) = \begin{pmatrix} I_{d_1} & 0_{d_1 \times d_2} \\ -I_{d_1} & 0_{d_1 \times d_2}  \\ 0_{d_2 \times d_1} & I_{d_2}  \\ 0_{d_2 \times d_1} & -I_{d_2} \\ 0_{d_3 \times d_1} & 0_{d_3 \times d_2} \end{pmatrix}~,~C_2(\lambda) &= \begin{pmatrix} 0_{d_1 \times d_{\gamma}} \\ -A_2(h^*(\lambda))   \end{pmatrix} ~,
 C_3(\lambda) = \begin{pmatrix} 0_{2(d_1+d_2) \times d_{\gamma}} \\ A_3  \end{pmatrix} ~,
\end{align*}
where $I_d$ denotes an identity matrix of dimension $d \times d$, $0_{a \times b}$ denotes a matrix of zeroes of dimension $a \times b$, $d_1$ denotes the dimension of the vector $[H(A_1(h^*(\lambda)) - P)]_{D=1}$, $d_2$ denotes the dimension of the vector $E$, $d_3$ denotes the row dimension of the matrix $A_3$, and $d_{\gamma}$ denotes the dimension of the vector $\gamma$. The estimated $\mu$ will be asymptotically normal provided $P$ and $E$ are.

Next, to show $\theta_0 \in \theta^{\Gamma}(\mathbf{\Gamma}^*(h),h)$ can be written as \eqref{eq:nulllinear}, note that this is equivalent to having $\mathbf{\Gamma}^*(h)$ is non-empty and $h^*(\lambda) \in \mathbf{H}^*(\lambda)$, which is the same as above, but along with the additional restriction that there exists $\gamma$ such that $\theta^{\Gamma}(\gamma,h) = \theta_0$. In turn, $\theta_0 \in \theta^{\Gamma}(\mathbf{\Gamma}^*(h),h)$ can be written as \eqref{eq:nulllinear} as in \eqref{eq:matrix_est} above, but where the matrices are augmented to include the additional restriction in $\theta^{\Gamma}(\gamma,h) = \theta_0$. In particular, in the case that $\theta^{\Gamma}(\gamma,h) = c'_{\theta} \gamma$ and has no estimated component to it when evaluated at $h = h^*(\lambda)$, the restriction $c'_{\theta} \gamma = \theta$ can be directly included in \eqref{eq:restriction_shape}. In contrast, in the case that $\theta^{\Gamma}(\gamma,h) = c'_{\theta}(h) \gamma$ and hence has an estimated element to it when evaluated at $h = h^*(\lambda)$, the restriction $c'_{\theta}(h^*(\lambda)) \gamma = \theta_0$ can be introduced similar to \eqref{eq:restriction_data_h} and appropriately augmenting the matrices. Specifically, in this case, we take
\begin{align*}
 \mu(\lambda) = \begin{pmatrix} 0 \\ [H(A_1(h^*(\lambda)) - P)]_{D=1} \\ E \end{pmatrix} ~&,~c(\lambda) = \begin{pmatrix} \theta_0 \\ -\theta_0 \\ 0_{2(d_1 + d_2) \times 1} \\ a_3  \end{pmatrix}~, \\
 C_1(\lambda) = \begin{pmatrix} 1 & 0_{1 \times d_1} & 0_{1 \times d_2} \\ -1 & 0_{1 \times d_1} & 0_{1 \times d_2} \\ 0_{d_1 \times 1} & I_{d_1} & 0_{d_1 \times d_2} \\ 0_{d_1 \times 1} & -I_{d_1} & 0_{d_1 \times d_2}  \\ 0_{d_2 \times 1} & 0_{d_2 \times d_1} & I_{d_2}  \\ 0_{d_2 \times 1} & 0_{d_2 \times d_1} & -I_{d_2} \\ 0_{d_3 \times 1} & 0_{d_3 \times d_1} & 0_{d_3 \times d_2} \end{pmatrix}~,~C_2(\lambda) &= \begin{pmatrix} c'_{\theta}(h^*(\lambda)) \\ 0_{d_1 \times d_{\gamma}} \\ -A_2(h^*(\lambda))   \end{pmatrix} ~,
 C_3(\lambda) = \begin{pmatrix} 0_{2(1 + d_1+d_2) \times d_{\gamma}} \\ A_3 \end{pmatrix} ~.
\end{align*}
However, note that $\hat{\mu}(\lambda)$ is not asymptotically normal here with a non-singular variance matrix. Nonetheless, this is permitted by the method of \cite{cox/etal:24} as its main result requires a weaker requirement where only $\hat{\mu}(\lambda) - \hat{C}_2(\lambda) \gamma$ with $\gamma^*$ defined in \eqref{eq:TS_V_min} is asymptotically normal with a non-singular variance matrix---see \citet[][Footnote 10]{cox/etal:24}.

\section{Numerically Verifying Full Rank of Jacobian}\label{Asec:rank_jacobian}

In this section, we describe how we verify that the Jacobian defined in Section \ref{sec:preschoolmodel} has full rank. Note that the Jacobian $\partial H / \partial \sigma$ corresponds to
\begin{align*}
   J(\sigma,\delta_c) = \begin{pmatrix}
\frac{\partial H_{c,0}}{\partial \zeta_p}(\sigma,\delta_c) & \frac{\partial H_{c,0}}{\partial \zeta_c}(\sigma,\delta_c) & \frac{\partial H_{c,0}}{\partial \zeta}(\sigma,\delta_c) & \frac{\partial H_{c,0}}{\partial \rho}(\sigma,\delta_c) \\
\frac{\partial H_{c,1}}{\partial \zeta_p}(\sigma,\delta_c) & \frac{\partial H_{c,1}}{\partial \zeta_c}(\sigma,\delta_c) & \frac{\partial H_{c,1}}{\partial \zeta}(\sigma,\delta_c) & \frac{\partial H_{c,1}}{\partial \rho} (\sigma,\delta_c)\\
\frac{\partial H_{p,0}}{\partial \zeta_p}(\sigma,\delta_c) & \frac{\partial H_{p,0}}{\partial \zeta_c}(\sigma,\delta_c) & \frac{\partial H_{p,0}}{\partial \zeta}(\sigma,\delta_c) & \frac{\partial H_{p,0}}{\partial \rho}(\sigma,\delta_c) \\
\frac{\partial H_{p,1}}{\partial \zeta_p}(\sigma,\delta_c) & \frac{\partial H_{p,1}}{\partial \zeta_c}(\sigma,\delta_c) & \frac{\partial H_{p,1}}{\partial \zeta}(\sigma,\delta_c) & \frac{\partial H_{p,1}}{\partial \rho}(\sigma,\delta_c)
\end{pmatrix}~.
\end{align*}
To verify that it has full rank on $(\sigma,\delta) \in \mathbf{R}^2 \times \mathbf{R}_{++} \times (-1,1) \times [0,1] \equiv \mathbf{T}$, we compute its determinant on a grid of points in $\mathbf{T}$ and check it is not equal to zero up to numerical error. In particular, we take a fine grid given by $\mathbf{T}_{\text{grid}} = \{\Phi^{-1}(\zeta) : \zeta \in \{0.05, 0.10, \ldots , 0.90, 0.95\}\}^2 \times \{\Phi^{-1}(\zeta) : \zeta \in \{0.55, 0.6, \ldots , 0.90, 0.95\}\} \times \{-0.9,-0.85,\ldots,0.85,0.9\} \times \{0,0.05,\ldots,0.95,1\}$ and check if $J(\sigma,\delta_c)$ has full rank for $\sigma,\delta_c \in \mathbf{T}_{\text{grid}}$. To perform the above exercise, it is useful to obtain analytical formulas for the derivatives in the Jacobian. For this purpose, note that if $(\tilde{u}_1,\tilde{u}_2)$ are distributed as a bivariate standard normal distribution with correlation parameter $\rho$, then $(\tilde{u}_p,(\tilde{u}_p-\tilde{u}_c) / 2\tilde{\rho})$ and $(\tilde{u}_c,(\tilde{u}_c-\tilde{u}_p) / 2\tilde{\rho})$ are distributed as a bivariate normal distribution with correlation parameter $\tilde{\rho} \equiv \sqrt{(1-\rho)/2} \in (0,1)$. This observation allows first rewriting the components in $H$ by
\begin{align*}
   H_{c,0}(\theta) & = \delta_c \Phi\left(\zeta_c + \zeta,\frac{\zeta_c + \zeta - \zeta_p}{2 \tilde{\rho}}; \tilde{\rho}\right) + (1-\delta_c) \Phi\left(\zeta_c,\frac{\zeta_c - \zeta_p}{2 \tilde{\rho}}; \tilde{\rho}\right) ~,\\
   H_{c,1}(\theta) & = \delta_c \Phi\left(\zeta_c + \zeta,\frac{\zeta_c - \zeta_p}{2 \tilde{\rho}}; \tilde{\rho}\right) + (1-\delta_c) \Phi\left(\zeta_c,\frac{\zeta_c - \zeta_p - \zeta}{2 \tilde{\rho}}; \tilde{\rho}\right)~, \\
   H_{p,0}(\theta) & = \delta_c \Phi\left(\zeta_p,\frac{\zeta_p - \zeta_c - \zeta}{2 \tilde{\rho}}; \tilde{\rho}\right) + (1-\delta_c) \Phi\left(\zeta_p,\frac{\zeta_p - \zeta_c}{2 \tilde{\rho}}; \tilde{\rho}\right)~, \\
   H_{p,1}(\theta) & = \delta_c \Phi\left(\zeta_p + \zeta,\frac{\zeta_p - \zeta_c}{2 \tilde{\rho}}; \tilde{\rho}\right) + (1-\delta_c) \Phi\left(\zeta_p + \zeta,\frac{\zeta_p + \zeta - \zeta_c}{2 \tilde{\rho}}; \tilde{\rho}\right)~.
\end{align*}
By exploiting the following properties of a bivariate normal distribution
 \begin{align*}
   \Phi_1(a,b;\rho) = \Phi\left(\frac{b - \rho a}{ \sqrt{1 - \rho^2}}\right) \phi(a)~&;~\Phi_2(a,b;\rho) = \Phi\left(\frac{a - \rho b}{ \sqrt{1 - \rho^2}}\right) \phi(b)~;\\
 \text{ and }  \Phi_\rho(a,b;\rho) &= \phi\left(\frac{b - \rho a}{ \sqrt{1 - \rho^2}}\right) \phi(a)~,
 \end{align*}
where $\Phi_j$ denote the derivative of $\Phi$ with respect to the $j$th argument and $\Phi_\rho$ denotes the derivative of $\Phi$ with respect to the correlation parameter, we can then straightforwardly obtain analytical expressions for the derivatives in $J(\sigma,\delta_c)$, which we use when computing the Jacobian in our above exercise.

\section{Derivation of Marginal Benefits and Costs}\label{Asec:Mderivation}

The derivation of the marginal benefit and cost expressions reported in Section \ref{sec:policy} follows \citet[][Appendix C.2]{kline/walters:16}. We present it here in terms of our notation for completeness. It is useful to first write $E[Y]$ and $P(D=d)$ in terms of  $E[Y(D(z_p,z_c))|X=x]$, $P[D(z_p,z_c)=d|X=x]$ and $(\delta_{p|x},\delta_{c|x})$ as follows:
\begin{align}
 E[Y] = &~\sum_{x \in \{0,1\}} P(X=x) E[Y|X=x]~,  \label{eq:E_Y}\\
 P(D=d) = &~ \sum_{x \in \{0,1\}} P(X=x)P(D=d|X=x)~, \label{eq:E_D}
\end{align}
where, as $Z_p$ and $Z_c$ are randomly assigned with probability $\delta_{p|x}$ and $\delta_{c|x}$ for $X=x \in \{0,1\}$, we have
\begin{align*}
 E[Y|X=x] =~&  \delta_{p|x} \delta_{c|x} E[Y(D(1,1))|X=x] + \delta_{p|x} (1- \delta_{c|x}) E[Y(D(1,0))|X=x] \nonumber \\
 &+ (1-\delta_{p|x})\delta_{c|x} E[Y(D(0,1))|X=x] \nonumber \\
 &+ (1-\delta_{p|x}) (1-\delta_{c|x}) E[Y(D(0,0))|X=x]~, \\ 
 P(D=d|X=x) = &~ \delta_{p|x} \delta_{c|x} P(D(1,1)=d|X=x) + \delta_{p|x} (1- \delta_{c|x}) P(D(1,0)=d|X=x) \nonumber \\
 &+ (1-\delta_{p|x})\delta_{c|x} P(D(0,1)=d|X=x) \nonumber \\
 &+ (1-\delta_{p|x}) (1-\delta_{c|x}) P(D(0,0)=d|X=x)~.
\end{align*}

For the marginal benefit and costs in the non-adjusting case, it is assumed that $\delta_{c|x}$ does not adjust to $\delta_{p|x}$. Differentiating \eqref{eq:E_Y} and \eqref{eq:E_D} with respect to $\delta_{p|x}$ for $x \in \{0,1\}$ we obtain
\begin{align}
 \sum_{x \in \{0,1\}}\frac{\partial E[Y]}{\partial \delta_{p|x}} &\equiv \sum_{x \in \{0,1\}} P(X=x) \frac{\partial E[Y|X=x] }{ \partial \delta_{p|x}} \\
 &= \sum_{x \in \{0,1\}} P(X=x) \Delta_{Y,p|x}~, \label{eq:E_Y_der} \\
\sum_{x \in \{0,1\}} \frac{ \partial P(D=d) }{ \partial \delta_{p|x}} &\equiv \sum_{x \in \{0,1\}} P(X=x) \frac{\partial P(D=d|X=x) }{ \partial \delta_{p|x} } \\
&= \sum_{x \in \{0,1\}} P(X=x) \Delta_{1\{D=d\},p|x}~. \label{eq:E_D_der}
\end{align}
where  recall $\Delta_{R,p|x} = E[R(1,Z_c) - R(0,Z_c)|X=x]$ for random variable $R$. Noting that the marginal benefit and cost are defined by differentiating \eqref{eq:Benefit} and \eqref{eq:Cost} with respect to $\delta_{p|x}$, i.e.
\begin{align}
 MB &= \sum_{x \in \{0,1\}} (1- \tau) \phi_{\text{ben}} \frac{\partial E[Y] }{ \partial \delta_{p|x} }~, \label{eq:MB_true} \\
 MC &= \sum_{x \in \{0,1\}} \phi_p  \frac{\partial P(D=p) }{ \partial \delta_{p|x}} + \phi_c \frac{\partial P(D=c) }{ \partial \delta_{p|x} } - \tau \phi_{\text{ben}} \frac{\partial E[Y] }{ \partial \delta_{p|x} } ~, \label{eq:MC_true}
\end{align}
and plugging in \eqref{eq:E_Y_der} and \eqref{eq:E_D_der} in these expressions, we then obtain those in \eqref{eq:MB_nonrat} and \eqref{eq:MC_nonrat}.

For the marginal benefit and costs in the adjusting case, it is assumed that $\delta_{c|x}$ adjusts to $\delta_{p|x}$ such that $d P(D=c|X=x) / d \delta_{p|x} = 0$. Differentiating \eqref{eq:E_Y} and \eqref{eq:E_D} with respect to $\delta_{p|x}$ while allowing $\delta_{c|x}$ to adjust to $\delta_{p|x}$ we obtain
\begin{align*}
 \sum_{x \in \{0,1\}}\frac{\partial E[Y]}{\partial \delta_{p|x}} &\equiv \sum_{x \in \{0,1\}} P(X=x) \frac{\partial E[Y|X=x] }{ \partial \delta_{p|x}} \\
 &= \sum_{x \in \{0,1\}} P(X=x) \left( \Delta_{Y,p|x} + \frac{d \delta_{c|x} }{d \delta_{p|x} } \Delta_{Y,c|x} \right) \\
\sum_{x \in \{0,1\}}\frac{ \partial P(D=d) }{ \partial \delta_{p|x}} &\equiv \sum_{x \in \{0,1\}} P(X=x) \frac{\partial P(D=d|X=x) }{ \partial \delta_{p|x} } \\
&= \sum_{x \in \{0,1\}} P(X=x) \left( \Delta_{1\{D=d\},p|x} + \frac{d \delta_{c|x} }{d \delta_{p|x} }\Delta_{1\{D=d\},c|x}  \right)
\end{align*}
where, in addition to above, recall $\Delta_{R,c|x} = E[R(Z_p,1) - R(Z_p,0)|X=x]$ for random variable $R$. Moreover, taking $d P(D=c|X=x) / d \delta_{p|x} = 0$, we obtain $d \delta_{c|x} / d \delta_{p|x} =  -\Delta_{1\{D=c\},p|x} / \Delta_{1\{D=c\},c|x}$. Plugging these expressions into \eqref{eq:MB_true} and \eqref{eq:MC_true}, we then obtain those in \eqref{eq:MB_rat} and \eqref{eq:MC_rat}.

\section{Computational Programs for MVPF parameters}\label{Asec:comp_prog}

In this section, we describe how our estimation and confidence interval procedures in Section \ref{sec:estimation_inference} can be adapted for the MVPF parameter defined in \eqref{eq:MVPF}. The particular complication here is that the numerator and denominator may each change sign, requiring case-by-case handling.

For the purposes of the estimation procedure, let $\theta_{MB}(\gamma,h)$ and $\theta_{MC}(\gamma,h)$ denote the marginal benefit and cost parameter written in terms of $\gamma$ and $h$, where note that they are linear in terms of $\gamma$ given $h$. Moreover, as in Section \ref{sec:estimation}, let
\begin{align*}
 \mathbf{\Gamma}^*_{+,+}(\widehat{h}^*(\lambda)) &= \{\gamma \in \mathbf{\Gamma}(\widehat{h}^*(\lambda)) : \widehat{Q}(\gamma,\lambda) \leq Q^* + \kappa,~\theta_{MB}(\gamma,\widehat{h}^*(\lambda)) \geq 0,~\theta_{MC}(\gamma,\widehat{h}^*(\lambda)) \geq 0\}~, \\
 \mathbf{\Gamma}^*_{-,-}(\widehat{h}^*(\lambda)) &= \{\gamma \in \mathbf{\Gamma}(\widehat{h}^*(\lambda)) : \widehat{Q}(\gamma,\lambda) \leq Q^* + \kappa,~\theta_{MB}(\gamma,\widehat{h}^*(\lambda)) \leq 0,~\theta_{MC}(\gamma,\widehat{h}^*(\lambda)) \leq 0\}~, \\
 \mathbf{\Gamma}^*_{+,-}(\widehat{h}^*(\lambda)) &= \{\gamma \in \mathbf{\Gamma}(\widehat{h}^*(\lambda)) : \widehat{Q}(\gamma,\lambda) \leq Q^* + \kappa,~\theta_{MB}(\gamma,\widehat{h}^*(\lambda)) \geq 0,~\theta_{MC}(\gamma,\widehat{h}^*(\lambda)) \leq 0\}~, \\
 \mathbf{\Gamma}^*_{-,+}(\widehat{h}^*(\lambda)) &= \{\gamma \in \mathbf{\Gamma}(\widehat{h}^*(\lambda)) : \widehat{Q}(\gamma,\lambda) \leq Q^* + \kappa,~\theta_{MB}(\gamma,\widehat{h}^*(\lambda)) \leq 0,~\theta_{MC}(\gamma,\widehat{h}^*(\lambda)) \geq 0\}~,
\end{align*}
and
\begin{align}
 \widehat{\mathbf{L}}^*_{+,+} &= \{\lambda \in \mathbf{L} : \min_{ \gamma \in \mathbf{\Gamma}(\widehat{h}^*(\lambda)), \theta_{MB}(\gamma,\widehat{h}^*(\lambda)) \geq 0,~\theta_{MC}(\gamma,\widehat{h}^*(\lambda)) \geq 0} \widehat{Q}(\gamma,\lambda) \leq Q^* + \kappa \}~, \label{eq:L_hat_mvpf1}\\
 \widehat{\mathbf{L}}^*_{-,-} &= \{\lambda \in \mathbf{L} : \min_{ \gamma \in \mathbf{\Gamma}(\widehat{h}^*(\lambda)), \theta_{MB}(\gamma,\widehat{h}^*(\lambda)) \leq 0,~\theta_{MC}(\gamma,\widehat{h}^*(\lambda)) \leq 0} \widehat{Q}(\gamma,\lambda) \leq Q^* + \kappa \}~,\\
 \widehat{\mathbf{L}}^*_{+,-} &= \{\lambda \in \mathbf{L} : \min_{ \gamma \in \mathbf{\Gamma}(\widehat{h}^*(\lambda)), \theta_{MB}(\gamma,\widehat{h}^*(\lambda)) \geq 0,~\theta_{MC}(\gamma,\widehat{h}^*(\lambda)) \leq 0} \widehat{Q}(\gamma,\lambda) \leq Q^* + \kappa \}~,\\
 \widehat{\mathbf{L}}^*_{-,+} &= \{\lambda \in \mathbf{L} : \min_{ \gamma \in \mathbf{\Gamma}(\widehat{h}^*(\lambda)), \theta_{MB}(\gamma,\widehat{h}^*(\lambda)) \leq 0,~\theta_{MC}(\gamma,\widehat{h}^*(\lambda)) \geq 0} \widehat{Q}(\gamma,\lambda) \leq Q^* + \kappa \}~\label{eq:L_hat_mvpf2}
\end{align}
denote the subsets of \eqref{eq:A_hat} and \eqref{eq:L_hat} that additionally condition on the sign of the marginal benefit and cost parameters, which underlies the definition of the MVPF parameter in \eqref{eq:MVPF}. The estimated identified set for \eqref{eq:MVPF} can then be defined by
\begin{align}
 \widehat{\Theta}_{MVPF} = \bigcup_{\widehat{\mathbf{L}}^*_{+,+}} [\widehat{\theta}_{L,+}(\widehat{h}^*(\lambda)),\widehat{\theta}_{U,+}(\widehat{h}^*(\lambda))] \bigcup_{\widehat{\mathbf{L}}^*_{-,-}} [\widehat{\theta}_{L,-}(\widehat{h}^*(\lambda)),\widehat{\theta}_{U,-}(\widehat{h}^*(\lambda))] \bigcup_{\widehat{\mathbf{L}}^*_{+,-}} \{+\infty\} \bigcup_{\widehat{\mathbf{L}}^*_{-,+}} \{-\infty\}~,
\end{align}
where $\widehat{\theta}_{L,+}(\widehat{h}^*(\lambda))$ ($\widehat{\theta}_{L,-}(\widehat{h}^*(\lambda))$) and $\widehat{\theta}_{U,+}(\widehat{h}^*(\lambda))$ ($\widehat{\theta}_{U,-}(\widehat{h}^*(\lambda))$) are computed by solving minimization and maximization problems in \eqref{eq:optimization} using $\mathbf{\Gamma}^*_{+,+}(\widehat{h}^*(\lambda))$ ($\mathbf{\Gamma}^*_{-,-}(\widehat{h}^*(\lambda))$) defined above. In turn, the estimation procedure follows the algorithm in Section \ref{sec:estimation} except in Step 2 we need to compute four additional minimizations that account for the signs of the marginal benefits and costs to compute the sets in \eqref{eq:L_hat_mvpf1}-\eqref{eq:L_hat_mvpf2}, and in Step 4 we need to compute two separate minimization and maximization problems to obtain bounds that account for the signs of the marginal benefits and costs. However, these programs continue to be structured and correspond to convex quadratic and quadratically-constrained quadratic programs. We note that the latter programs in Step 4 correspond to having linear-fractional objectives, but which can be converted to linear objectives along with an additional variable satisfying linear restrictions using the so-called Charnes Cooper transformation \citep[][Section 4.3.2]{boyd/andenberghe:04}.

For the purposes of the confidence interval procedure, the null hypothesis $H_0(\lambda): \theta_0 \in \theta^{\Gamma}(\mathbf{\Gamma}^*(h^*(\lambda)),h^*(\lambda)),~h^*(\lambda) \in \mathbf{H}^*(\lambda)$, similar to \eqref{eq:nulllinear}, corresponds to testing whether $\gamma$ satisfies a linear system of inequalities while including the linear condition that (i) $\theta_{MB}(\gamma,h) = \theta_0 \cdot \theta_{MC}(\gamma,h)$ for $\theta_0 \in (-\infty,\infty)$; (ii)  $\theta_{MB}(\gamma,h) \geq 0,~ \theta_{MC}(\gamma,h) \leq 0$ for $\theta_0 = \infty$; and (iii) $\theta_{MB}(\gamma,h) \leq 0,~ \theta_{MC}(\gamma,h) \geq 0$ for $\theta_0 = -\infty$. In turn, by deriving the versions of $\mu(\lambda)$, $C_1(\lambda)$, $C_2(\lambda)$ and $c(\lambda)$ that account for this additional condition in each of these cases, we can then test the hypothesis for the MVPF parameter by analogously using the algorithm in Section \ref{sec:inference} and construct confidence intervals by collecting the values we do not reject. To this end, using the notation in Section \ref{Asec:Bformulation}, note that $\theta_{MB}(\gamma,h) = c'_{MB}(h) \gamma$ and $\theta_{MC}(\gamma,h) = c_{0,MC}(h) + c'_{MC}(h) \gamma$, i.e. both the numerator and denominator include elements that are estimated when evaluated at $h = h^*(\lambda)$ and the denominator includes an additional intercept component that will also be estimated given that the cost parameter includes elements that depend only on choice probabilities from the selection model. In turn, observe that we can take in the case of (i), $\mu(\lambda)$, $C_1(\lambda)$, $C_2(\lambda)$, $C_3(\lambda)$ and $c(\lambda)$ to be
\begin{align*}
 \mu(\lambda) = \begin{pmatrix} -\theta_0 c_{0,MC}(h^*(\lambda)) \\ H(A_1(h^*(\lambda)) - P)]_{D=1} \\ E \end{pmatrix} ~&,~c(\lambda) = \begin{pmatrix} 0_{2(1 + d_1 + d_2) \times 1} \\ a_3  \end{pmatrix}~, \\
 C_1(\lambda) = \begin{pmatrix} 1 & 0_{1 \times d_1} & 0_{1 \times d_2} \\ -1 & 0_{1 \times d_1} & 0_{1 \times d_2} \\ 0_{d_1 \times 1} & I_{d_1} & 0_{d_1 \times d_2} \\ 0_{d_1 \times 1} & -I_{d_1} & 0_{d_1 \times d_2}  \\ 0_{d_2 \times 1} & 0_{d_2 \times d_1} & I_{d_2}  \\ 0_{d_2 \times 1} & 0_{d_2 \times d_1} & -I_{d_2} \\ 0_{d_3 \times 1} & 0_{d_3 \times d_1} & 0_{d_3 \times d_2} \end{pmatrix}~,~C_2(\lambda) &= \begin{pmatrix} c'_{MB}(h^*(\lambda)) - \theta_0 c'_{MC}(h^*(\lambda))  \\ 0_{d_1 \times d_{\gamma}} \\ -A_2(h^*(\lambda))   \end{pmatrix} ~, \\ 
 C_3(\lambda) &= \begin{pmatrix} 0_{2(1 + d_1+d_2) \times d_{\gamma}} \\ A_3  \end{pmatrix} ~;
\end{align*}
in the case of (ii), to be
\begin{align*}
 \mu(\lambda) = \begin{pmatrix} 0 \\ c_{0,MC}(h^*(\lambda)) \\ H(A_1(h^*(\lambda)) - P)]_{D=1} \\ E \end{pmatrix} ~&,~c(\lambda) = \begin{pmatrix} 0_{2(1 + d_1 + d_2) \times 1} \\ a_3  \end{pmatrix}~, \\
 C_1(\lambda) = \begin{pmatrix} -1 & 0 & 0_{1 \times d_1} & 0_{1 \times d_2} \\ 0 & 1 & 0_{1 \times d_1} & 0_{1 \times d_2} \\ 0_{d_1 \times 1} & 0_{d_1 \times 1} & I_{d_1} & 0_{d_1 \times d_2} \\ 0_{d_1 \times 1} & 0_{d_1 \times 1} & -I_{d_1} & 0_{d_1 \times d_2}  \\ 0_{d_2 \times 1} & 0_{d_2 \times 1} & 0_{d_2 \times d_1} & I_{d_2}  \\ 0_{d_2 \times 1} & 0_{d_2 \times 1} & 0_{d_2 \times d_1} & -I_{d_2} \\ 0_{d_3 \times 1} & 0_{d_3 \times 1} & 0_{d_3 \times d_1} & 0_{d_3 \times d_2} \end{pmatrix}~,~C_2(\lambda) &= \begin{pmatrix} c'_{MB}(h^*(\lambda)) \\ c'_{MC}(h^*(\lambda))  \\ 0_{d_1 \times d_{\gamma}} \\ -A_2(h^*(\lambda))   \end{pmatrix} ~, \\ 
 C_3(\lambda) &= \begin{pmatrix} 0_{2(1 + d_1+d_2) \times d_{\gamma}} \\ A_3  \end{pmatrix} ~;
\end{align*}
and, in the case of (iii), to be
\begin{align*}
 \mu(\lambda) = \begin{pmatrix} 0 \\ c_{0,MC}(h^*(\lambda)) \\ H(A_1(h^*(\lambda)) - P)]_{D=1} \\ E \end{pmatrix} ~&,~c(\lambda) = \begin{pmatrix} 0_{2(1 + d_1 + d_2) \times 1} \\ a_3  \end{pmatrix}~, \\
 C_1(\lambda) = \begin{pmatrix} 1 & 0 & 0_{1 \times d_1} & 0_{1 \times d_2} \\ 0 & -1 & 0_{1 \times d_1} & 0_{1 \times d_2} \\ 0_{d_1 \times 1} & 0_{d_1 \times 1} & I_{d_1} & 0_{d_1 \times d_2} \\ 0_{d_1 \times 1} & 0_{d_1 \times 1} & -I_{d_1} & 0_{d_1 \times d_2}  \\ 0_{d_2 \times 1} & 0_{d_2 \times 1} & 0_{d_2 \times d_1} & I_{d_2}  \\ 0_{d_2 \times 1} & 0_{d_2 \times 1} & 0_{d_2 \times d_1} & -I_{d_2} \\ 0_{d_3 \times 1} & 0_{d_3 \times 1} & 0_{d_3 \times d_1} & 0_{d_3 \times d_2} \end{pmatrix}~,~C_2(\lambda) &= \begin{pmatrix} c'_{MB}(h^*(\lambda)) \\ c'_{MC}(h^*(\lambda))  \\ 0_{d_1 \times d_{\gamma}} \\ -A_2(h^*(\lambda))   \end{pmatrix} ~, \\ 
 C_3(\lambda) &= \begin{pmatrix} 0_{2(1 + d_1+d_2) \times d_{\gamma}} \\ A_3  \end{pmatrix} ~.
\end{align*}

\section{Proofs}\label{Asec:proofs}

\subsection{Proof of Proposition \ref{prop:linearprog}}\label{Asec:proof_linearprog}

Given Assumption \ref{ass:PM}, $\mathbf{\Gamma}(h)$ corresponds to a convex set as it is given by a system of linear inequalities and so does $\mathbf{\Gamma}^*(h)$ as it is a subset of $\mathbf{\Gamma}(h)$ with additional linear restrictions. Moreover, note that $\theta^{\Gamma}$ is a linear function in $\gamma$. In turn, as the image of a convex set under a continuous function is also convex, it directly follows that the image of a non-empty $\mathbf{\Gamma}^*(h)$ under $\theta^{\Gamma}$ is a non-empty convex set on the real line, i.e. an interval with endpoints given in \eqref{eq:optimization}.

\subsection{Proof of Proposition \ref{prop:equivalence}}\label{Asec:proof_equivalence}

In order to show $\theta(\mathbf{M}^*_{\text{np}}(h),h) = \theta(\mathbf{M}^*(h),h)$, we need to show that $\theta(\mathbf{M}^*(h),h) \subseteq \theta(\mathbf{M}^*_{\text{np}}(h),h)$ and $\theta(\mathbf{M}^*_{\text{np}}(h),h) \subseteq \theta(\mathbf{M}^*(h),h)$. As $\mathbf{M}^*(h) \subseteq \mathbf{M}^{*}_{\text{np}}(h)$, it directly follows that $\theta(\mathbf{M}^*(h),h) \subseteq \theta(\mathbf{M}^*_{\text{np}}(h),h)$. To show $\theta(\mathbf{M}^*_{\text{np}}(h),h) \subseteq \theta(\mathbf{M}^*(h),h)$, we show that for every $\theta_0 \in \theta(\mathbf{M}^*_{\text{np}}(h),h)$, there exists $\tilde{m} \in \mathbf{M}^{*}(h)$ such that $\theta(\tilde{m},h) = \theta_0$. To this end, as $\theta_0 \in \theta(\mathbf{M}^{*}_{\text{np}}(h),h)$, there exists a $m^{\dagger} \in \mathbf{M}^{*}_{\text{np}}(h)$ such that $\theta(m^{\dagger},h) = \theta_0$. We use $m^{\dagger}$ to construct $\tilde{m}$ satisfying $\tilde{m} \in \mathbf{M}^{*}(h)$ and $\theta(\tilde{m},h) = \theta_0$. Specifically, we take
\begin{align*}
\tilde{m}_{d|x} = \sum_{k=0}^K \tilde{\gamma}_{k,d|x} b_k(h)~,
\end{align*}
for $d \in \mathcal{D}$, $x \in \mathcal{X}$, where $b_k(h)$ satisfies \eqref{eq:b_constant} with a partition $\mathbb{U}(h)$ satisfying Definition \ref{def:P}, and where the coefficients equal
\begin{align}\label{eq:alpha_m_relation}
\tilde{\gamma}_{k,d|x} = E[m^{\dagger}_{d|x}(u) | U \in \mathcal{U}_k, X=x]~.
\end{align}

To show $\theta(\tilde{m},h) = \theta_0$, we can equivalently show $\theta(m^\dagger,h) = \theta^{\Gamma}(\tilde{\gamma},h)$, where note that $\theta$ in \eqref{eq:theta} under Assumption \ref{ass:PM} with \eqref{eq:b_constant} and $\mathbb{U}(h)$ satisfying Definition \ref{def:P} implies that $\theta^{\Gamma}$ corresponds to
\begin{align*}
 \theta^{\Gamma}(\gamma,h) = \sum_{x \in \mathcal{X}} \sum_{l \in \mathcal{L}} \sum_{d \in \mathcal{D}} w_{d,l|x}(h) \sum_{k=0}^K 1\{\mathcal{U}_k \in \mathbb{U}^{\theta}_{d,l|x}(h)\}  \gamma_{k,d|x}   \int\limits_{\mathcal{U}_k} dF_x ~.
\end{align*}
It follows that $\theta(m^\dagger,h) = \theta^{\Gamma}(\tilde{\gamma},h)$ as
\begin{align*}
\theta(m^\dagger,h) &= \sum_{x \in \mathcal{X}} \sum_{l \in \mathcal{L}} \sum_{d \in \mathcal{D}} w_{d,l|x}(h)\int\limits_{\mathcal{U}^{\theta}_{d,l|x}(g)} m^\dagger_{d|x}(u) dF_x \\
&=  \sum_{x \in \mathcal{X}} \sum_{l \in \mathcal{L}} \sum_{d \in \mathcal{D}} w_{d,l|x}(h) \sum_{\mathcal{U} \in \mathbb{U}^{\theta}_{d,l|x}(h)} \int\limits_{\mathcal{U}} m^\dagger_{d|x}(u) dF_x \\
&=  \sum_{x \in \mathcal{X}} \sum_{l \in \mathcal{L}} \sum_{d \in \mathcal{D}} w_{d,l|x}(h) \sum_{k=0}^K 1\{\mathcal{U}_k \in \mathbb{U}^{\theta}_{d,l|x}(h)\}  \tilde{\gamma}_{k,d|x}   \int\limits_{\mathcal{U}_k} dF_x  \\
&= \theta^{\Gamma}(\tilde{\gamma},h)~,
\end{align*}
where in the second equality $\mathbb{U}^{\theta}_{d,l|x}(h) = \{\mathcal{U} \in \mathbb{U}(h) : \mathcal{U} \subseteq \mathcal{U}_{d,l|x}^{\theta}(g)\}$, which follows given that $\mathbb{U}(h)$ satisfies Definition \ref{def:P}, and third equality by substituting in the relation between $m^\dagger$ and $\tilde{\gamma}$ from \eqref{eq:alpha_m_relation}.

To show $\tilde{m} \in \mathbf{M}^{*}(h)$, we can equivalently show $\tilde{\gamma} \in \mathbf{\Gamma}^*(h) = \{\gamma \in \mathbf{\Gamma}(h) : \gamma \text{ satisfies } \eqref{eq:Y_moments_alpha}\}$. As it is assumed that $\mathbf{\Gamma}(h)$ satisfies \eqref{eq:NP_condition}, we have $\tilde{\gamma} \in \mathbf{\Gamma}(h)$ as $\tilde{m} \in \mathbf{M}_{\text{np}}(h)$. To show that $\tilde{\gamma}$ satisfies \eqref{eq:Y_moments_alpha}, note that \eqref{eq:Y_moments_alpha} under \eqref{eq:b_constant} with $\mathbb{U}(h)$ satisfying Definition \ref{def:P} can be stated as
\begin{align*}
\sum_{k=0}^K 1\{\mathcal{U}_k \in \mathbb{U}^{\text{sm}}_{d,z|x}(h)\} \gamma_{k,d|x} \int\limits_{\mathcal{U}_k} d F_x = E_{d|z,x}~.
\end{align*}
Given that $m^{\dagger}$ satisfies \eqref{eq:Y_moments}, it then follows that $\tilde{\gamma}$ satisfies this as
\begin{align*}
E_{d|z,x} = \int\limits_{\mathcal{U}^{\text{sm}}_{d,z|x}(g)} m^{\dagger}_{d|x} d F_x = \sum_{\mathcal{U} \in \mathbb{U}^{\text{sm}}_{d,z|x}(h)} \int\limits_{\mathcal{U}} m^{\dagger}_{d|x} d F_x =  \sum_{k=0}^K 1\{\mathcal{U}_k \in \mathbb{U}^{\text{sm}}_{d,z|x}(h)\} \gamma_{k,d|x} \int\limits_{\mathcal{U}_k} d F_x~,
\end{align*}
where the second equality follows from $\mathbb{U}(h)$ satisfying Definition \ref{def:P}, and the third equality follows by substituting in the relation between $m^\dagger$ and $\tilde{\gamma}$ from \eqref{eq:alpha_m_relation}. This completes the proof.

\subsection{Proof of Proposition \ref{prop:arum}}\label{Asec:proof_arum}

Note that the moments here are
 \begin{align*}
P(D=1|Z=z,X=x) &= \int 1\{\tilde{u}_1 < \tilde{g}_{1,z|x},~ \tilde{u}_2 - \tilde{u}_1 > \tilde{g}_{2,z|x} - \tilde{g}_{1,z|x}\}d\tilde{F}_x \equiv H_{1|x}(\tilde{g}_{1,z|x},\tilde{g}_{2,z|x})~, \\
P(D=2|Z=z,X=x) &= \int 1\{\tilde{u}_2 < \tilde{g}_{2,z|x},~ \tilde{u}_1 - \tilde{u}_2 > \tilde{g}_{1,z|x} - \tilde{g}_{2,z|x}\}d\tilde{F}_x \equiv H_{2|x}(\tilde{g}_{1,z|x},\tilde{g}_{2,z|x})~.
\end{align*}
for a given $z \in \mathcal{Z}$ and $x \in \mathcal{X}$. We show the function $H_x(v_1,v_2) \equiv (H_{1|x}(v_1,v_2),H_{2|x}(v_1,v_2))$ is injective on the domain $\mathbf{R}^2$ for each $x \in \mathcal{X}$ given which it follows that if there exists a $(\tilde{g}_{1,z|x},\tilde{g}_{2,z|x}) \in \mathbf{R}^2$ that solves the above system of equations for a given $z \in \mathcal{Z}$ and $x \in \mathcal{X}$ then it must be unique. From this, it will follow there exists at most one $\tilde{g}$ that satisfies \eqref{eq:D_moments_arum1} and \eqref{eq:D_moments_arum2} and hence the result.

To do so, we use \citet[][Corollary 1]{berry/etal:13} that shows if $(H_{0|x},H_{1|x},H_{2|x})$, where $H_{0|x}(v_1,v_2) \equiv 1 - H_{1|x}(v_1,v_2) - H_{2|x}(v_1,v_2)$, satisfies a so-called weak substitutes and connected strict substitution condition---defined in Assumptions 2 and 3 in \cite{berry/etal:13}, respectively---for all $(v_1,v_2) \in \mathbf{R}^2$ and each $x \in \mathcal{X}$ then $H_x$ is injective on $\mathbf{R}^2$ for each $x \in \mathcal{X}$. For each $x \in \mathcal{X}$, as $F_x$ is assumed to have a density, these conditions are satisfied in our setup if
\begin{align*}
H_{12|x}(v_1,v_2) &< 0~, \\
H_{21|x}(v_1,v_2) &< 0~, \\
H_{01|x}(v_1,v_2),~H_{02|x}(v_1,v_2) &< 0~,
\end{align*}
where
\begin{align*}
H_{ij|x}(v_1,v_2) &= \frac{\partial H_{i|x}}{\partial v_j}(v_1,v_2)
\end{align*}
for $i \in \{0,1,2\}$ and $j \in \{1,2\}$. Indeed, this directly follows by taking the derivatives of $H_{1|x}$ and $H_{2|x}$ with respect to $v_1$ and $v_2$, respectively, and the assumption that $F_x$ has a strictly positive density. This completes the proof.

\subsection{Proof of Proposition \ref{prop:sequential}}\label{Asec:proof_sequential}

For each $z \in \mathcal{Z}$ and $x \in \mathcal{X}$, we can identify $g_{1,z|x}$ using \eqref{eq:D_moments_sequential1} as follows
\begin{align*}
  g_{1,z|x} = 1 - P(D=0|Z=z,X=x)~.
\end{align*}
which exists provided the right hand side of the above equation is greater than 0 given the requirement that $g_{1,z|x} > 0$. As $g_{1,z|x}$ is identified and $F_x$ is known, we have a single unknown $g_{2,z|x}$ in \eqref{eq:D_moments_sequential2}. As $F_x$ is assumed to be strictly increasing in the second dimension for $u_1 > 0$, it implies that the right hand side of \eqref{eq:D_moments_sequential2} is strictly decreasing in $g_{2,z|x}$ for $g_{1,z|x} > 0$ and, in turn, if there exists $g_{2,z|x}$ satisfying \eqref{eq:D_moments_sequential2} then it must be unique. Doing the same across  $z \in \mathcal{Z}$ and $x \in \mathcal{X}$ implies that there exists at most a unique $g$ that solves \eqref{eq:D_moments_sequential1} and \eqref{eq:D_moments_sequential2}, which completes the proof.

\subsection{Proof of Proposition \ref{prop:dynamic}}\label{Asec:proof_dynamic}

For a given $z \in \mathcal{Z}$ and $x \in \mathcal{X}$, we can identify $g_{1,z|x}$ using \eqref{eq:D_moments_dynamic2}-\eqref{eq:D_moments_dynamic3} as follows
\begin{align*}
  g_{1,z|x} = P(D=(1,0)|Z=z,X=x) + P(D=(1,1)|Z=z,X=x)~,
\end{align*}
which exists provided the right hand side of the above equation is between 0 and 1 given the requirement that $g_{1,z|x} \in (0,1)$. As $g_{1,z|x}$ is identified and $F_x$ is known, we have a single unknown $g_{3,z|x}$ in \eqref{eq:D_moments_dynamic3}. As $F_x(u_1,1,u_3)$ is assumed to be strictly increasing in the third dimension for $u_1 > 0$, it implies that the right hand side of \eqref{eq:D_moments_dynamic3} is strictly increasing in $g_{3,z|x}$ for $g_{1,z|x} \in (0,1)$ and, in turn, if there exists $g_{3,z|x}$ satisfying \eqref{eq:D_moments_dynamic3} then it must be unique. Similarly, to identify $g_{2,z|x}$, observe that we have a single unknown $g_{2,z|x}$ in \eqref{eq:D_moments_dynamic1}. As it is assumed that $\partial F_x(u_1,u_2,1) / \partial u_2 < 1$ for $u_1 < 1$, it implies that the right hand side of \eqref{eq:D_moments_dynamic1} is strictly increasing in $g_{2,z|x}$ for $g_{1,z|x} \in (0,1)$ and, in turn, if there exists $g_{2,z|x}$ satisfying \eqref{eq:D_moments_dynamic1} then it must be unique. Doing the same across  $z \in \mathcal{Z}$ and $x \in \mathcal{X}$ implies that there exists at most a unique $g$ that solves \eqref{eq:D_moments_dynamic1}-\eqref{eq:D_moments_dynamic3}, which completes the proof.

\subsection{Proof of Proposition \ref{prop:double}}\label{Asec:proof_double}

For a given $x \in \mathcal{X}$, as $g_{1,z'_1|x}$ is known for some $z'_1 \in \mathcal{Z}$ and $F_x$ is known, we have a single unknown $g_{2,z_2|x}$ in the moment
\begin{align*}
P(D=1|Z=(z'_1,z_2),X=x) = F_x(g_{1,z'_1|x},g_{2,z_2|x})
\end{align*}
for $z_2 \in \mathcal{Z}_2$. As $F_x$ is strictly increasing in the second dimension, we have that if there exists a $g_{2,z_2|x}$ that solves the above equation, then it must be unique. Suppose that $g_{2,z_2|x}$ exists for each $z_2 \in \mathcal{Z}_2$. In an analogous manner, taking some $z'_2 \in \mathcal{Z}_2$, we have that as $g_{2,z'_2|x}$ is identified and $F_x$ is known, there is a single unknown $g_{1,z_1|x}$ in the moment
\begin{align*}
P(D=1|Z=(z_1,z'_2),X=x) = F_x(g_{1,z_1|x},g_{2,z'_2|x})
\end{align*}
for $z_1 \in \mathcal{Z}_1 \setminus \{z'_1\}$. As $F_x$ is also strictly increasing in the first dimension, we again have that if there exists a $g_{1,z_1|x}$ that solves the above equation, then it must be unique. Doing the same across $x \in \mathcal{X}$ shows that if there exists a $g$ that solves \eqref{eq:D_moments_double} then it must be unique, which completes the proof.

\subsection{Proof of Proposition \ref{prop:firststage_HS}}\label{Asec:proof_firststage_HS}

For each $x \in \mathcal{X}$ taking $\delta_{c|x} \in [0,1]$ as given, our objective is to show that for $H(\sigma_x,\delta_{c|x}) = P_x$, where $P_x = (P_{c,0|x},P_{c,1|x},P_{p,0|x},P_{p,1|x})$, there exists a unique $\sigma_x \in \mathbf{R}^2 \times \mathbf{R}_{++} \times (-1,1) \equiv \mathbf{S}$ that satisfies this system of equations. To this end, we use a global univalence theorem from \citet[][Corollary 1.4]{ambrosetti/prodi:95} that shows that if (i) $H$ is locally invertible, and (ii) $H$ is continuous and proper with a connected image, then the result follows---see \citet[][Theorem 3.2]{chern/etal:24} and \citet[][Theorem 2]{depaula/etal:24} who use this result to argue point identification in alternative settings.

Condition (i) follows from the assumption as it is supposed that its Jacobian has full rank on the parameter space. To show Condition (ii), note that $H$ is continuous by definition given the normal distribution assumption, and since $\mathbf{S}$ is a connected set, its image under $H$ is a connected set as $H$ is continuous. To argue that $H$ is proper, we need to show that for every $K \subset H(\mathbf{S}) = (0,1)^4$ that is compact, we have that $H^{-1}(K)$ is compact. To this end, let $K$ be a compact set $K \subset (0,1)^4$. As it is compact, it is closed and bounded. We need to show that $H^{-1}(K)$ is closed and bounded as well. As $H$ is continuous and as a pre-image of a closed set under a continuous function is also closed, we have that $H^{-1}(K)$ is closed. To show that it is also bounded, suppose it is not. This means there exists a sequence of $\{(\zeta_{1,k},\zeta_{2,k},\zeta_k,\rho_k)\}_{k=1}^{\infty}$ in $H^{-1}(K)$ such that $|\zeta_{1,k}| \to \infty$, $|\zeta_{2,k}| \to \infty$, or $|\zeta_{k}| \to \infty$. However, this would imply $H_j \to 0$ or $H_j \to 1$ for some $j$ and hence the boundary of $(0,1)$ is contained in $K$ for the coordinate $j$. This is a contradiction and hence it must be that $H^{-1}(K)$ is bounded.

\subsection{Proof of Proposition \ref{prop:prop_recycle}}\label{Asec:prop_recycle}

\begin{proposition}\label{prop:prop_recycle}
For  each $\lambda \in \mathbf{L}_0$, suppose $TS(\lambda)$ and $cv(1-\alpha,\lambda)$ are such that
\begin{align}\label{eq:test_size_lambda}
 \lim_{N \to \infty}\text{Prob}\left\{TS(\lambda) \geq cv(1-\alpha,\lambda)\right\} \leq \alpha~
\end{align}
and $\widehat{\mathbf{L}}_0$ is such that
\begin{align}\label{eq:L_0_consistent}
 \lim_{N \to \infty}\text{Prob}\left\{\widehat{\mathbf{L}}_0 \subseteq \mathbf{L}_0\right\} = 1~,
\end{align}
where $N$ denotes sample size. We then have that
\begin{align}\label{eq:test_size}
 \lim_{N \to \infty} \text{Prob}\left\{\min_{\lambda \in \mathbf{L}} TS(\lambda) \geq \min_{\lambda \in \widehat{\mathbf{L}}_0} cv(1-\alpha,\lambda)\right\} \leq \alpha~.
\end{align}
\end{proposition}

\noindent \textsc{Proof}: Let $cv^*(1-\alpha)$ be the $(1-\alpha)$-quantile of $\min_{\lambda \in \mathbf{L}} TS(\lambda)$, and let $cv^*(1-\alpha,\lambda)$ be the $(1-\alpha)$-quantile of $TS(\lambda)$ for each $\lambda \in \mathbf{L}$. As the quantile of a minimum is weakly smaller than the minimum of quantiles, note that
\begin{align}\label{eq:min_crit}
 cv^*(1-\alpha) \leq \min_{\lambda \in \mathbf{L}}cv^*(1-\alpha,\lambda)~.
\end{align}
Moreover it follows from the definition of $cv^*(1-\alpha,\lambda)$ and \eqref{eq:test_size_lambda} that
\begin{align*}
 cv^*(1-\alpha,\lambda) \leq cv(1-\alpha,\lambda)
\end{align*}
for each $\lambda \in \mathbf{L}_0$, and hence along with \eqref{eq:min_crit} that
\begin{align*}
 cv^*(1-\alpha) \leq \min_{\lambda \in \mathbf{L}} cv^*(1-\alpha,\lambda) \leq \min_{\lambda \in \mathbf{L}_0} cv^*(1-\alpha,\lambda) \leq  \min_{\lambda \in \mathbf{L}_0} cv(1-\alpha,\lambda)
\end{align*}
with probability approaching 1 as $N \to \infty$. It then follows along with the definition of $cv^*(1-\alpha)$ that
\begin{align}\label{eq:size_0}
 \text{Prob}\left\{\min_{\lambda \in \mathbf{L}} TS(\lambda) \geq \min_{\lambda \in \mathbf{L}_0} cv(1-\alpha,\lambda)\right\} \leq \text{Prob}\left\{\min_{\lambda \in \mathbf{L}} TS(\lambda) \geq cv^*(1-\alpha)\right\} \leq \alpha
\end{align}
with probability approaching 1 as $N \to \infty$. To complete the proof, note then that
\begin{align*}
 \text{Prob}\left\{\min_{\lambda \in \mathbf{L}} TS(\lambda) \geq \min_{\lambda \in \widehat{\mathbf{L}}_0} cv(1-\alpha,\lambda)\right\} &\leq \text{Prob}\left\{\min_{\lambda \in \mathbf{L}} TS(\lambda) \geq \min_{\lambda \in \mathbf{L}_0} cv(1-\alpha,\lambda)\right\} \\
 &~~~~~~~~~+ \text{Prob}\left\{\min_{\lambda \in \widehat{\mathbf{L}}_0} cv(1-\alpha,\lambda) < \min_{\lambda \in \mathbf{L}_0} cv(1-\alpha,\lambda)\right\} \\
 &\leq \text{Prob}\left\{\min_{\lambda \in \mathbf{L}} TS(\lambda) \geq \min_{\lambda \in \mathbf{L}_0} cv(1-\alpha,\lambda)\right\} + \text{Prob}\left\{ \widehat{\mathbf{L}}_0 \not\subseteq \mathbf{L}_0 \right\} \\
 &\leq \alpha
\end{align*}
with probability approaching 1 as $N \to \infty$, where the first and second inequalities follow trivially and the final inequality follows from \eqref{eq:L_0_consistent} and \eqref{eq:size_0}.

\normalsize

\section{Appendix Tables}\label{Asec:tables}

\begin{landscape}
\begin{table}[!t]
\begin{center}
\caption{Various treatment effect parameters and their restatements in terms of \eqref{eq:theta}}
\label{tab:theta_exp}
\scalebox{0.7}{
\def\arraystretch{1.75}
\makebox[\textwidth]{\begin{threeparttable}
\input{theta_exp.tex}
\begin{tablenotes}[flushleft]
\setlength\labelsep{0pt}
\item When $|\mathcal{L}|=1$, the index $l$ is redundant and $\mathcal{L}$ can be any arbitrary set with a single element. For $\text{PRTE}^{\delta,\delta'}$, $l\equiv (l_1,l_2)$, where $l_1 \in \mathcal{Z}$ and $l_2 \in \{\delta,\delta'\}$.
\end{tablenotes}
\end{threeparttable} }}
\end{center}
\end{table}
\end{landscape}

\begin{landscape}
\begin{table}[!h]
\begin{center}
\caption{Shape restrictions on $m$ that correspond to linear restrictions on $\gamma$}
\label{tab:alpha_shape}
\scalebox{0.7}{
\def\arraystretch{1.75}
\makebox[\textwidth]{\begin{threeparttable}
\input{alpha_shape_derivation.tex}
\begin{tablenotes}[flushleft]
\setlength\labelsep{0pt}
\item $\mathcal{U}_{\text{grid}}$ denotes a grid of points in $[0,1]^J$. The restrictions in terms of $\gamma$ when the restrictions are considered over $\mathcal{U}_{\text{grid}}$ can be straightforwardly derived by plugging in the parametrization from Assumption \ref{ass:PM} into the assumptions on the MTRs evaluated at the points in the grid. For the restrictions under \eqref{eq:b_constant}, a grid is not necessarily required as the parametrization takes constant values and the assumptions can be characterized equivalently in terms of linear restrictions on $\gamma$. We only report the restrictions for Assumptions B, M$_{d',d''}$, BV$_{d',d''}$ and S in this case as these assumptions correspond to those that can be imposed on the MTRs in Proposition \ref{prop:equivalence}.
\end{tablenotes}
\end{threeparttable} }}
\end{center}
\end{table}
\end{landscape}

\newpage
\normalsize
\bibliography{references.bib}

\end{document}

%% file: table_descriptive.tex
 \begin{tabular}{ L{5cm}  C{4cm} C{4cm} }  
 \toprule  
 \multicolumn{3}{c}{Panel (a): Experimental Impacts}                                         \\ \cline{1-3}  
 \multirow{2}{*}{$\text{ITT}^D$}  &  \multicolumn{2}{c}{0.689}                          \\              
                                   &  \multicolumn{2}{c}{[0.018]}                     \\ \cline{2-3}  
 \multirow{2}{*}{$\text{ITT}^Y$}  &  \multicolumn{2}{c}{0.159}                          \\              
                                   &  \multicolumn{2}{c}{[0.029]}                     \\ \cline{2-3}  
 \multirow{2}{*}{$\text{IV}$}     &  \multicolumn{2}{c}{0.230}                             \\              
                                   &  \multicolumn{2}{c}{[0.043]}                        \\              
 \midrule                                                                                                     
 \multicolumn{3}{c}{Panel (b): Preschool choices by Head Start offer status}                 \\ \cline{1-3}  
                                         & Offered                & Not offered              \\ \cline{1-3}      
 \multirow{1}{*}{Home care}              & 0.094        & 0.548          \\ \cline{2-3}  
 \multirow{1}{*}{Competing preschool}    & 0.082        & 0.318          \\ \cline{2-3}  
 \multirow{1}{*}{Head Start}             & 0.823        & 0.135          \\ \hline       
 Sample size                             & 2290            & 1337              \\              
 \bottomrule 
 \end{tabular}

%% file: table_IVdecomposition.tex
 \begin{tabular}{ L{4cm}  R{1.1cm} L{1.75cm} R{1cm} R{1cm} R{1cm} R{1cm} R{1cm} R{1cm} R{1cm} R{1cm} }  
 \toprule  
 Assumptions on MTRs            & & & (1)        & (2)        & (3)        & (4)        & (5)        & (6)        & (7)        & (8)        \\ \hline  
 Parametric ($K$)               & & &  1         &  1         & 2          & NP         & 1          & 2          & NP         & 1           \\         
 Separability                   & & & \checkmark & \checkmark & \checkmark & \checkmark &            &            &            & \checkmark  \\         
 Bounded Variation              & & &            & \checkmark & \checkmark & \checkmark & \checkmark & \checkmark & \checkmark & \checkmark  \\         
 Positive Selection             & & &            &            &            &            &            &            &            & \checkmark  \\         
 Positive Effects               & & &            &            &            &            &            &            &            & \checkmark  \\         
 \midrule                                                                                                                                                                         
 Parameter                      & & &            &            &            &            &            &            &            &              \\ \hline  
                                & & Lower CI & \multicolumn{8}{c}{\scriptsize 0.61}  \\  
 $\text{S}_{np}$               & & Est.     & \multicolumn{8}{c}{0.66}  \\  
                                & & Upper CI & \multicolumn{8}{c}{ \scriptsize 0.70} \\ \cline{1-11}  
 \multicolumn{1}{l}{\multirow{8}{*}{$\text{LATE}_{np}$}} & & Lower CI & \scriptsize -0.10 & \scriptsize 0.07 & \scriptsize 0.07 & \scriptsize 0.13 & \scriptsize 0.08 & \scriptsize 0.13 & \scriptsize 0.13  & \scriptsize 0.10    \\  
                                                        & \multicolumn{1}{l}{\multirow{2}{*}{$\delta_{c|x} = 0$}}   &  Est. LB   & \multicolumn{1}{r}{\multirow{2}{*}{0.27}} & \multicolumn{1}{r}{\multirow{2}{*}{0.30}} & 0.25 & 0.22 & 0.25 & 0.22 & 0.22 & \multicolumn{1}{r}{\multirow{2}{*}{0.27}} \\  
                                                        &                                                            &  Est. UB   &                                                        &                                                        & 0.42 & 0.46 & 0.44 & 0.46 & 0.46 &                                                                                                                                  \\  
                                  & & Upper CI & \scriptsize 0.60 & \scriptsize 0.61 & \scriptsize 0.61 & \scriptsize 0.56 & \scriptsize 0.63 & \scriptsize 0.56 & \scriptsize 0.56  & \scriptsize 0.56   \\   
 \cline{2-11} 
                                                         & & Lower CI & \scriptsize -0.10 & \scriptsize 0.04 & \scriptsize 0.06 & \scriptsize 0.11 & \scriptsize 0.06 & \scriptsize 0.11 & \scriptsize 0.11  & \scriptsize 0.06    \\  
                                                        & \multicolumn{1}{l}{\multirow{2}{*}{$\delta_{c|x} \in [0,1]$}} & Est. LB &  0.22 & 0.27 & 0.23 & 0.21 &  0.23 & 0.22 & 0.22 & 0.29  \\  
                                                        &                                                                 & Est. UB &  0.57 & 0.39 & 0.51 & 0.59 &  0.48 & 0.58 & 0.59 & 0.34   \\  
                                  & & Upper CI & \scriptsize 0.80 & \scriptsize 0.73 & \scriptsize 0.77 & \scriptsize 0.75 & \scriptsize 0.72 & \scriptsize 0.75 & \scriptsize 0.72  & \scriptsize 0.73   \\   
 \cline{1-11} 
  \multicolumn{1}{l}{\multirow{8}{*}{$\text{LATE}_{cp}$}} & & Lower CI & \scriptsize -0.70 & \scriptsize -0.20 & \scriptsize -0.20 & \scriptsize -0.20 & \scriptsize -0.20 & \scriptsize -0.20 & \scriptsize -0.20  & \scriptsize -0.20    \\  
                                                        & \multicolumn{1}{l}{\multirow{2}{*}{$\delta_{c|x} = 0$}}   &  Est. LB   & \multicolumn{1}{r}{\multirow{2}{*}{0.14}} & \multicolumn{1}{r}{\multirow{2}{*}{0.08}} & -0.17 & -0.20 & -0.18 & -0.20 & -0.20 & \multicolumn{1}{r}{\multirow{2}{*}{0.13}} \\  
                                                        &                                                            &  Est. UB   &                                                        &                                                        & 0.17 & 0.20 & 0.17 & 0.20 & 0.20 &                                                                                                                                  \\  
                                  & & Upper CI & \scriptsize 0.80 & \scriptsize 0.20 & \scriptsize 0.20 & \scriptsize 0.20 & \scriptsize 0.20 & \scriptsize 0.20 & \scriptsize 0.20  & \scriptsize 0.20   \\   
 \cline{2-11} 
                                                         & & Lower CI & \scriptsize -0.70 & \scriptsize -0.20 & \scriptsize -0.20 & \scriptsize -0.20 & \scriptsize -0.20 & \scriptsize -0.20 & \scriptsize -0.20  & \scriptsize -0.20    \\  
                                                        & \multicolumn{1}{l}{\multirow{2}{*}{$\delta_{c|x} \in [0,1]$}} & Est. LB &  -0.44 & -0.05 & -0.20 & -0.20 &  -0.19 & -0.20 & -0.20 & 0.06  \\  
                                                        &                                                                 & Est. UB &  0.24 & 0.15 & 0.20 & 0.20 &  0.20 & 0.20 & 0.20 & 0.12   \\  
                                  & & Upper CI & \scriptsize 0.80 & \scriptsize 0.20 & \scriptsize 0.20 & \scriptsize 0.20 & \scriptsize 0.20 & \scriptsize 0.20 & \scriptsize 0.20  & \scriptsize 0.20   \\   
 \bottomrule 
 \end{tabular}

%% file: table_policy.tex
 \begin{tabular}{ L{4cm}  R{1.1cm} L{1.75cm} R{1cm} R{1cm} R{1cm} R{1cm} R{1cm} R{1cm} R{1cm} R{1cm} }  
 \toprule  
 Assumptions on MTRs             & & & (1)        & (2)        & (3)        & (4)        & (5)        & (6)        & (7)        & (8)        \\ \hline  
 Parametric ($K$)                & & &  1         &  1         & 2          & NP         & 1          & 2          & NP         & 1           \\         
 Separability                    & & & \checkmark & \checkmark & \checkmark & \checkmark &            &            &            & \checkmark  \\         
 Bounded Variation               & & &            & \checkmark & \checkmark & \checkmark & \checkmark & \checkmark & \checkmark & \checkmark  \\         
 Positive Selection              & & &            &            &            &            &            &            &            & \checkmark  \\         
 Positive Effects                & & &            &            &            &            &            &            &            & \checkmark  \\         
 \midrule                                                                                                                                                                         
 Parameter                       & & &            &            &            &            &            &            &            &             \\ \hline  
                                  & & Lower CI  & \multicolumn{8}{c}{\scriptsize 1.20}  \\  
 $\text{MVPF}_{\text{non-adj}}$ & & Est.      & \multicolumn{8}{c}{2.86}  \\  
                                  & & Upper CI  & \multicolumn{8}{c}{ \scriptsize 9.20} \\ \cline{1-11}  
 \multicolumn{1}{l}{\multirow{8}{*}{$\text{MVPF}_{\text{adj}}$}} & & Lower CI  & \scriptsize $-\infty$ & \scriptsize 0.30 & \scriptsize 0.30 & \scriptsize 0.20 & \scriptsize 0.40 & \scriptsize 0.40 & \scriptsize 0.40  & \scriptsize 0.50    \\  
                                                        & \multicolumn{1}{l}{\multirow{2}{*}{$\delta_{c|x} = 0$}}  & Est. LB      & \multicolumn{1}{r}{\multirow{2}{*}{3.57}} & \multicolumn{1}{r}{\multirow{2}{*}{2.61}} & 1.60 & 0.97 & 1.77 & 1.10 & 0.97 & \multicolumn{1}{r}{\multirow{2}{*}{2.09}} \\  
                                                        &                                                           & Est. UB      &                                                        &                                                        & 9.50 & 25.46 & 10.06 & 18.09 & 25.19 &                                                         \\  
                                   & & Upper CI & \scriptsize $\infty$ & \scriptsize $\infty$ & \scriptsize $\infty$ & \scriptsize $\infty$ & \scriptsize $\infty$ & \scriptsize $\infty$ & \scriptsize $\infty$  & \scriptsize $\infty$    \\   
 \cline{2-11} 
                                                         & & Lower CI  & \scriptsize $-\infty$ & \scriptsize $-\infty$ & \scriptsize $-\infty$ & \scriptsize $-\infty$ & \scriptsize $-\infty$ & \scriptsize $-\infty$ & \scriptsize $-\infty$  & \scriptsize 0.40    \\  
                                                        & \multicolumn{1}{l}{\multirow{2}{*}{$\delta_{c|x} \in [0,1]$}} & Est. LB &  0.73 & 0.98 & 0.56 & $-\infty$ &  0.59 & 0.32 & $-\infty$ & 2.15  \\  
                                                        &                                                                 & Est. UB &  36.13 & 6.18 & 22.82 & $\infty$ &  14.10 & $\infty$ & $\infty$ & 3.08   \\  
                                   & & Upper CI & \scriptsize $\infty$ & \scriptsize $\infty$ & \scriptsize $\infty$ & \scriptsize $\infty$ & \scriptsize $\infty$ & \scriptsize $\infty$ & \scriptsize $\infty$  & \scriptsize $\infty$    \\   
 \bottomrule 
 \end{tabular}

%% file: theta_exp.tex
 \begin{tabular}{ L{5.5cm} L{6.5cm} C{2cm} C{11cm} C{4cm}}  
 \toprule 
 Parameter of interest & Expression & \multicolumn{3}{c}{Rewriting in terms of \eqref{eq:theta}} \\ \cline{3-5}
   & & $\mathcal{L}$  & \multicolumn{1}{c}{$w_{d,l|x}$} & \multicolumn{1}{c}{$\mathcal{U}^{\theta}_{d,l|x}$} \\ \hline
 Average Treatment Effect, $\text{ATE}(d',d'')$ & $E[Y(d') - Y(d'')]$ & $|\mathcal{L}|=1$ & $P(X=x) (1\{d=d'\} - 1\{d=d''\})$ & $[0,1]^J$ \\ \cline{2-5}
Average Treatment Effect on the Treated, $\text{ATT}(d',d'')$ & $E[Y(d') - Y(d'')|D=d']$ & $\mathcal{Z}$ & $\displaystyle\dfrac{P(X=x,Z=l) (1\{d=d'\} - 1\{d=d''\})}{P(D=d')}$ & $\mathcal{U}^{\text{sm}}_{d',l|x}$ \\ \cline{2-5}
Local Average Treatment Effect, $\text{LATE}_{\{d_z,z \in \mathcal{Z}'\}}(d',d'')$ & $E[Y(d') - Y(d'')|(D(z)=d_z,z \in \mathcal{Z}')]$ & $|\mathcal{L}|=1$ & $\dfrac{P(X=x) (1\{d=d'\} - 1\{d=d''\})}{\displaystyle\sum\limits_{x' \in \mathcal{X}}P(X=x')\int\limits_{\bigcap\limits_{z \in \mathcal{Z}'} \mathcal{U}^{\text{sm}}_{d_z,z|x}} dF_{x'}}$ & $\displaystyle\bigcap\limits_{z \in \mathcal{Z}'} \mathcal{U}^{\text{sm}}_{d_z,z|x}$ \\ \cline{2-5}
Policy Relevant Treatment Effect, $\text{PRTE}^{\delta',\delta''}$ & $E[Y^{\delta'} - Y^{\delta''}|D^{\delta'}\neq D^{\delta''}]$ & $\mathcal{Z} \times \{\delta,\delta'\}$ &  $\dfrac{P(X=x,Z=l_1) (1\{l_2=\delta'\} - 1\{l_2=\delta''\})}{\displaystyle 1 - \sum\limits_{x' \in \mathcal{X}}\sum\limits_{z' \in \mathcal{Z}}\sum\limits_{d' \in \mathcal{D}}P(X=x',Z=z') \int\limits_{ \bigcap\limits_{\delta \in \{\delta',\delta''\}}\mathcal{U}^{\text{sm}}_{d',z'|x'}(g + \delta)} dF_{x'}  }$ & $\mathcal{U}^{\text{sm}}_{d,l_1|x}(g + l_2)$ \\ \cline{2-5}
Policy Relevant Treatment Effect given $(D^{\delta'},D^{\delta''})$, $\text{PRTE}^{\delta',\delta''}(d',d'')$ & $E[Y^{\delta'} - Y^{\delta''}|D^{\delta'} = d',D^{\delta''}=d'']$ & $\mathcal{Z}$ & $\dfrac{P(X=x,Z=l) (1\{d=d'\} - 1\{d=d''\})}{\displaystyle \sum\limits_{x' \in \mathcal{X}}\sum\limits_{z' \in \mathcal{Z}}P(X=x',Z=z') \int\limits_{\bigcap\limits_{\delta \in \{\delta',\delta''\} } \mathcal{U}^{\text{sm}}_{d',z'|x'}(g + \delta)} dF_{x'}  }$ & $\displaystyle \bigcap\limits_{\delta \in \{\delta',\delta''\} } \mathcal{U}^{\text{sm}}_{d',l|x}(g + \delta)$ \\ 
\bottomrule 
 \end{tabular}

%% file: alpha_shape_derivation.tex
 \begin{tabular}{ L{5cm} L{5cm}  L{10cm}  L{10cm}}  
 \toprule 
 Assumption & Restriction on MTRs & \multicolumn{2}{c}{In terms of $\gamma$} \\ \cline{3-4}
 & & \multicolumn{1}{c}{Over $\mathcal{U}_{\text{grid}}$} &  \multicolumn{1}{c}{Under \eqref{eq:b_constant}} \\ \hline
 B (Boundedness) & $\underline{m} \leq m_{d|x} \leq \bar{m}$, where $\underline{m}$ and $\bar{m}$ are known constants.  & $$\underline{m} \leq \sum\limits_{k=0}^{K} \gamma_{k,d|x} b_{k}(h)(u)  \leq \bar{m}$$ for $u \in \mathcal{U}_{\text{grid}}$, $d \in \mathcal{D}$ and $x \in \mathcal{X}$.  &  $$\underline{m} \leq \gamma_{k,d|x} \leq \bar{m}$$ for$0 \leq k \leq K$, $d \in \mathcal{D}$ and $x \in \mathcal{X}$  \\  \cline{2-4}
 M$_{d',d''}$ (Monotonicity between $d'$ and $d''$) & $m_{d'|x} - m_{d''|x} \geq 0$. & $$\sum\limits_{k=0}^{K} (\gamma_{k,d'|x} - \gamma_{k,d''|x}) b_{k}(h)(u) \geq 0$$ for $u \in \mathcal{U}_{\text{grid}}$ and $x \in \mathcal{X}$. &  $$\gamma_{k,d'|x} - \gamma_{k,d''|x} \geq 0$$ for $0 \leq k \leq K$ and $x \in \mathcal{X}$. \\ \cline{2-4}
 BV$_{d',d''}$ (Bounded Variation between $d'$ and $d''$) & $|m_{d'|x} - m_{d''|x}| \leq \bar{m}_{d',d''}$, where $\bar{m}_{d',d''}$ is a known constant  & $$\underline{m}_{d',d''} \leq \sum\limits_{k=0}^{K} (\gamma_{k,d'|x} - \gamma_{k,d''|x}) b_{k}(h)(u) \leq \bar{m}_{d',d''}$$ for $u \in \mathcal{U}_{\text{grid}}$ and $x \in \mathcal{X}$.  &  $$\underline{m}_{d',d''} \leq \gamma_{k,d'|x} - \gamma_{k,d''|x} \leq \bar{m}_{d',d''}$$ for $0 \leq k \leq K$ and $x \in \mathcal{X}$. \\  \cline{2-4}
 M$_j$ (Monotonicity in $U_j$) & $m_{d|x}$ is increasing (decreasing) in $u_j$   & $$\sum\limits_{k=0}^{K} \gamma_{k,d|x}  (b_{k}(h)(u') - b_{k}(h)(u'')) \geq 0$$ for $u',u'' \in \mathcal{U}_{\text{grid}}$, $d \in \mathcal{D}$ and $x \in \mathcal{X}$, where $u'_j \geq u''_j$ and $u'_i = u''_i$ for $i \neq j$. &  \\ \cline{2-4}
 S (Separability between $U$ and $X$) & $m_{d|x}(u) = m^U_{d}(u) + m_{d}^{X}(x)$, where $m^U_{d}$ and $m^{X}_{d}$ are unknown functions & \sloppy Introduce auxiliary variables $(\gamma_{k,d} : 0 \leq k \leq K,~d \in \mathcal{D})$ and $(\gamma_{d|x} : d \in \mathcal{D},~x \in \mathcal{X})$ such that $$\sum\limits_{k=0}^{K}\gamma_{k,d|x} b_{k}(h)(u) = \sum\limits_{k=0}^{K}\gamma_{k,d} b_{k}(h)(u) + \gamma_{d|x}$$ for $u \in \mathcal{U}_{\text{grid}}$, $d \in \mathcal{D}$ and $x \in \mathcal{X}$.  & \sloppy Introduce auxiliary variables $(\gamma_{k,d} : 0 \leq k \leq K,~d \in \mathcal{D})$ and $(\gamma_{d|x} : d \in \mathcal{D},~x \in \mathcal{X})$ such that $$\gamma_{k,d|x} = \gamma_{k,d} + \gamma_{d|x}$$ for $0 \leq k \leq K$, $d \in \mathcal{D}$ and $x \in \mathcal{X}$.  \\
\bottomrule 
 \end{tabular} 

%% file: MTEmultiple.bbl
\begin{thebibliography}{60}
\expandafter\ifx\csname natexlab\endcsname\relax\def\natexlab#1{#1}\fi
\expandafter\ifx\csname url\endcsname\relax
  \def\url#1{\texttt{#1}}\fi
\expandafter\ifx\csname urlprefix\endcsname\relax\def\urlprefix{URL }\fi
\providecommand{\eprint}[2][]{\url{#2}}

\bibitem[{Abdulkadiro{\u{g}}lu et~al.(2020)Abdulkadiro{\u{g}}lu, Pathak,
  Schellenberg and Walters}]{abdulkadirouglu/etaL:20}
\textsc{Abdulkadiro{\u{g}}lu, A.}, \textsc{Pathak, P.~A.},
  \textsc{Schellenberg, J.} and \textsc{Walters, C.~R.} (2020).
\newblock Do parents value school effectiveness?
\newblock \textit{American Economic Review}, \textbf{110} 1502--1539.

\bibitem[{Ambrosetti and Prodi(1995)}]{ambrosetti/prodi:95}
\textsc{Ambrosetti, A.} and \textsc{Prodi, G.} (1995).
\newblock \textit{A primer of nonlinear analysis}.
\newblock 34, Cambridge University Press.

\bibitem[{Angrist and Imbens(1995)}]{angrist/imbens:95}
\textsc{Angrist, J.~D.} and \textsc{Imbens, G.~W.} (1995).
\newblock Two-stage least squares estimation of average causal effects in
  models with variable treatment intensity.
\newblock \textit{Journal of the American statistical Association}, \textbf{90}
  431--442.

\bibitem[{Arteaga(2021)}]{arteaga:21}
\textsc{Arteaga, C.} (2021).
\newblock Parental incarceration and children's educational attainment.
\newblock \textit{The Review of Economics and Statistics} 1--45.

\bibitem[{Bai et~al.(2025{\natexlab{a}})Bai, Huang, Moon, Santos, Shaikh and
  Vytlacil}]{bai/etal:2025}
\textsc{Bai, Y.}, \textsc{Huang, S.}, \textsc{Moon, S.}, \textsc{Santos, A.},
  \textsc{Shaikh, A.~M.} and \textsc{Vytlacil, E.~J.} (2025{\natexlab{a}}).
\newblock Inference for treatment effects conditional on generalized principal
  strata using instrumental variables.
\newblock \textit{arXiv preprint arXiv:2411.05220}.

\bibitem[{Bai et~al.(2025{\natexlab{b}})Bai, Huang and
  Tabord-Meehan}]{bai/tabord:25}
\textsc{Bai, Y.}, \textsc{Huang, S.} and \textsc{Tabord-Meehan, M.}
  (2025{\natexlab{b}}).
\newblock Sharp testable implications of encouragement designs.
\newblock \textit{arXiv preprint arXiv:2411.09808}.

\bibitem[{Berry et~al.(2013)Berry, Gandhi and Haile}]{berry/etal:13}
\textsc{Berry, S.}, \textsc{Gandhi, A.} and \textsc{Haile, P.} (2013).
\newblock Connected substitutes and invertibility of demand.
\newblock \textit{Econometrica}, \textbf{81} 2087--2111.

\bibitem[{Bhuller and Sigstad(2024)}]{bhuller/sigstad:24}
\textsc{Bhuller, M.} and \textsc{Sigstad, H.} (2024).
\newblock 2sls with multiple treatments.
\newblock \textit{Journal of Econometrics}, \textbf{242} 105785.

\bibitem[{Boyd and Vandenberghe(2004)}]{boyd/andenberghe:04}
\textsc{Boyd, S.} and \textsc{Vandenberghe, L.} (2004).
\newblock \textit{Convex optimization}.
\newblock Cambridge university press.

\bibitem[{Brinch et~al.(2017)Brinch, Mogstad and Wiswall}]{brinch/etal:17}
\textsc{Brinch, C.~N.}, \textsc{Mogstad, M.} and \textsc{Wiswall, M.} (2017).
\newblock Beyond late with a discrete instrument.
\newblock \textit{Journal of Political Economy}, \textbf{125} 985--1039.

\bibitem[{Bugni et~al.(2015)Bugni, Canay and Shi}]{bugni/etal:15}
\textsc{Bugni, F.~A.}, \textsc{Canay, I.~A.} and \textsc{Shi, X.} (2015).
\newblock Specification tests for partially identified models defined by moment
  inequalities.
\newblock \textit{Journal of Econometrics}, \textbf{185} 259--282.

\bibitem[{Chernozhukov et~al.(2024)Chernozhukov, Fern{\'a}ndez-Val, Han and
  W{\"u}thrich}]{chern/etal:24}
\textsc{Chernozhukov, V.}, \textsc{Fern{\'a}ndez-Val, I.}, \textsc{Han, S.} and
  \textsc{W{\"u}thrich, K.} (2024).
\newblock Estimating causal effects of discrete and continuous treatments with
  binary instruments.
\newblock \textit{arXiv preprint arXiv:2403.05850}.

\bibitem[{Chernozhukov et~al.(2023)Chernozhukov, Fern{\'a}ndez-Val and
  Luo}]{chern/etal:23}
\textsc{Chernozhukov, V.}, \textsc{Fern{\'a}ndez-Val, I.} and \textsc{Luo, S.}
  (2023).
\newblock Distribution regression with sample selection and uk wage
  decomposition.

\bibitem[{Chesher et~al.(2013)Chesher, Rosen and Smolinski}]{chesher/etal:13}
\textsc{Chesher, A.}, \textsc{Rosen, A.~M.} and \textsc{Smolinski, K.} (2013).
\newblock An instrumental variable model of multiple discrete choice.
\newblock \textit{Quantitative Economics}, \textbf{4} 157--196.

\bibitem[{Chetty et~al.(2011)Chetty, Friedman, Hilger, Saez, Schanzenbach and
  Yagan}]{chetty/etal:11}
\textsc{Chetty, R.}, \textsc{Friedman, J.~N.}, \textsc{Hilger, N.},
  \textsc{Saez, E.}, \textsc{Schanzenbach, D.~W.} and \textsc{Yagan, D.}
  (2011).
\newblock How does your kindergarten classroom affect your earnings? evidence
  from project star.
\newblock \textit{The Quarterly journal of economics}, \textbf{126} 1593--1660.

\bibitem[{Chetverikov et~al.(2018)Chetverikov, Santos and
  Shaikh}]{chetverikov/etal:18}
\textsc{Chetverikov, D.}, \textsc{Santos, A.} and \textsc{Shaikh, A.~M.}
  (2018).
\newblock The econometrics of shape restrictions.
\newblock \textit{Annual Review of Economics}, \textbf{10} 31--63.

\bibitem[{Cornelissen et~al.(2018)Cornelissen, Dustmann, Raute and
  Sch{\"o}nberg}]{cornelissen/etal:18}
\textsc{Cornelissen, T.}, \textsc{Dustmann, C.}, \textsc{Raute, A.} and
  \textsc{Sch{\"o}nberg, U.} (2018).
\newblock Who benefits from universal child care? estimating marginal returns
  to early child care attendance.
\newblock \textit{Journal of Political Economy}, \textbf{126} 2356--2409.

\bibitem[{Cox et~al.(2025)Cox, Shi and Shimizu}]{cox/etal:24}
\textsc{Cox, G.~F.}, \textsc{Shi, X.} and \textsc{Shimizu, Y.} (2025).
\newblock Testing inequalities linear in nuisance parameters.
\newblock \eprint{2510.27633},
  \urlprefix\url{https://arxiv.org/abs/2510.27633}.

\bibitem[{De~Paula et~al.(2024)De~Paula, Rasul and Souza}]{depaula/etal:24}
\textsc{De~Paula, A.}, \textsc{Rasul, I.} and \textsc{Souza, P.~C.} (2024).
\newblock Identifying network ties from panel data: Theory and an application
  to tax competition.
\newblock \textit{Review of Economic Studies} rdae088.

\bibitem[{Dutz et~al.(2021)Dutz, Huitfeldt, Lacouture, Mogstad, Torgovitsky and
  van Dijk}]{dutz/etal:21}
\textsc{Dutz, D.}, \textsc{Huitfeldt, I.}, \textsc{Lacouture, S.},
  \textsc{Mogstad, M.}, \textsc{Torgovitsky, A.} and \textsc{van Dijk, W.}
  (2021).
\newblock Selection in surveys: Using randomized incentives to detect and
  account for nonresponse bias.
\newblock Tech. rep., National Bureau of Economic Research.

\bibitem[{Goff(2025)}]{goff:25}
\textsc{Goff, L.} (2025).
\newblock When does iv identification not restrict outcomes?
\newblock \textit{arXiv preprint arXiv:2406.02835}.

\bibitem[{Gu et~al.(2022)Gu, Russell and Stringham}]{gu/etal:22}
\textsc{Gu, J.}, \textsc{Russell, T.} and \textsc{Stringham, T.} (2022).
\newblock Counterfactual identification and latent space enumeration in
  discrete outcome models.
\newblock \textit{Available at SSRN 4188109}.

\bibitem[{Han and Vytlacil(2017)}]{han/vytlacil:17}
\textsc{Han, S.} and \textsc{Vytlacil, E.~J.} (2017).
\newblock Identification in a generalization of bivariate probit models with
  dummy endogenous regressors.
\newblock \textit{Journal of Econometrics}, \textbf{199} 63--73.

\bibitem[{Heckman et~al.(2010)Heckman, Moon, Pinto, Savelyev and
  Yavitz}]{heckman/etal:10}
\textsc{Heckman, J.}, \textsc{Moon, S.~H.}, \textsc{Pinto, R.},
  \textsc{Savelyev, P.} and \textsc{Yavitz, A.} (2010).
\newblock Analyzing social experiments as implemented: A reexamination of the
  evidence from the highscope perry preschool program.
\newblock \textit{Quantitative economics}, \textbf{1} 1--46.

\bibitem[{Heckman et~al.(2016)Heckman, Humphries and
  Veramendi}]{heckman/etal:16}
\textsc{Heckman, J.~J.}, \textsc{Humphries, J.~E.} and \textsc{Veramendi, G.}
  (2016).
\newblock Dynamic treatment effects.
\newblock \textit{Journal of econometrics}, \textbf{191} 276--292.

\bibitem[{Heckman and Pinto(2018)}]{heckman/pinto:18}
\textsc{Heckman, J.~J.} and \textsc{Pinto, R.} (2018).
\newblock Unordered monotonicity.
\newblock \textit{Econometrica}, \textbf{86} 1--35.

\bibitem[{Heckman et~al.(2006)Heckman, Urzua and Vytlacil}]{heckman/etal:06}
\textsc{Heckman, J.~J.}, \textsc{Urzua, S.} and \textsc{Vytlacil, E.} (2006).
\newblock Understanding instrumental variables in models with essential
  heterogeneity.
\newblock \textit{The review of economics and statistics}, \textbf{88}
  389--432.

\bibitem[{Heckman et~al.(2008)Heckman, Urzua and Vytlacil}]{heckman/etal:08}
\textsc{Heckman, J.~J.}, \textsc{Urzua, S.} and \textsc{Vytlacil, E.} (2008).
\newblock Instrumental variables in models with multiple outcomes: the general
  unordered case.
\newblock \textit{Annales d'Économie et de Statistique} 151--174.

\bibitem[{Heckman and Vytlacil(2005)}]{heckman/vytlacil:05}
\textsc{Heckman, J.~J.} and \textsc{Vytlacil, E.} (2005).
\newblock Structural equations, treatment effects, and econometric policy
  evaluation 1.
\newblock \textit{Econometrica}, \textbf{73} 669--738.

\bibitem[{Heckman and Vytlacil(1999)}]{heckman/vytlacil:99}
\textsc{Heckman, J.~J.} and \textsc{Vytlacil, E.~J.} (1999).
\newblock Local instrumental variables and latent variable models for
  identifying and bounding treatment effects.
\newblock \textit{Proceedings of the national Academy of Sciences}, \textbf{96}
  4730--4734.

\bibitem[{Hendren(2016)}]{hendren:16}
\textsc{Hendren, N.} (2016).
\newblock The policy elasticity.
\newblock \textit{Tax Policy and the Economy}, \textbf{30} 51--89.

\bibitem[{Hull(2020)}]{hull:20}
\textsc{Hull, P.} (2020).
\newblock Estimating hospital quality with quasi-experimental data.
\newblock \textit{Available at SSRN 3118358}.

\bibitem[{Imbens and Angrist(1994)}]{imbens/angrist:94}
\textsc{Imbens, G.~W.} and \textsc{Angrist, J.~D.} (1994).
\newblock Identification and estimation of local average treatment effects.
\newblock \textit{Econometrica}, \textbf{62} 467--475.

\bibitem[{Kamat(2021)}]{kamat:21}
\textsc{Kamat, V.} (2021).
\newblock Identifying the effects of a program offer with an application to
  head start.
\newblock \textit{arXiv preprint arXiv:1711.02048}.

\bibitem[{Kamat and Norris(2022)}]{kamat/norris:22}
\textsc{Kamat, V.} and \textsc{Norris, S.} (2022).
\newblock Estimating welfare effects in a nonparametric choice model: The case
  of school vouchers.
\newblock \textit{arXiv preprint arXiv:2002.00103}.

\bibitem[{Kamat et~al.(2022)Kamat, Norris and Pecenco}]{kamat/etal:22}
\textsc{Kamat, V.}, \textsc{Norris, S.} and \textsc{Pecenco, M.} (2022).
\newblock Examiner designs with multiple treatments: Conviction, incarceration
  and policy effects.
\newblock Tech. rep.

\bibitem[{Kirkeboen et~al.(2016)Kirkeboen, Leuven and
  Mogstad}]{kirkeboen/etal:16}
\textsc{Kirkeboen, L.~J.}, \textsc{Leuven, E.} and \textsc{Mogstad, M.} (2016).
\newblock Field of study, earnings, and self-selection.
\newblock \textit{The Quarterly Journal of Economics}, \textbf{131} 1057--1111.

\bibitem[{Kline and Walters(2016)}]{kline/walters:16}
\textsc{Kline, P.} and \textsc{Walters, C.~R.} (2016).
\newblock Evaluating public programs with close substitutes: The case of head
  start.
\newblock \textit{The Quarterly Journal of Economics}, \textbf{131} 1795--1848.

\bibitem[{Kline and Walters(2019)}]{kline/walters:19}
\textsc{Kline, P.} and \textsc{Walters, C.~R.} (2019).
\newblock On heckits, late, and numerical equivalence.
\newblock \textit{Econometrica}, \textbf{87} 677--696.

\bibitem[{Lee and Salani{\'e}(2018)}]{lee/salanie:18}
\textsc{Lee, S.} and \textsc{Salani{\'e}, B.} (2018).
\newblock Identifying effects of multivalued treatments.
\newblock \textit{Econometrica}, \textbf{86} 1939--1963.

\bibitem[{Lee and Salani{\'e}(2023)}]{lee/salanie:23}
\textsc{Lee, S.} and \textsc{Salani{\'e}, B.} (2023).
\newblock Filtered and unfiltered treatment effects with targeting instruments.
\newblock \textit{arXiv preprint arXiv:2007.10432}.

\bibitem[{Manski(1997)}]{manski:97}
\textsc{Manski, C.~F.} (1997).
\newblock Monotone treatment response.
\newblock \textit{Econometrica: Journal of the Econometric Society} 1311--1334.

\bibitem[{Manski(2008)}]{manski:08}
\textsc{Manski, C.~F.} (2008).
\newblock \textit{Identification for prediction and decision}.
\newblock Harvard University Press.

\bibitem[{Manski and Pepper(2000)}]{manski/pepper:00}
\textsc{Manski, C.~F.} and \textsc{Pepper, J.~V.} (2000).
\newblock Monotone instrumental variables: With an application to the returns
  to schooling.
\newblock \textit{Econometrica}, \textbf{68} 997--1010.

\bibitem[{Manski and Pepper(2009)}]{manski/pepper:09}
\textsc{Manski, C.~F.} and \textsc{Pepper, J.~V.} (2009).
\newblock {More on monotone instrumental variables}.
\newblock \textit{The Econometrics Journal}, \textbf{12} S200--S216.

\bibitem[{Manski and Pepper(2018)}]{manski/pepper:18}
\textsc{Manski, C.~F.} and \textsc{Pepper, J.~V.} (2018).
\newblock How do right-to-carry laws affect crime rates? coping with ambiguity
  using bounded-variation assumptions.
\newblock \textit{Review of Economics and Statistics}, \textbf{100} 232--244.

\bibitem[{Mogstad et~al.(2018)Mogstad, Santos and
  Torgovitsky}]{mogstad/etal:18}
\textsc{Mogstad, M.}, \textsc{Santos, A.} and \textsc{Torgovitsky, A.} (2018).
\newblock Using instrumental variables for inference about policy relevant
  treatment parameters.
\newblock \textit{Econometrica}, \textbf{86} 1589--1619.

\bibitem[{Mogstad et~al.(2021)Mogstad, Torgovitsky and
  Walters}]{mogstad/etal:21}
\textsc{Mogstad, M.}, \textsc{Torgovitsky, A.} and \textsc{Walters, C.~R.}
  (2021).
\newblock The causal interpretation of two-stage least squares with multiple
  instrumental variables.
\newblock \textit{American Economic Review}, \textbf{111} 3663--98.

\bibitem[{Mogstad et~al.(2024)Mogstad, Torgovitsky and
  Walters}]{mogstad/etal:24}
\textsc{Mogstad, M.}, \textsc{Torgovitsky, A.} and \textsc{Walters, C.~R.}
  (2024).
\newblock Policy evaluation with multiple instrumental variables.
\newblock \textit{Journal of Econometrics}, \textbf{243} 105718.

\bibitem[{Mountjoy(2022)}]{mountjoy:22}
\textsc{Mountjoy, J.} (2022).
\newblock Community colleges and upward mobility.
\newblock \textit{Available at SSRN 3373801}.

\bibitem[{Navjeevan et~al.(2023)Navjeevan, Pinto and
  Santos}]{navjeevan/etaL:23}
\textsc{Navjeevan, M.}, \textsc{Pinto, R.} and \textsc{Santos, A.} (2023).
\newblock Identification and estimation in a class of potential outcomes
  models.
\newblock \textit{arXiv preprint arXiv:2310.05311}.

\bibitem[{Nelsen(2007)}]{nelsen:07}
\textsc{Nelsen, R.~B.} (2007).
\newblock \textit{An introduction to copulas}.
\newblock Springer science \& business media.

\bibitem[{Pinto(2021)}]{pinto:19}
\textsc{Pinto, R.} (2021).
\newblock Beyond intention to treat: Using the incentives in moving to
  opportunity to identify neighborhood effects.
\newblock \textit{NBER Working Paper}.

\bibitem[{Poirier(1980)}]{poirier:80}
\textsc{Poirier, D.~J.} (1980).
\newblock Partial observability in bivariate probit models.
\newblock \textit{Journal of econometrics}, \textbf{12} 209--217.

\bibitem[{Puma et~al.(2010)Puma, Bell, Cook, Heid, Shapiro, Broene, Jenkins,
  Fletcher, Quinn, Friedman et~al.}]{puma_etal:10}
\textsc{Puma, M.}, \textsc{Bell, S.}, \textsc{Cook, R.}, \textsc{Heid, C.},
  \textsc{Shapiro, G.}, \textsc{Broene, P.}, \textsc{Jenkins, F.},
  \textsc{Fletcher, P.}, \textsc{Quinn, L.}, \textsc{Friedman, J.}
  \textsc{et~al.} (2010).
\newblock Head start impact study. final report.
\newblock \textit{Administration for Children \& Families}.

\bibitem[{Rose and Shem-Tov(2021)}]{rose/shem:21}
\textsc{Rose, E.~K.} and \textsc{Shem-Tov, Y.} (2021).
\newblock How does incarceration affect reoffending? estimating the
  dose-response function.
\newblock \textit{Journal of Political Economy}, \textbf{129} 3302--3356.

\bibitem[{Tebaldi et~al.(2023)Tebaldi, Torgovitsky and Yang}]{tebaldi/etal:23}
\textsc{Tebaldi, P.}, \textsc{Torgovitsky, A.} and \textsc{Yang, H.} (2023).
\newblock Nonparametric estimates of demand in the california health insurance
  exchange.
\newblock \textit{Econometrica}, \textbf{91} 107--146.

\bibitem[{Train(2009)}]{train:09}
\textsc{Train, K.~E.} (2009).
\newblock \textit{Discrete choice methods with simulation}.
\newblock Cambridge university press.

\bibitem[{Tsuda(2023)}]{tsuda:23}
\textsc{Tsuda, T.} (2023).
\newblock Identification of the marginal treatment effect with multivalued
  treatments.
\newblock \textit{arXiv preprint arXiv:2209.11444}.

\bibitem[{Vytlacil(2002)}]{vytlacil:02}
\textsc{Vytlacil, E.} (2002).
\newblock Independence, monotonicity, and latent index models: An equivalence
  result.
\newblock \textit{Econometrica}, \textbf{70} 331--341.

\end{thebibliography}
